\documentstyle[preprint,aps,eqsecnum]{revtex}

\tighten

\begin{document}

\draft

\preprint{DOE/ER/40561-238-INT95-18-02}

\title{
                    Mean-field description of ground-state properties
                                 of drip-line nuclei: \\
                             Pairing and continuum effects
         }
\author  {
           J. Dobaczewski,$^{1,2}$
           W. Nazarewicz,$^{1-4}$
       and T.R. Werner\,$^{1,5}$
}
\address {
           $^1$Institute of Theoretical Physics, Warsaw University,
               ul. Ho\.za 69, 00-681 Warsaw, Poland\\[1mm]
           $^2$Institute for Nuclear Theory, University of Washington,\\
Seattle, Washington  98195, U.S.A.\\[1mm]
           $^3$Department of Physics \& Astronomy, University of Tennessee,\\
               Knoxville, Tennessee 37996, U.S.A.\\[1mm]
           $^4$Physics Division,  Oak Ridge National Laboratory, \\
               P.O. Box 2008, Oak Ridge, Tennessee 37831, U.S.A.\\[1mm]
           $^5$Joint Institute for Heavy-Ion Research, Oak Ridge,
               Tennessee 37831, USA
}
\author  {     and \\[1mm]
           J.F. Berger,$^6$
           C.R. Chinn,$^7$
       and J. Decharg\'e\,$^6$
}
\address {
           $^6$Centre d'Etudes de Bruy\`eres-le-Ch\^atel, B.P. No. 12,
               91680 Bruy\`eres-le-Ch\^atel, France\\[1mm]
           $^7$Department of Physics \& Astronomy, Vanderbilt University,
               Nashville, Tennessee 37235, U.S.A.
}

\maketitle

\begin{abstract}
Ground-state properties of exotic even-even nuclei with extreme
neutron-to-proton ratios are described in the framework of the
self-consistent mean-field theory with pairing formulated in
coordinate space.  This theory properly accounts for the
influence of the particle continuum, which is particularly
important for weakly bound systems. The pairing properties of
nuclei far from stability are studied with several interactions
emphasizing different aspects, such as the range and density
dependence of the effective interaction.  Measurable
consequences of spatially extended pairing fields are presented,
and the sensitivity of the theoretical predictions to model
details is discussed.
\end{abstract}


\narrowtext

\section{Introduction}
\label{sec1}

One of the most exciting challenges in today's nuclear structure
is the physics of exotic nuclei far from the line of
$\beta$-stability.  What makes this subject particularly
interesting (and difficult) is the unique combination of weak
binding and the proximity of the particle continuum, both
implying the large diffuseness of the nuclear surface and
extreme spatial dimensions characterizing the outermost nucleons
\cite{[Roe92],[Mue93],[Han93],[Naz95]}.

{}For the weakly bound nuclei the decay channels have to be
considered explicitly.  Due to the virtual scattering of
nucleons from bound orbitals to unbound scattering states, the
traditional shell-model technology becomes inappropriate. The
proper tool is the continuum shell model \cite{[Fan61],[Phi77]}
which correctly accounts for the coupling to resonances; the
single-particle basis of the continuum shell model consists of
both bound and unbound states.  The explicit coupling between
bound states and the continuum and the presence of low-lying
low-$\ell$ scattering states invites strong interplay between
various aspects of nuclear structure and reaction theory.

Particularly exciting are new phenomena on the neutron-rich
side. Because neutrons do not carry an electric charge, the
neutron drip line is located very far from the valley of
$\beta$-stability. Consequently, the neutron drip-line systems
(i.e., those close to the  neutron drip line) are characterized
by unusually large $N$/$Z$ ratios.  The outer zone  of these
nuclei are expected to constitute essentially a new form of a
many-body system: low-density neutron matter (neutron halos and
skins).

Except for the lightest nuclei, the bounds of neutron stability
are not known experimentally.  Theoretically, because of their
sensitivity to various theoretical details (e.g., approximations
used, parameter values, interactions) predicted drip lines are
strongly model-dependent.  The placement of the one-neutron drip
line, defined by the condition $S_n(Z,N) = B_n(Z,N)-B_n(Z,N-1) =
0$, is solely determined by the {binding energy  difference}
between two neighboring isotopes.  Analogously, the vanishing
two-neutron separation energy, $S_{2n}(Z,N) =
B_n(Z,N)-B_n(Z,N-2)$, defines the position of the two-neutron
drip line.  Since experimental masses (binding energies) near
the neutron drip lines are unknown, in order to extrapolate far
from stability, the large-scale mass calculations are usually
used (see, e.g.,
\cite{[Hau88],[Mol95],[Abo95],[Smo93],[Dob95a]}).  However,
since their techniques and parameters are optimized to reproduce
known atomic masses, it is by no means obvious whether the
particle number dependence obtained from global calculations at
extreme values of $N/Z$ is correct.  Apart from strong
theoretical and experimental interest in nuclear physics aspects
of exotic nuclei, calculations for nuclei far from stability
have strong astrophysical implications, especially in the
context of  the r-process mechanism \cite{[How93],[Che95]}.

In previous work \cite{[Naz94]} several aspects of nuclear
structure at the limits of extreme isospin were  discussed by
means of the macroscopic-microscopic approach.  In the present
study, ground-state properties of drip-line systems and the
sensitivity of predictions to effective forces are investigated
by means of the self-consistent Hartree-Fock-Bogoliubov (HFB)
approach.

The paper is organized as follows.  Section~\ref{forces}
discusses the effective interactions employed in this study.
Since pairing correlations are crucial for the behavior of
drip-line systems, particular attention is paid to the
particle-particle (p-p, pairing) component of the  interaction.
After a short review of general properties of effective
pairing interactions, with emphasis on the density dependence,
the pairing forces investigated in our work, namely contact
forces (delta interaction, density-dependent delta interaction,
and Skyrme interaction) and the finite-range Gogny force, are
described.

The basic ingredients of the HFB formalism in the coordinate
representation (single-quasiparticle orbitals, time-reversal
symmetry, canonical states, and various densities)  are defined
in Sec.{\ }\ref{sec2a}.  In contrast to the
single-quasiparticle wave functions which often contain a
scattering (outgoing) component, canonical states
(Sec.{\ }\ref{sec2ab}) are always localized, even if they have
positive average energy.  The interpretation of particle and
(especially) pair densities in terms of single-particle and
correlation probabilities is given in Sec.{\ }\ref{sec2ac}.
This interpretation is essential when relating the calculated
HFB densities and fields  to various { experimental
observables}.

The structure of the HFB equations is analyzed in
Sec.{\ }\ref{sec2b}.  Here, various functions entering the equations
of motion (i.e., mass parameters and mean field potentials)
are introduced for both p-h and p-p channels
(Secs.{\ }\ref{sec2ba} and \ref{sec2bap}).

The advantage of using the coordinate-space HFB formalism for
weakly bound systems is that in this method the particle
continuum is treated properly. This important point is discussed
in detail in Sec.{\ }\ref{sec3}. In particular, the difference
between the single-particle Hartree-Fock (HF) spectra and
canonical HFB spectra (Sec.{\ }\ref{sec3d}), the asymptotic
properties of the HFB states (Sec.{\ }\ref{sec2bcd})
and densities (Sec.{\ }\ref{sec3c}), and the
effect of the pairing coupling to positive-energy states
(Sec.{\ }\ref{sec3e}) are carefully explained.

The robust predictions of the formalism for various experimental
observables (pairing gaps and pair transfer amplitudes, masses
and separation energies, radii, shell gaps, and shell structure)
are reviewed in Sec.{\ }\ref{sec4}, where experimental
fingerprints of  the surface-peaked pairing fields and the
quenching of shell effects far from stability are also given.
Section~\ref{sec6} contains the main conclusions of the paper.
The technical details (i.e., the form of a mean-field Gogny
Hamiltonian and the discussion of the energy cut-off in the
Skyrme model) are collected in the Appendices.

\section{Effective interactions in the p-p channel}
\label{forces}

The uniqueness of drip-line nuclei for studies of effective
interactions is due to the very special role played by the pairing force.
This is seen from approximate HFB relations between the Fermi
level, $\lambda$, pairing gap, $\Delta$, and the particle
separation energy, $S$ \cite{[Bei75a]}:
   \begin{equation}\label{Sn0}
   S  \approx  -\lambda - \Delta.
   \end{equation}
Since for drip-line nuclei $S$ is very small, $\lambda +
\Delta$$\approx$0. Consequently, the single-particle field
characterized by $\lambda$ (determined by the p-h component of
the effective interaction) and the pairing field $\Delta$
(determined by the p-p part of the effective interaction) are
equally important. In other words, contrary to the situation
encountered close to the line of beta stability, the pairing
component can no longer be treated as a {\em residual}
interaction; i.e., a small perturbation important only in the
neighborhood of the Fermi surface.

Surprisingly, rather little is known about the basic properties
of the p-p force.  In most calculations, the pairing Hamiltonian
has been approximated by the  state-independent seniority
pairing force, or schematic multipole pairing interaction
\cite{[Lan64]}. Such oversimplified forces, usually treated by
means of the BCS approximation, perform remarkably well when
applied to nuclei in the neighborhood of the stability valley
(where, as pointed out above,  pairing can be considered as a
small correction). As a result, considerable effort was devoted
in the past to optimizing the p-h part of the interaction, while
leaving the p-p component aside.

Up to now, the microscopic theory of the pairing interaction has
only seldom been applied in realistic calculations for finite
nuclei (see Ref.{\ }\cite{[Del95]} for a recent example).  A
``first-principle" derivation of pairing interaction from the
bare $NN$ force using the renormalization procedure ($G$-matrix
technique), still encounters many problems such as, e.g.,
treatment of core polarization \cite{[Kuc91],[Kad87]}.  Hence,
phenomenological pairing interactions are usually introduced.
Two important open questions asked in this context are: (i) the
role of finite range, and (ii) the importance of density
dependence.  Since the realistic effective interactions are
believed to have a finite range, the first question seems purely
academic. However, the remarkable success of zero-range Skyrme
forces suggests that, in many cases, the finite-range effect can
be mocked up by an explicit velocity dependence. To what extent
this is true for the pairing channel remains to be seen. One
obvious advantage of using finite-range forces is the automatic
cut-off of high-momentum components; for the zero-range forces
this is solved by restricting the pair scattering  to a limited
energy range and by an appropriate renormalization of the
pairing coupling constant (see Appendix{\ }\ref{appB}).

The answer to the question on the density dependence is  much
less clear.  Early  calculations \cite{[Bru60],[Eme60]} for
nuclear matter predicted a very weak $^1S_0$ pairing at the
saturation point ($k_F$=1.35\,fm$^{-1}$).  Consequently, it was
concluded that strong pairing correlations in finite nuclei had
to be due to interactions at the nuclear surface. This led to
the surface delta interaction (SDI) \cite{[Gre65]}, a highly
successful residual interaction between valence nucleons.  Of
course, the SDI is an extreme example of surface interaction.
More realistic density-dependent pairing forces are variants of
the density-dependent delta interaction (DDDI) introduced in
the Migdal theory of finite Fermi systems \cite{[Mig67]}.

Since the effective interactions commonly used in the HF
calculations are bound to be density dependent in order to
reproduce the compressibility of the infinite nuclear matter
\cite{[RS80]} (an explicit density dependence is also said to
account for three- and higher-body components of the
interaction),  it seems natural to introduce the density
dependence in the p-p channel as well \cite{[Boc67]}.

Interestingly, the presence (absence) of the density dependence in
the pairing channel has  consequences for the spatial properties
of pairing densities and fields.  As early recognized
\cite{[Sap65]}, the density-independent p-p force gives rise to a
pairing field that has a volume character.  For instance, the
commonly used contact delta interaction,
   \begin{equation}\label{DIDI}
   V^{\delta}(\bbox{r},\bbox{r}')
        = V_0 \delta(\bbox{r}-\bbox{r}'),
   \end{equation}
leads to volume pairing.  By adding a density-dependent
component, the pairing field becomes surface-peaked.  A simple
modification of force (\ref{DIDI}) is the DDDI
\cite{[Boc67],[Cha76],[Kad78]}
   \begin{equation}\label{DDDI}
      V^{\delta\rho}(\bbox{r},\bbox{r}') =
      V_0\delta(\bbox{r}-\bbox{r}')
      \left\{1-\left[\rho(\bbox{r})/\rho_c\right]^\gamma\right\},
   \end{equation}
where $\rho(\bbox{r})$ is the isoscalar nucleonic density, and
$V_0, \rho_c$ and $\gamma$ are constants. If $\rho_c$ is chosen
such that it is close to the saturation density,
$\rho_c$$\approx$$\rho(\bbox{r}=0)$, both the resulting pair
density  and the pairing potential $\Delta(\bbox{r})$
(see Secs.{\ }\ref{sec2ac} and \ref{sec2ba}) are small in the
nuclear interior.  By varying the magnitude of the
density-dependent term, the transition from volume pairing
[$\rho_c$$\gg$$\rho(0)$] to surface pairing can be probed.

What are the experimental arguments in favor of surface pairing?
Probably the strongest evidence is the odd-even staggering in
differential radii, explained in terms of the direct coupling
between the proton density and the neutron pairing tensor
\cite{[Zaw87],[Reg88],[Fay94],[Hor94]}.  Other experimental
observables which strongly reflect the spatial character of
pairing are the particle widths and energies of deep-hole states
\cite{[Bul80],[Bel87]}, strongly influenced by the
pairing-induced coupling to the particle continuum, and the pair
transfer form factors, directly reflecting the shape of the pair
density.  Because of strong surface effects, the properties of
weakly bound nuclei are sensitive to the density dependence of
pairing.  In particular,  the same type of force is used to
describe the spatial extension of loosely bound light systems
\cite{[Ber91],[Len91],[Ost92],[Boh93]}.  (The measurable
fingerprints of surface pairing in neutron-rich systems are
further discussed in Sec.{\ }\ref{sec4}.) In this context, it
is also worth mentioning that the self-consistent model with the
DDDI has recently been used to describe the nuclear charge radii
\cite{[Taj93b]} and the moments of inertia of superdeformed
nuclei \cite{[Ter95]}. In the latter case, the inclusion of a
density dependence in the p-p channel turned out to be crucial
for the reproduction of experimental data around $^{194}$Hg.

In a series of papers \cite{[Zve84],[Zve85],[Smi88]} the
Quasiparticle Lagrangian method (QLM) \cite{[Kho82]} based on
the single-particle Green function approach in the coordinate
representation \cite{[Shl75]} has been applied to the
description of nuclear superfluidity.  The resulting pairing
interaction, based on the Landau-Migdal ansatz \cite{[Mig67]},
has zero-range and contains two-body and three-body components,
thus leading to a density-dependent contact force similar to
that of Eq.{\ }(\ref{DDDI}). (Note that in the approximations
of Ref.{\ }\cite{[Zve84]}, the {\em neutron} pairing
interaction is proportional to the {\em proton} density and vice
versa.) However, in practical QLM calculations
\cite{[Zve84],[Zve85],[Smi88]}, a pure density-independent delta
force was used.

A better understanding of the density dependence of the nuclear pairing
interaction is important for theories of superfluidity in
neutron stars.  As pointed out in Ref.{\ }\cite{[Pet95]}, it
is impossible at present to deduce the magnitude of the pairing gaps
in neutron stars with sufficient accuracy. Indeed, calculations
of $^1S_0$ pairing gaps in pure neutron matter, or symmetric
nuclear matter, based on bare $NN$ interactions \cite{[Che93]}
suggest a strong dependence on the force used; in general, the
singlet-$S$ pairing is very small at the saturation point.  On
the other hand, nuclear matter calculations with an effective
finite-range interaction, namely the Gogny force \cite{[Kuc89]},
yield rather large values of the pairing gap at saturation
($\Delta$$\simeq$0.7\,MeV).  (For relativistic HFB calculations
for symmetric nuclear matter, see Ref.{\ }\cite{[Kuc91]}. The
pairing properties of the Skyrme force in nuclear matter were
investigated in Ref.{\ }\cite{[Tak94]}. See also
Ref.{\ }\cite{[Bal95]} for schematic calculations of pairing
properties in nuclear matter based on the Green function method
with a contact interaction, and
Ref.{\ }\cite{[Deb94]} for a semi-classical description
of neutron superfluidity in neutron  stars using the Gogny force.)

In this study, several self-consistent models based upon the HFB
approaches are used.  The effective interactions employed, and
other model parameters, are briefly discussed below.

The spherical HFB-Skyrme calculations have been carried out in
spatial coordinates following the method introduced in
Ref.{\ }\cite{[Dob84]} and discussed in detail in
Secs.{\ }\ref{sec2a}-\ref{sec3}.  Several effective Skyrme
interactions are investigated.  These are:  (i) the Skyrme
parametrization SkP introduced in Ref.{\ }\cite{[Dob84]} (SkP
has exactly the same form in the particle-hole (p-h) and pairing
channels); (ii) Skyrme interaction SkP$^{\delta}$ of
Ref.{\ }\cite{[Dob95c]} (in the p-h channel, this force is the
SkP Skyrme parametrization, while its pairing component is given
by delta interaction, Eq.{\ }(\ref{DIDI}); (iii) the Skyrme
interaction SkP$^{\delta\rho}$ of Ref.{\ }\cite{[Dob95c]} (in
the p-h channel, this force is the SkP Skyrme parametrization,
while its pairing component is given by Eq.{\ }(\ref{DDDI});
(iv) the force SIII$^{\delta}$ (in the p-h channel, this is the
SIII Skyrme parametrization \cite{[Bei75]}; its pairing
component is given by the delta force of
Ref.{\ }\cite{[Dob95c]}); (v) the force SkM$^{\delta}$ (in the
p-h channel, this is the SkM$^*$ Skyrme parametrization
\cite{[Bar82]}, and its pairing part is given by the delta force
with the parameters of Ref.{\ }\cite{[Dob95c]}).

Apart from other parameters, the above Skyrme forces differ in
their  values of the effective mass for symmetric nuclear
matter, $m^*/m$.  Namely, $m^*/m$ is 0.76, 0.79, and 1 for SIII,
SkM$^*$, and SkP,  respectively.  All HFB-Skyrme results have
been obtained using the pairing phase space as determined in
Ref.{\ }\cite{[Dob84]} (see also discussion in
Appendix{\ }\ref{appB}).

A set of spherical HFB calculations has also been performed
using the finite-range density-dependent Gogny interaction D1S
of Ref.{\ }\cite{[Dec80]}.  In this effective interaction
\cite{[Gog73]} the central part consists of four  terms
parametrized with finite-range Gaussians (see
Appendix{\ }\ref{appA}).  Spin-orbit and density-dependent
terms of zero range are also included as in the Skyrme
parametrizations.  The  pairing field is calculated from the D1S
force, i.e., the same interaction is used for a microscopic
description for both the mean field and the pairing channels.
However, by a specific choice of the exchange contribution, the
pairing component of the D1S is density-independent.  It is also
interesting to note that the pairing component of the D1S is
repulsive at short distances and attractive at long ranges
\cite{[Kuc91],[Rum94]}.  For the D1S force, the effective mass
for infinite nuclear matter is $m^*/m$=0.70.

The parameters of the D1S interaction were chosen to reproduce
certain global properties of a set of spherical nuclei and of
nuclear matter \cite{[Ber89]}.  The HFB+Gogny results presented
here were obtained by expanding the HFB wave functions in a
harmonic oscillator basis containing up to 19 shells.

\section{Independent-quasiparticle states}
\label{sec2a}

The HFB approach is a variational method which uses
independent-quasiparticle states as trial wave functions.  These
states are particularly convenient when used in a variational
theory, because, due to the Wick theorem \cite{[Wic50]}, one can
easily calculate for them the average values of an arbitrary
many-body Hamiltonian. Even if the exact eigenstates of such a
Hamiltonian can be rather remote from any one of the
independent-quasiparticle states, one can argue \cite{[Bal92b]}
that one may obtain in this way fair estimates of at least
one-body observables.

An independent-quasiparticle state is defined as a vacuum of
quasiparticle operators which are linear combinations of
particle creation and annihilation operators. This linear
combination is called the Bogoliubov transformation
\cite{[Bog58],[Bog59a],[Bog59b]}.  According to the Thouless
theorem \cite{[Tho60]}, every independent-quasiparticle state
$|\Psi\rangle$, which is not orthogonal to the vacuum state
$|0\rangle$, i.e., $\langle0|\Psi\rangle$$\neq$0, can be
presented in the form
   \begin{equation}\label{eq101}
       |\Psi\rangle = \exp\left\{-\frac{1}{2}\sum_{\mu\nu}
                      Z^+_{\mu\nu}a^+_\mu a^+_\nu\right\} |0\rangle ,
   \end{equation}
where the Thouless matrix $Z$ is antisymmetric, $Z^+=-Z^*$, and
in general complex.  The phase of state (\ref{eq101}) is fixed
by the condition $\langle0|\Psi\rangle$=1; the norm is given by
$\langle\Psi|\Psi\rangle$=$\det(1+Z^+Z)^{1/2}$.  In the
following, state $|\Psi\rangle$ will  represent the
$I^\pi$=0$^+$ ground state of the even-even system.

We refer to standard textbooks \cite{[RS80]} for a discussion of
the properties of the Bogoliubov transformation. Here we start our
discussion from the trial wave function (\ref{eq101}) which is
parametrized by the matrix elements of $Z$.  This form of the
independent-quasiparticle state is very convenient in
variational applications because variations with respect to all
matrix elements $Z_{\mu\nu}$=$-Z_{\nu\mu}$ are independent of
one another.

Instead of using the matrix representation corresponding to a
set of single-particle creation operators $a^+_\mu$ numbered by
the discrete index $\mu$, one may use the spatial coordinate
representation.  This is particularly useful when discussing
spatial properties of the variational wave functions and the
coupling to the particle continuum. Therefore, in the following,
we shall consider the operators creating a particle in the space
point $\bbox{r}$ and having the projection of spin
$\sigma$=$\pm\frac{1}{2}$,
   \begin{equation}\label{eq103}
   a^+_{\bbox{r}\sigma} = \sum_\mu \psi^*_\mu(\bbox{r}\sigma)
   a^+_\mu,
   \end{equation}
where $\psi_\mu(\bbox{r}\sigma)$ is the wave function of the
$\mu$-th single-particle state. To simplify the following
expressions, we consider only one type of particle. A
generalization to systems described by a product of neutron and
proton wave functions is straightforward, while that involving
the mixing in the isospin degree of freedom is discussed in
Ref.{\ }\cite{[Roh95]}.

The inverse relation with respect to (\ref{eq103}) is given by
   \begin{equation}\label{eq104}
   a^+_\mu = \int \text{d}^3\bbox{r} \sum_\sigma
                \psi_\mu(\bbox{r}\sigma) a^+_{\bbox{r}\sigma}.
   \end{equation}
Equations (\ref{eq103}) and (\ref{eq104}) assume that the wave
functions $\psi_\mu(\bbox{r}\sigma)$ form an orthonormal and
complete set. In practical calculations, the basis has to be
truncated and the completeness is realized only approximately.
The choice of the single-particle wave functions used (size of
the set and, in particular, the  asymptotic behavior) is of
crucial importance to the phenomena discussed in this study.

In coordinate space, the Thouless state (\ref{eq101}) has
the form
   \begin{equation}\label{eq105}
       |\Psi\rangle = \exp\left\{-\frac{1}{2}
                      \int \text{d}^3\bbox{r}\text{d}^3\bbox{r}'
                      \sum_{\sigma\sigma'}
                      Z^+(\bbox{r}\sigma,\bbox{r}'\sigma')
                      a^+_{\bbox{r}\sigma} a^+_{\bbox{r}'\sigma'}
                      \right\} \! |0\rangle ,
   \end{equation}
and is defined by the antisymmetric complex function,
$Z^+(\bbox{r}\sigma,\bbox{r}'\sigma')$=
$-Z^*(\bbox{r}\sigma,\bbox{r}'\sigma')$, of space-spin
coordinates.  Already, at this point, we see that any variational
method employing an attractive effective interaction
for a {\em bound finite system} must lead
to functions which are localized in space,
   \begin{equation}\label{eq106}
                      \lim_{|\bbox{r}|\rightarrow\infty}
                      Z(\bbox{r}\sigma,\bbox{r}'\sigma')=0,
                \mbox{\hspace{3ex}for any~} \bbox{r}', \sigma',
                \mbox{and~} \sigma.
   \end{equation}
Recall that in the coordinate space, values of  the function
$Z(\bbox{r}\sigma,\bbox{r}'\sigma')$ at different space-spin
points are the variational parameters, and that any arbitrarily
small value of this function at large distance,
$|\bbox{r}|\rightarrow\infty$, would create at this point a
non-zero probability density. Whether this would be
energetically favorable depends upon the number of particles in
the system and on the interaction used in the variational
method. Apart from exotic phenomena such as halos, and apart
from infinite matter such as in the neutron-star crust, we
assume that the attractiveness of the interaction always favors
compact, localized probability densities, and hence we require
the localization condition (\ref{eq106}) for the variational
parameters $Z(\bbox{r}\sigma,\bbox{r}'\sigma')$.

An expansion of the variational function
$Z(\bbox{r}\sigma,\bbox{r}'\sigma')$ in terms of the
single-particle wave functions is a straightforward consequence
of transformations (\ref{eq103}) and (\ref{eq104}),
   \begin{equation}\label{eq107}
       Z(\bbox{r}\sigma,\bbox{r}'\sigma')
         = \sum_{\mu\nu} \psi^*_\mu(\bbox{r} \sigma )Z_{\mu\nu}
                         \psi^*_\nu(\bbox{r}'\sigma') .
   \end{equation}
The localization condition, Eq.~(\ref{eq106}),  can, therefore,
be guaranteed in the most economic way by requiring that {\em
all} single-particle wave functions $\psi_\mu(\bbox{r}\sigma)$
vanish at large distances.  Of course, this is only a matter of
convenience and manageability, because any localized function
can be expanded in any complete basis.  It is, however, obvious
that such an expansion converges very slowly if the basis has
inappropriate asymptotic properties. For example, one can expect
that a plane-wave expansion of
$Z(\bbox{r}\sigma,\bbox{r}'\sigma')$ would require an infinite
number of basis states $\psi_\mu(\bbox{r}\sigma)$, and in
practice, any reduction to a finite basis would lead to serious
errors. A discussion pertaining to asymptotic properties of
functions in spatial coordinates, and the choice of an
appropriate single-particle basis, will be a pivotal point in
our study.

\subsection{Time-reversal}
\label{sec2aa}

The present study is entirely restricted to an analysis of
ground-state phenomena, and therefore, we use only time-even
variational independent-quasiparticle wave functions.  The
time-reversal operator can be represented as a product of the
spin-flip operator and the complex conjugation; i.e., $\hat
T$=$-i\hat\sigma_y\hat K$ \cite{[Mes61]}.  The explicit
time-reversed creation operators then have the form
   \begin{mathletters}\begin{eqnarray}
   \hat T^+a^+_{\bbox{r}\sigma}\hat T &=& -2\sigma a^+_{\bbox{r},-\sigma} ,
                                \label{eq108a} \\
   \hat T^+a^+_\mu\hat T &=& \int \text{d}^3\bbox{r} \sum_\sigma
                [2\sigma\psi^*_\mu(\bbox{r},-\sigma)] a^+_{\bbox{r}\sigma}.
                                \label{eq108b}
   \end{eqnarray}\end{mathletters}%
We now suppose that the set of basis states represented by the
creation operators $a^+_\mu$ is closed with respect to time
reversal, and that the state $\hat T^+a^+_\mu\hat T$ is actually
proportional (up to a phase factor $s_{\bar\mu}$=$-s_{\mu}$,
$|s_{\mu}|$=1) to another basis state denoted by a bar over the
Greek symbol, i.e.,
   \begin{mathletters}\begin{eqnarray}
   \hat T^+a^+_\mu\hat T &=& s_{\bar\mu}a^+_{\bar\mu},  \label{eq109a} \\
   s_{\bar\mu}\psi_{\bar\mu}(\bbox{r}\sigma)
     &=&  2\sigma\psi^*_\mu(\bbox{r},-\sigma).  \label{eq109b}
   \end{eqnarray}\end{mathletters}%
In this way, the single-particle basis is assumed to be composed
of pairs of time-reversed states denoted by indices $\mu$ and
$\bar\mu$.  In what follows, we use the convention that
$\bar{\bar\mu}$$\equiv$$\mu$, and that the sums over either
$\mu$ or $\bar\mu$ are always performed over {\em all} basis
states.  The phase factors $s_{\mu}$ depend on relative phases
chosen for the $\mu$-th and $\bar\mu$-th states of the basis; it
is convenient to keep them unspecified in all theoretical
formulae and to make a definite suitable choice of the phase
convention only in a specific final application.

\subsection{Canonical basis}
\label{sec2ab}

A requirement of the time-reversal symmetry of the quasiparticle vacuum
(\ref{eq101}) or (\ref{eq105}),
$\hat T|\Psi\rangle$=$|\Psi\rangle$, leads to the following
conditions:
   \begin{mathletters}\begin{eqnarray}
   Z_{\mu\nu} &=& s^*_{\mu}s^*_{\nu}Z^*_{\bar\mu\bar\nu},
                              \label{eq110a} \\
   Z(\bbox{r}\sigma,\bbox{r}'\sigma')
     &=&  4\sigma\sigma'Z^*(\bbox{r},-\sigma,\bbox{r}',-\sigma').
                              \label{eq110b}
   \end{eqnarray}\end{mathletters}%
These properties allow the introduction of more suitable forms of
$Z_{\mu\nu}$ and $Z(\bbox{r}\sigma,\bbox{r}'\sigma')$; namely,
   \begin{mathletters}\begin{eqnarray}
   \tilde Z_{\mu\nu} &:=& s_{\mu}Z_{\bar\mu\nu} ,
                              \label{eq111a} \\
   \tilde Z(\bbox{r}\sigma,\bbox{r}'\sigma')
     &:=&  2\sigma Z(\bbox{r},-\sigma,\bbox{r}'\sigma') .
                              \label{eq111b}
   \end{eqnarray}\end{mathletters}%
The matrix $\tilde Z_{\mu\nu}$ and the function $\tilde
Z(\bbox{r}\sigma,\bbox{r}'\sigma')$ are both time-even and
hermitian,
   \begin{mathletters}\begin{eqnarray}
   \tilde Z^*_{\mu\nu} &=& \tilde Z_{\nu\mu} ,
                              \label{eq112a} \\
   \tilde Z^*(\bbox{r}\sigma,\bbox{r}'\sigma')
          &=& \tilde Z(\bbox{r}'\sigma',\bbox{r}\sigma) ,
                              \label{eq112b}
   \end{eqnarray}\end{mathletters}%
and therefore they can be considered as usual operators in the
corresponding Hilbert spaces. In particular, the function
$\tilde Z^*(\bbox{r}\sigma,\bbox{r}'\sigma')$ can be
diagonalized by solving the following integral eigenequation:
   \begin{equation}\label{eq113}
                      \int \text{d}^3\bbox{r}' \sum_{\sigma'}
                      \tilde Z(\bbox{r}\sigma,\bbox{r}'\sigma')
                      \breve\psi_{\mu}(\bbox{r}'\sigma')
              = z_\mu \breve\psi_{\mu}(\bbox{r}\sigma),
   \end{equation}
where $z_\mu$ are real eigenvalues, $z_\mu$=$z_{\bar\mu}$.  The
eigenfunctions $\breve\psi_{\mu}(\bbox{r}\sigma)$ form the
single-particle basis, usually referred to as the {\em canonical
basis}. Canonical states, together with the eigenvalues $z_\mu$,
completely define the quasiparticle vacuum $|\Psi\rangle$.
(Here and in the following we use the checked symbols, e.g.,
$\breve\psi_{\mu}$ and $\breve{a}^+_{\mu}$, to denote objects
pertaining to the canonical basis.)

Two important remarks concerning the canonical basis are now in
order.  First, the localization condition (\ref{eq106}) directly
results in the fact that {\em all} canonical-basis
single-particle wave functions
$\breve\psi_{\mu}(\bbox{r}\sigma)$ are localized in space; i.e.,
vanish at large distances, $|\bbox{r}|\rightarrow\infty$.
Therefore, as discussed previously, a choice of the localized
wave functions for the basis states $\psi_{\mu}(\bbox{r}\sigma)$
may allow for a rapid convergence in the expansion
   \begin{mathletters}\begin{eqnarray}
   \breve\psi_{\mu}(\bbox{r}\sigma) &=& \sum_\nu D_{\nu\mu}
       \psi_{\nu}(\bbox{r}\sigma) , \label{eq114a} \\
    \breve a^+_{\mu}                 &=& \sum_\nu D_{\nu\mu}
        a^+_{\nu}                 . \label{eq114b}
   \end{eqnarray}\end{mathletters}%
Second, since $\tilde Z(\bbox{r}\sigma,\bbox{r}'\sigma')$ and
$\tilde Z_{\mu\nu}$ are related by
[cf.{\ }Eq.{\ }(\ref{eq107})]
   \begin{equation}\label{eq115}
       \tilde Z(\bbox{r}\sigma,\bbox{r}'\sigma')
         = \sum_{\mu\nu} \psi_\mu(\bbox{r} \sigma )\tilde Z_{\mu\nu}
                         \psi^*_\nu(\bbox{r}'\sigma') ,
   \end{equation}
a diagonalization of
$\tilde{Z}(\bbox{r}\sigma,\bbox{r}'\sigma')$,
Eq.{\ }(\ref{eq113}), is equivalent to a diagonalization of
$\tilde Z_{\mu\nu}$,
   \begin{equation}\label{eq117}
   \sum_\nu\tilde Z_{\mu\nu}D_{\nu\tau} = z_{\tau}D_{\mu\tau}.
   \end{equation}
Therefore, in the canonical basis,
the Thouless state (\ref{eq101}) acquires the well-known
separable BCS-like  form
   \begin{eqnarray}
       |\Psi\rangle &=& \exp\left\{\frac{1}{2}\sum_{\mu\nu}
                        \tilde Z_{\mu\nu}s_\mu a^+_{\bar\mu} a^+_\nu\right\}
                        |0\rangle
                                            \nonumber \\[1ex]
                    &=& \exp\left\{\sum_{\nu>0}
                        z_{\nu}s_\nu \breve a^+_{\bar\nu}
                                     \breve a^+_\nu\right\}
                        |0\rangle
                                            \nonumber \\
                    &=& \prod_{\nu>0}\left(1+
                        z_{\nu}s_\nu \breve a^+_{\bar\nu}
                                     \breve a^+_\nu\right)
                        |0\rangle, \label{eq116}
   \end{eqnarray}
where the symbol $\nu>0$ denotes the sum over one-half of the
basis states with only one state (either one) of each
time-reversed pair ($\nu,\bar\nu$) included, and
$\breve{a}^+_\nu$ is the creation operator in the canonical
basis.

\subsection{Density matrices and the correlation probability}
\label{sec2ac}

According to the Wick theorem \cite{[Wic50],[RS80]} for the
independent-quasiparticle state, Eqs.{\ }(\ref{eq101}) or
(\ref{eq105}), an average value of any operator can be expressed
through average values of bifermion operators,
   \begin{mathletters}\begin{eqnarray}
   \rho(\bbox{r}\sigma,\bbox{r}'\sigma') &=&
   \langle\Psi|a^+_{\bbox{r}'\sigma'}a_{\bbox{r}\sigma}|\Psi\rangle,
                                        \label{eq118a} \\
   \tilde\rho(\bbox{r}\sigma,\bbox{r}'\sigma') &=& -2\sigma'
   \langle\Psi|a_{\bbox{r}',-\sigma'}a_{\bbox{r}\sigma}|\Psi\rangle.
                                        \label{eq118b}
   \end{eqnarray}\end{mathletters}%
{}Functions $\rho(\bbox{r}\sigma,\bbox{r}'\sigma')$ and
$\tilde\rho(\bbox{r}\sigma,\bbox{r}'\sigma')$ are called the
particle and pairing density matrices, respectively. For a
time-reversal invariant state $|\Psi\rangle$, both density
matrices are time-even and hermitian:
   \begin{mathletters}\label{7ab}\begin{eqnarray}
   \rho(\bbox{r}  \sigma, \bbox{r}'  \sigma') &=& 4\sigma \sigma'
   \rho(\bbox{r} -\sigma, \bbox{r}' -\sigma')^{*},
                                        \label{7} \\
   \tilde\rho(\bbox{r}  \sigma, \bbox{r}'  \sigma') &=& 4\sigma \sigma'
   \tilde\rho(\bbox{r} -\sigma, \bbox{r}' -\sigma')^{*}.
                                        \label{7b}
   \end{eqnarray}\end{mathletters}%
Therefore, the pairing density matrix
$\tilde\rho(\bbox{r}\sigma,\bbox{r}'\sigma')$ is more convenient
to use than the standard pairing tensor
$\kappa(\bbox{r}\sigma,\bbox{r}'\sigma')$ \cite{[RS80]},
   \begin{equation}\label{eq119}
   \kappa(\bbox{r}\sigma,\bbox{r}'\sigma') =
   2\sigma'\tilde\rho(\bbox{r}\sigma,\bbox{r}',-\sigma'),
   \end{equation}
which is an antisymmetric function of the space-spin arguments.

The formulae expressing $\rho(\bbox{r}\sigma,\bbox{r}'\sigma')$
and $\tilde\rho(\bbox{r}\sigma,\bbox{r}'\sigma')$ in terms of
the function $\tilde Z(\bbox{r}\sigma,\bbox{r}'\sigma')$ can be
easily derived from those for the density matrix and the pairing
tensor \cite{[RS80]}, and they read
   \begin{mathletters}\label{eq120}\begin{eqnarray}
         \rho &=& (1+\tilde Z^2)^{-1}\tilde Z^2,   \label{eq120a} \\
   \tilde\rho &=& (1+\tilde Z^2)^{-1}\tilde Z.     \label{eq120b}
   \end{eqnarray}\end{mathletters}%
As a result, the density matrices obey the following relations:
   \begin{mathletters}\label{eq121}\begin{eqnarray}
         \tilde\rho\cdot\rho &=& \rho\cdot\tilde\rho ,  \label{eq121a} \\
   \rho\cdot\rho &+& \tilde\rho\cdot\tilde\rho = \rho . \label{eq121b}
   \end{eqnarray}\end{mathletters}%
In the above equations, the matrix multiplications and
inversions should be understood in the operator sense; i.e.,
they involve the integration over space and summation over spin
variables. For instance:
   \begin{equation}\label{eq124}
     (\tilde\rho\cdot\rho)(\bbox{r}\sigma,\bbox{r}'\sigma')
                 =    \int \text{d}^3\bbox{r}'' \sum_{\sigma''}
                      \tilde\rho(\bbox{r}\sigma,\bbox{r}''\sigma'')
                            \rho(\bbox{r}''\sigma'',\bbox{r}'\sigma').
   \end{equation}

{\em Local} HFB densities, i.e., the density matrices for equal
spatial arguments, $\bbox{r}'$=$\bbox{r}$, have very
well-defined physical interpretations. To see this, let us
assume that $\psi_{\bbox{x}s}(\bbox{r}\sigma)$ is a normalized
single-particle wave function (wave packet) concentrated in a
small volume $V_{\bbox{x}}$ around the point
$\bbox{r}$=$\bbox{x}$ and having the spin $s$=$\sigma$.  The
corresponding creation operator
   \begin{equation}\label{eq125}
   a^+_{\bbox{x}s} = \int \text{d}^3\bbox{r} \sum_\sigma
                \psi_{\bbox{x}s}(\bbox{r}\sigma) a^+_{\bbox{r}\sigma},
   \end{equation}
together with its hermitian conjugate, define the operator
   \begin{equation}\label{eq126}
     \hat N_{\bbox{x}s} =  a^+_{\bbox{x}s}a_{\bbox{x}s}  ,
   \end{equation}
which measures the number of particles in the vicinity of the
point $\bbox{x}$. Since
   \begin{equation}\label{eq127}
     \hat N_{\bbox{x}s}^2 = \hat N_{\bbox{x}s}   ,
   \end{equation}
$\hat N_{\bbox{x}s}$ can be regarded as a projection operator
which projects out the component of the many-body wave function
that contains one spin-$s$ fermion in the volume $V_{\bbox{x}}$.
Therefore, its average value gives {\em the probability to find
a particle with spin $s$ in this volume}:
   \begin{equation}\label{eq128}
     {\cal{P}}_{1}
      (\bbox{x}s) = \langle\Psi|\hat N_{\bbox{x}s}|\Psi\rangle =
        V_{\bbox{x}}\rho(\bbox{x}s,\bbox{x}s)   .
   \end{equation}

In a very similar way, the probability of finding a fermion in
$V_{\bbox{x}}$ having opposite spin can be obtained by
considering the time-reversed wave function
$2\sigma\psi^*_{\bbox{x}s}(\bbox{r},-\sigma)$,
cf.{\ }Eq.{\ }(\ref{eq108b}).  This gives
   \begin{equation}\label{eq129}
     {\cal{P}}_{1}
        (\bbox{x},-s) = \langle\Psi|\hat N_{\bbox{x},-s}|\Psi\rangle =
        V_{\bbox{x}}\rho(\bbox{x},-s,\bbox{x},-s)   .
   \end{equation}
Due to time-reversal symmetry, probabilities (\ref{eq128})
and (\ref{eq129}) are equal.

We may now ask the question, ``What is {\em the probability of
finding a pair of fermions with opposite spin projections} in the
volume $V_{\bbox{x}}$,  ${\cal{P}}_{2}(\bbox{x})$?". If one
considers two {\em independent} measurements, where in the first
one is found the spin-$s$ fermion, and in another one the
spin-($-s$) fermion, ${\cal{P}}_{2}(\bbox{x})$ is equal to the
product of individual probabilities; i.e.,
${\cal{P}}_{1}(\bbox{x},s){\cal{P}}_{1}(\bbox{x},-s)$.  On the
other hand, if one wants to find in $V_{\bbox{x}}$ both fermions
{\em simultaneously}, one should project out from $|\Psi\rangle$
a corresponding two-fermion component. In this case,
${\cal{P}}_{2}(\bbox{x})$ becomes the expectation value of the
product of the projection operators $\hat N_{\bbox{x}s}$ and
$\hat N_{\bbox{x},-s}$; i.e.,
${\cal{P}}_{2}(\bbox{x})=\langle\Psi|\hat N_{\bbox{x}s} \hat
N_{\bbox{x},-s}|\Psi\rangle$.  Using the Wick theorem, this
average value is
    \begin{equation}\label{eq132a}\begin{array}{rl}
    {\cal{P}}_{2}(\bbox{x})
     =&  V^2_{\bbox{x}}      \rho(\bbox{x}s,\bbox{x}s)
                            \rho(\bbox{x},-s,\bbox{x},-s) + \\
      &  V^2_{\bbox{x}}\tilde\rho(\bbox{x}s,\bbox{x}s)
                      \tilde\rho(\bbox{x},-s,\bbox{x},-s),
    \end{array}\end{equation}
or in terms of the time-even spin-averaged densities:
    \begin{equation}\label{eq132b}
    {\cal{P}}_{2}(\bbox{x})
     =  \frac{1}{4}V^2_{\bbox{x}}      \rho(\bbox{x})^2
     +  \frac{1}{4}V^2_{\bbox{x}}\tilde\rho(\bbox{x})^2,
    \end{equation}
for
   \begin{mathletters}\label{eq320}\begin{eqnarray}
            \rho(\bbox{r}) &=& \sum_\sigma
                   \rho(\bbox{r}\sigma,\bbox{r}\sigma),
                                               \label{eq320a} \\
      \tilde\rho(\bbox{r}) &=& \sum_\sigma
             \tilde\rho(\bbox{r}\sigma,\bbox{r}\sigma).
                                               \label{eq320b}
   \end{eqnarray}\end{mathletters}%
Since the first terms in Eqs.{\ }(\ref{eq132a}) and
(\ref{eq132b}) describe the probability of finding the two
fermions in independent measurements, the second terms in these
equations should be interpreted as the probability of finding
{\em the correlated pair} at point $\bbox{x}$.

The above arguments allow us to give a transparent physical
interpretation to the local HFB densities. Namely, as usual,
$\rho(\bbox{r})$ represents the probability density of finding a
particle at the given point.  On the other hand,
$\tilde\rho(\bbox{r})^2$ gives the correlation probability
density; i.e., the probability of finding a  pair of fermions
{\em in excess} of the probability of finding two uncorrelated
fermions.

It is important to note that kinematic conditions (\ref{eq121}),
which result from the fact that $|\Psi\rangle$ is an
independent-quasiparticle state, Eq.{\ }(\ref{eq105}), {\em do
not}  directly constrain the local values of the particle and
pairing density matrices.  In particular, there is no obvious
kinematic relation between the probability of finding two
independent particles at a given point of space, and the
probability of finding a correlated pair at the same point.  In
particular, the first one can be small, while the second can be
large (see discussion in Secs.{\ }\ref{sec2aca} and
\ref{sec3c}).  This result means that in such a situation the
experiments probing the presence of two particles will always
find these two particles as correlated pairs without a
``background'' characteristic of two independent particles.

Relations (\ref{eq120}) imply that all three functions:
$\tilde Z(\bbox{r}\sigma,\bbox{r}'\sigma')$,
$\rho(\bbox{r}\sigma,\bbox{r}'\sigma')$, and
$\tilde\rho(\bbox{r}\sigma,\bbox{r}'\sigma')$
are diagonal in the canonical basis, cf.{\ }Eq.{\ }(\ref{eq113}).
Using the standard
notation for the eigenvalues of $\rho$ and $\tilde\rho$, one obtains
   \begin{mathletters}\label{eq122}\begin{eqnarray}
                      \int \text{d}^3\bbox{r}' \sum_{\sigma'}
                      \rho(\bbox{r}\sigma,\bbox{r}'\sigma')
                      \breve\psi_{\mu}(\bbox{r}'\sigma')
          &=& v^2_\mu \breve\psi_{\mu}(\bbox{r}\sigma) ,  \label{eq122a} \\
                      \int \text{d}^3\bbox{r}' \sum_{\sigma'}
                      \tilde\rho(\bbox{r}\sigma,\bbox{r}'\sigma')
                      \breve\psi_{\mu}(\bbox{r}'\sigma')
      &=& u_\mu v_\mu \breve\psi_{\mu}(\bbox{r}\sigma) ,  \label{eq122b}
   \end{eqnarray}\end{mathletters}%
where the real factors $v_\mu$ and $u_\mu$ are given by
   \begin{equation}\label{eq123a}
    v_\mu = v_{\bar\mu} = \frac{z_\mu}{\sqrt{1+z^2_\mu}} \quad,\quad
    u_\mu = u_{\bar\mu} = \frac{  1  }{\sqrt{1+z^2_\mu}} .
   \end{equation}
A completeness of the canonical basis leads to standard
expressions for the density matrices:
   \begin{mathletters}\label{eq321}\begin{eqnarray}
                      \rho(\bbox{r}\sigma,\bbox{r}'\sigma')
          &=& \sum_\mu v^2_\mu     \breve\psi_{\mu}^*(\bbox{r}\sigma)
                                   \breve\psi_{\mu}(\bbox{r}'\sigma')
                                                  ,  \label{eq321a} \\
                      \tilde\rho(\bbox{r}\sigma,\bbox{r}'\sigma')
          &=& \sum_\mu u_\mu v_\mu \breve\psi_{\mu}^*(\bbox{r}\sigma)
                                   \breve\psi_{\mu}(\bbox{r}'\sigma')
                                                  ,  \label{eq321b}
   \end{eqnarray}\end{mathletters}%
Equation (\ref{eq122a}) represents the traditional definition of
the canonical states as the eigenstates of the HFB density
matrix.  It also shows that the canonical states are the {\em
natural states}
\cite{[Low55],[Van93],[Ant93],[Ant94],[Ant95],[Pol95]} for the
density matrix corresponding to the independent-quasiparticle
many-body state $|\Psi\rangle$, Eq.{\ }(\ref{eq118a}), and the
eigenvalues $v^2_\mu$ are the corresponding natural occupation
numbers.

One may now easily repeat the previous analysis of probabilities
of finding a particle, or a pair of particles, in the
canonical-basis single-particle state
$\breve\psi_{\mu}(\bbox{r}\sigma)$. The result, analogous to
Eqs.{\ }(\ref{eq132a}) and (\ref{eq132b}), is
${\cal{P}}_1(\mu)$=$v^2_\mu$, and ${\cal{P}}_2(\mu)$=$u^2_\mu
v^2_\mu$+$v^4_\mu$.  In this case, due to the normalization
condition $u^2_\mu$+$v^2_\mu$=1,
${\cal{P}}_1(\mu)$=${\cal{P}}_2(\mu)$. This result  means that
the particles in the canonical states with indices $\mu$ and
$\bar{\mu}$ are extremely correlated spatially; i.e., the
probability of finding the canonical pair, $u^2_\mu v^2_\mu$, is
directly dependent on the probability of finding two independent
canonical fermions, $v^4_\mu$.  However, as discussed above, a
similar direct  relation between ${\cal{P}}_1(\bbox{x})$ and
${\cal{P}}_2(\bbox{x})$ does not exist.  In particular
${\cal{P}}_1(\bbox{x})$$\ne$${\cal{P}}_2(\bbox{x})$.

\subsubsection{Examples of particle and pairing densities}
\label{sec2aca}

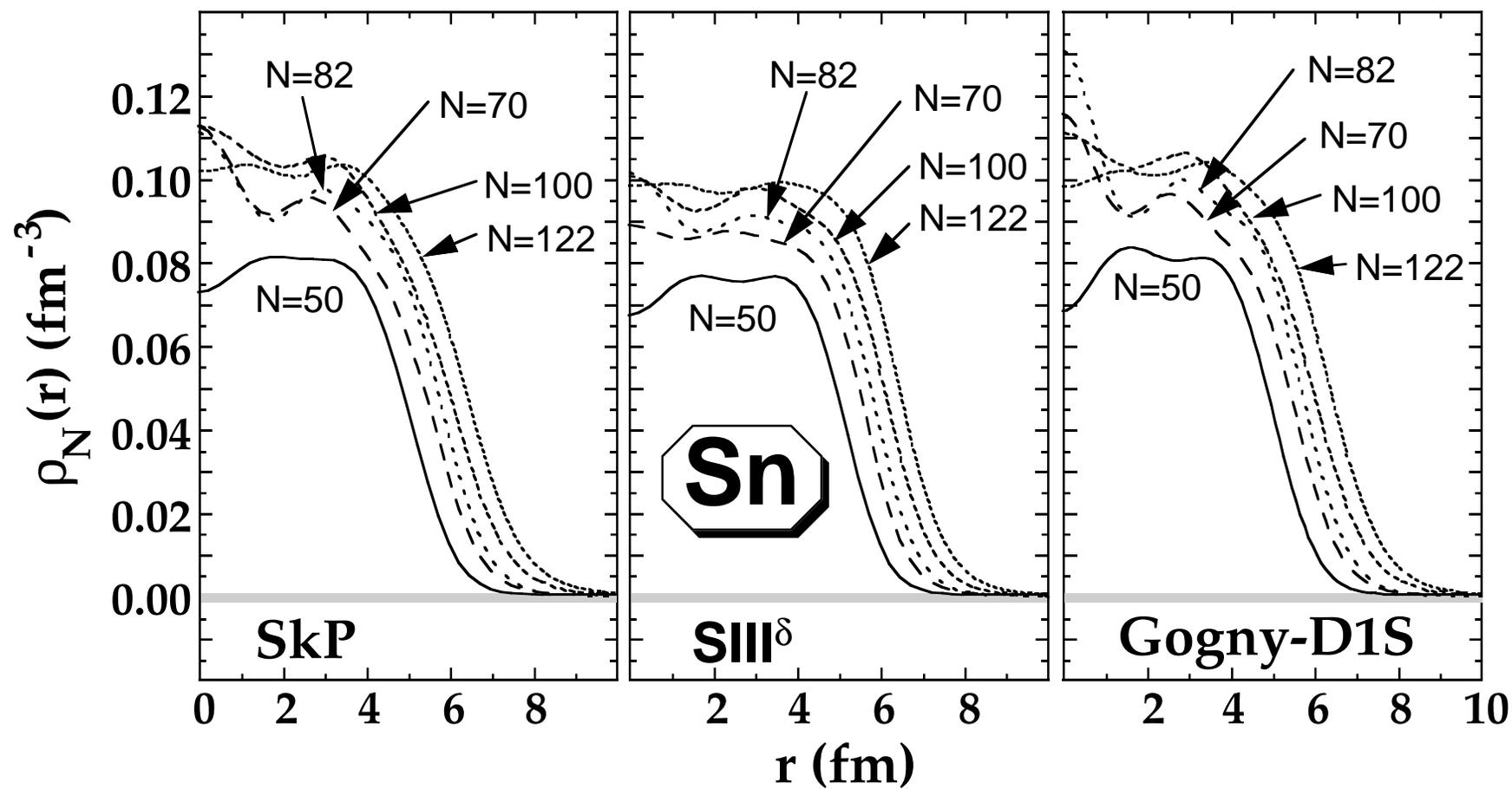
\begin{figure}[ht]
\caption[FF]{
Self-consistent spherical neutron densities $\rho_N(r)$
calculated with the SkP, SIII$^\delta$, and D1S interactions for
selected tin isotopes across the $\beta$-stability valley.
}
\label{FIG16}
\end{figure}

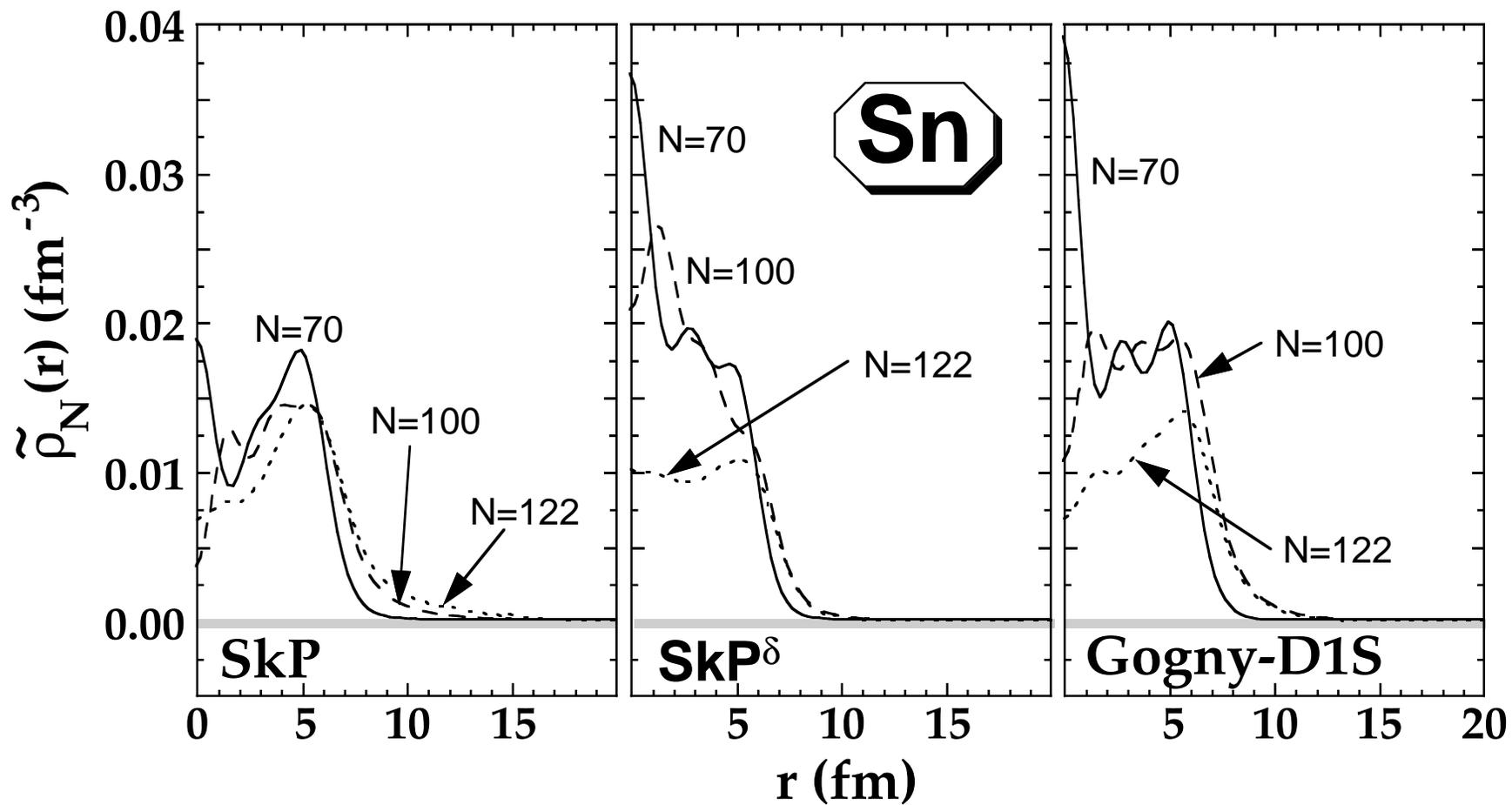
\begin{figure}[ht]
\caption[FF]{
Self-consistent spherical neutron pairing densities
$\tilde\rho_N(r)$ calculated with the SkP, SkP$^\delta$, and D1S
interactions for selected tin isotopes across the
$\beta$-stability valley.
}
\label{FIG17}
\end{figure}

{}Figures {\ }\ref{FIG16} and \ref{FIG17} display the particle
and pairing local spherical neutron HFB densities $\rho(r)$ and
$\tilde\rho(r)$, Eq.{\ }(\ref{eq320}), as functions of the
radial coordinate $r$=$|\bbox{r}|$.  Results are shown for
several tin isotopes across the stability valley.  {}For
particle densities, the results obtained with the SkP and
SkP$^\delta$ interactions are almost indistinguishable.
Therefore, Fig.{\ }\ref{FIG16} (middle panel) shows results
for the SIII$^\delta$ interaction. For pairing densities,
compared in Fig.{\ }\ref{FIG17} are results for SkP,
SkP$^\delta$, and D1S effective interactions.

The particle densities obtained with these three effective
interactions are qualitatively very similar. One can see that
adding  neutrons results in a simultaneous increase of the
central neutron density, and of the density in the surface
region. The relative magnitude of the two effects is governed by
a balance between the volume and the surface asymmetry energies
of effective interactions. Since all three forces  considered
have been fitted in a similar way to bulk nuclear properties,
including the isospin dependence, the resulting balance between
the volume and the surface isospin effects is similar. Of
course, this does not exclude some differences which are seen
when a more detailed comparison is carried out.

The pairing densities shown in Fig.{\ }\ref{FIG17} reflect
different characters of the interactions used in the p-p
channel.  The contact force (the SkP$^\delta$ results) leads to
the pairing densities which are, in general, largest at the
origin and decrease towards the surface.  (This general trend is
slightly modified by shell fluctuations resulting from
contributions from orbitals near the Fermi level.)  At the
surface, the isospin dependence of SkP$^\delta$ is fairly weak.
For example, there is very little difference between the pairing
densities in $^{150}$Sn and $^{172}$Sn.  These results are
characteristic for the volume-type pairing correlations.

A different pattern appears for the SkP results, where the
density dependence renders the p-p interaction strongly peaked
at the surface. In this case, the pairing densities tend to
increase when going from the center of the nucleus towards its
surface. Again, the shell fluctuations are superimposed on top
of this general behavior.  In particular, the central bump in
the pairing density in $^{120}$Sn is due to a contribution from
the 3$s_{1/2}$ state.  A more pronounced dependence on the
neutron excess is seen in the surface region. Especially near
the drip line, the pairing density develops a long tail
extending towards large distances.

The results obtained for the finite-range interaction D1S
exhibit intermediate features between the surface and the volume
type of pairing correlations. In particular, in the nuclear
interior one observes a fairly large region of relatively
constant pairing density.  The overall magnitude of the pairing
densities is very similar in all three approaches.  In
particular, it is interesting to see that at the nuclear surface
($r$$\sim$5\,fm) all three pairing densities in $^{120}$Sn are
very close to 0.018\,fm$^{-3}$.

\section{Hartree-Fock-Bogoliubov equations}
\label{sec2b}

We begin this section by presenting basic definitions and
equations of the HFB approach. The HFB theory is discussed in
many textbooks and review articles (see
Refs.{\ }\cite{[Goo79],[RS80]}, for example), while its
aspects pertaining to the coordinate representation have been
presented in Ref.{\ }\cite{[Dob84]}. An earlier discussion of
the coordinate-representation HFB formalism has been given by
Bulgac, whose work is available only in preprint form
\cite{[Bul80]}.  Recently, similar methods have also been
applied to a description of light nuclei \cite{[Ost92],[Boh93]}.
It is also worth mentioning that  the Green function approach in
the coordinate representation (the Gor'kov method
\cite{[Gor58]}), is formally equivalent to HFB -- cf. discussion
in Refs.{\ }\cite{[Zve84],[Zve85]}. The only difference
between the methods lies in the explicit energy dependence of
the quasiparticle mass operator, an analog  to the p-h
single-particle HF Hamiltonian (see below).

\subsection{HFB energy and HFB potentials}
\label{sec2ba}

The two-body effective Hamiltonian of a nuclear system
can be written in the coordinate representation as
\widetext
      \begin{eqnarray}
       \hat H &=& \int\text{d}^3\bbox{r}\text{d}^3\bbox{r}'
                  \sum_{\sigma\sigma'}
                      T(\bbox{r}\sigma,\bbox{r}'\sigma')
                      a^+_{\bbox{r}\sigma} a_{\bbox{r}'\sigma'}
                                                      \label{eq135} \\
             &+& \frac{1}{4}
                  \int\text{d}^3\bbox{r}_1 \text{d}^3\bbox{r}_2
                      \text{d}^3\bbox{r}_1'\text{d}^3\bbox{r}_2'
                  \sum_{\sigma_1\sigma_2\sigma_1'\sigma_2'}
                      V(\bbox{r}_1 \sigma_1 ,\bbox{r}_2 \sigma_2 ;
                        \bbox{r}_1'\sigma_1',\bbox{r}_2'\sigma_2')
                      a^+_{\bbox{r}_1 \sigma_1 }a^+_{\bbox{r}_2 \sigma_2 }
                        a_{\bbox{r}_2'\sigma_2'}  a_{\bbox{r}_1'\sigma_1'}
                                                 \nonumber .
  \end{eqnarray}
The first term represents the kinetic energy, while the second
one is the two-body interaction. In the following, we assume that
$V(\bbox{r}_1\sigma_1,\bbox{r}_2 \sigma_2;
\bbox{r}_1'\sigma_1',\bbox{r}_2'\sigma_2')$ includes the
exchange terms.

The average energy of the Hamiltonian (\ref{eq135}) in a time-even
HFB vacuum (\ref{eq105}) reads
   \begin{eqnarray}
    E_{\text{HFB}} &=& \int\text{d}^3\bbox{r}\text{d}^3\bbox{r}'
                       \sum_{\sigma\sigma'}
                      T(\bbox{r}\sigma,\bbox{r}'\sigma')
                      \rho(\bbox{r}'\sigma',\bbox{r}\sigma)
                                                      \label{eq136} \\
             &+& \frac{1}{2}
                  \int\text{d}^3\bbox{r}_1 \text{d}^3\bbox{r}_2
                      \text{d}^3\bbox{r}_1'\text{d}^3\bbox{r}_2'
                  \sum_{\sigma_1\sigma_2\sigma_1'\sigma_2'}
                      V(\bbox{r}_1 \sigma_1 ,\bbox{r}_2 \sigma_2 ;
                        \bbox{r}_1'\sigma_1',\bbox{r}_2'\sigma_2')
                      \rho(\bbox{r}_1'\sigma_1',\bbox{r}_1\sigma_1)
                      \rho(\bbox{r}_2'\sigma_2',\bbox{r}_2\sigma_2)
                                                 \nonumber \\
             &-& \frac{1}{4}
                  \int\text{d}^3\bbox{r}_1 \text{d}^3\bbox{r}_2
                      \text{d}^3\bbox{r}_1'\text{d}^3\bbox{r}_2'
                  \sum_{\sigma_1\sigma_2\sigma_1'\sigma_2'}
                      4\sigma_1\sigma_2'
                      V(\bbox{r}_1,-\sigma_1 ,\bbox{r}_2 \sigma_2 ;
                        \bbox{r}_1'\sigma_1',\bbox{r}_2',-\sigma_2')
                \tilde\rho(\bbox{r}_1 \sigma_1 ,\bbox{r}_2 \sigma_2 )
                \tilde\rho(\bbox{r}_1'\sigma_1',\bbox{r}_2'\sigma_2')
                                                 \nonumber .
  \end{eqnarray}
The last two terms are the interaction energies in the
particle-hole (p-h) and in the particle-particle (p-p) channels,
respectively.  Equivalently, one can define the p-h and p-p
single-particle Hamiltonians,
$h(\bbox{r}\sigma,\bbox{r}'\sigma')$ =
$T(\bbox{r}\sigma,\bbox{r}'\sigma')$ +
$\Gamma(\bbox{r}\sigma,\bbox{r}'\sigma')$ and $\tilde
h(\bbox{r}\sigma,\bbox{r}'\sigma')$, respectively:
   \begin{mathletters}\label{eq141}\begin{eqnarray}
      \Gamma(\bbox{r}\sigma,\bbox{r}'\sigma')  &=&
                  \int\text{d}^3\bbox{r}_2
                      \text{d}^3\bbox{r}_2'
                  \sum_{\sigma_2\sigma_2'}
                      V(\bbox{r} \sigma ,\bbox{r}_2 \sigma_2 ;
                        \bbox{r}'\sigma',\bbox{r}_2'\sigma_2')
                      \rho(\bbox{r}_2'\sigma_2',\bbox{r}_2\sigma_2)
                             ,   \label{eq141a} \\
      \tilde h(\bbox{r}\sigma,\bbox{r}'\sigma')  &=&
                  \int\text{d}^3\bbox{r}_1'\text{d}^3\bbox{r}_2'
                  \sum_{\sigma_1'\sigma_2'}
                      2\sigma'\sigma_2'
                      V(\bbox{r}\sigma ,\bbox{r}',-\sigma' ;
                        \bbox{r}_1'\sigma_1',\bbox{r}_2',-\sigma_2')
                \tilde\rho(\bbox{r}_1'\sigma_1',\bbox{r}_2'\sigma_2')
                             ,   \label{eq141b}
   \end{eqnarray}\end{mathletters}%
which gives the HFB energy in the form:
   \begin{equation}\label{eq142}
      E_{\text{HFB}} = \frac{1}{2}
                       \int\text{d}^3\bbox{r}\text{d}^3\bbox{r}'
                       \sum_{\sigma\sigma'}\left(
                       T(\bbox{r}\sigma,\bbox{r}'\sigma')
                       \rho(\bbox{r}'\sigma',\bbox{r}\sigma)
                    +  h(\bbox{r}\sigma,\bbox{r}'\sigma')
                       \rho(\bbox{r}'\sigma',\bbox{r}\sigma)
              + \tilde h(\bbox{r}\sigma,\bbox{r}'\sigma')
                \tilde\rho(\bbox{r}'\sigma',\bbox{r}\sigma)\right) .
   \end{equation}
\narrowtext\noindent%
Additional terms coming from the density-dependence of the two-body
interaction $V$ have been for simplicity omitted in
Eqs.{\ }(\ref{eq141a}), (\ref{eq141b}), and (\ref{eq142}).
The last term in Eq.{\ }(\ref{eq142}),
   \begin{equation}\label{epair}
      E_{\text{pair}} = \frac{1}{2}
                       \int\text{d}^3\bbox{r}\text{d}^3\bbox{r}'
                       \sum_{\sigma\sigma'}
                   \tilde h(\bbox{r}\sigma,\bbox{r}'\sigma')
                \tilde\rho(\bbox{r}'\sigma',\bbox{r}\sigma),
   \end{equation}
represents the pairing energy.  We also define the average
magnitude of pairing correlations by the formula \cite{[Dob84]}
   \begin{equation}\label{eq157}
      \langle\Delta\rangle = - \frac{1}{N^\tau}
                       \int\text{d}^3\bbox{r}\text{d}^3\bbox{r}'
                       \sum_{\sigma\sigma'}
                \tilde h(\bbox{r}\sigma,\bbox{r}'\sigma')
                    \rho(\bbox{r}'\sigma',\bbox{r}\sigma) ,
   \end{equation}
where $N^\tau$ is the number of particles (neutrons or protons).

The p-h and p-p mean fields (\ref{eq141}) have particularly
simple forms for the Skyrme interaction \cite{[Dob84]}.  In
Appendix \ref{appA} we present the form of the p-h and p-p
mean-field Hamiltonians in the case of a local two-body
finite-range Gogny interaction.

\subsubsection{Examples of the p-h and p-p potentials}
\label{sec2baa}

In this section we aim at comparing the self-consistent
potentials obtained with the Skyrme and Gogny forces.  Such a
comparison cannot be carried out directly, because the
corresponding integral kernels
$h(\bbox{r}\sigma,\bbox{r}'\sigma')$ and $\tilde
h(\bbox{r}\sigma,\bbox{r}'\sigma')$ have different structure.
For the Skyrme interaction, they are proportional to
$\delta(\bbox{r}$$-$$\bbox{r}')$ and depend also on the
differential operators (linear momenta) \cite{[Dob84]}, while
for the Gogny interaction they are sums of terms proportional to
$\delta(\bbox{r}$$-$$\bbox{r}')$ and terms which are functions
of $\bbox{r}$ and $\bbox{r}'$ (Appendix \ref{appA}).

Therefore, for the purpose of the present comparison we
introduce operational prescriptions to calculate the local parts
of the integral kernels:
   \begin{mathletters}\label{local}\begin{eqnarray}
      U(\bbox{r}) &=& {\cal LOC}\left[
               \Gamma(\bbox{r}\sigma,\bbox{r'}\sigma')\right]
                                                 , \label{locala} \\
      \tilde{U}(\bbox{r}) &=& {\cal LOC}\left[
               \tilde{h}(\bbox{r}\sigma,\bbox{r'}\sigma')\right]
                                                 . \label{localb}
   \end{eqnarray}\end{mathletters}%
These formal definitions in practice amount to: (i) disregarding
the momentum-dependent terms of the kernels, (ii) considering
only terms with $\sigma$=$\sigma'$=1/2 (which by time reversal
symmetry are equal to those with $\sigma$=$\sigma'$=$-$1/2), and
(iii) taking into account {\em only} the term proportional to
$\delta(\bbox{r}$$-$$\bbox{r}')$, if such a term is present.
The expressions for ${U}(\bbox{r})$ and $\tilde{U}(\bbox{r})$
can be found in Appendix A of Ref.{\ }\cite{[Dob84]} (Skyrme
interaction) and in Appendix A (Gogny interaction).  In the
Skyrme calculations, the contribution of the Coulomb interaction
to $\tilde{U}(\bbox{r})$ has been neglected since it is
estimated to be small.

In the case of finite-range local interactions (such as Gogny or
Coulomb), the corresponding non-local pairing field
$\tilde{h}(\bbox{r}\sigma,\bbox{r'}\sigma')$ does not contain
the term proportional to $\delta(\bbox{r}$$-$$\bbox{r}')$ (see
Appendix{\ }\ref{appA}).  Consequently, the local field
$\tilde{U}(\bbox{r})$ cannot be extracted in a meaningful way.
For instance, the diagonal (i.e., $\bbox{r'}$=$\bbox{r}$) part
of of the D1S pairing field is positive; i.e., it is dominated
by the short-range repulsive component rather than the
long-range attractive part \cite{[Kuc91],[Rum94]}.

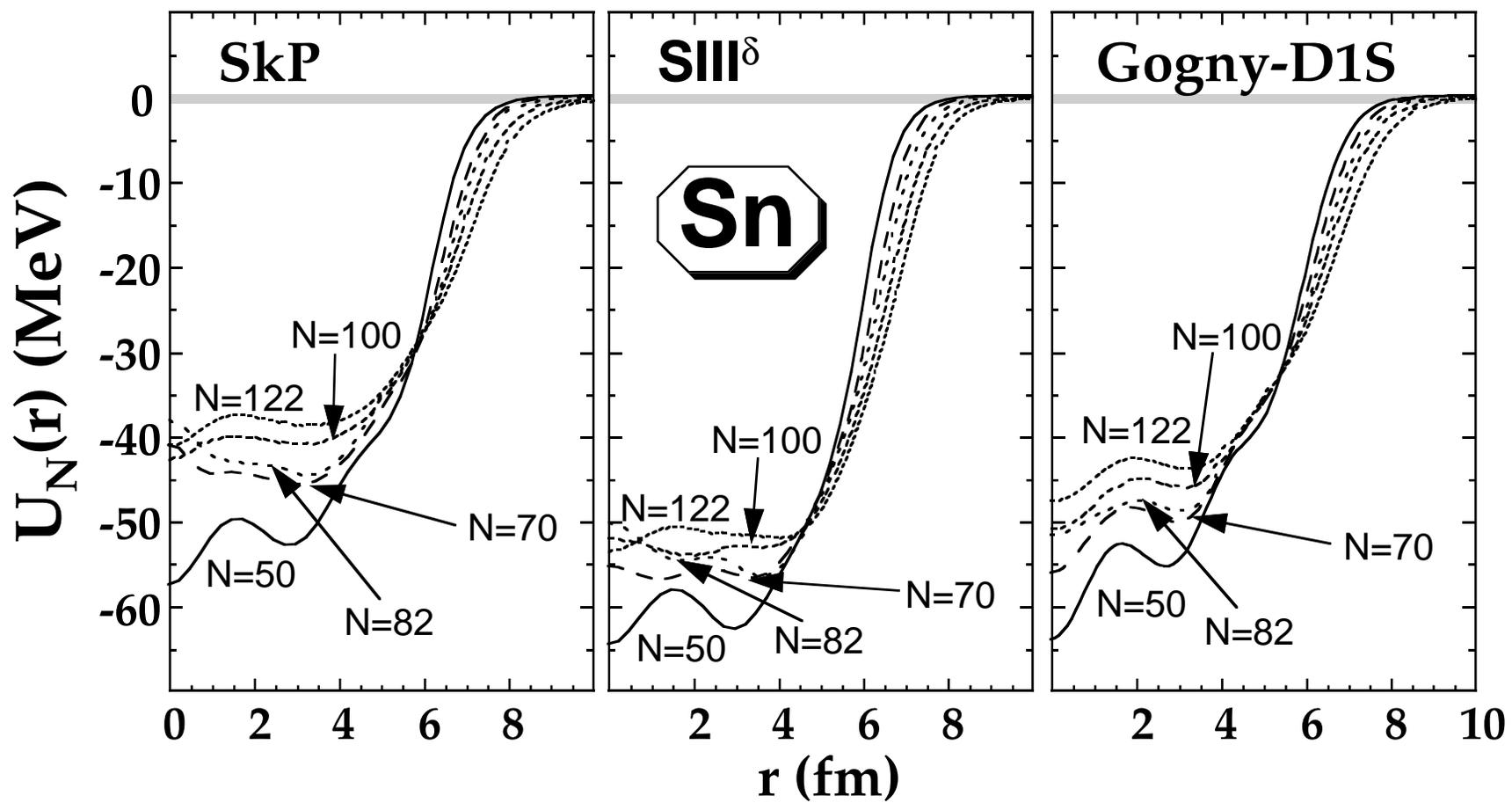
\begin{figure}[ht]
\caption[FF]{
Self-consistent spherical local neutron potentials
${U}_N(r)$ calculated with the SkP, SIII$^\delta$, and D1S
interactions for selected tin isotopes across the
$\beta$-stability valley.
}
\label{FIG01a}
\end{figure}

In the spherical case, the potentials ${U}(\bbox{r})$ and
$\tilde{U}(\bbox{r})$ depend on only one radial coordinate
$r$=$|\bbox{r}|$. This facilitates the qualitative comparison
between different forces.  Figure{\ }\ref{FIG01a} displays the
self-consistent spherical local p-h potentials ${U}(r)$,
Eq.{\ }(\ref{local}), for several tin isotopes, calculated
with SkP, SIII$^\delta$, and D1S interactions (the results with
SkP$^\delta$ are very close to those with SkP).  The terms
depending on the angular momentum, which result from a reduction
to the radial coordinate, are not included.  (The general
behavior of the self-consistent p-h potentials has already been
discussed many times in the literature,
e.g.{\ }\cite{[Fuk93],[Dob94],[Deb94a]}, and we include these
results only for completeness and for a comparison with the
corresponding p-p potentials, for which the detailed analysis
does not exist.)

Qualitatively, the results for ${U}(r)$ obtained with different
effective forces are quite similar, which reflects the fact that
all these interactions correctly describe global nuclear
properties. In particular, one sees that with increasing neutron
excess the neutron potentials become more shallow in the
interior and more wide in the outer region.  Interestingly, for
each of these three forces there exists a pivoting point at
which the potential does not depend on the neutron excess.  For
the three forces presented, this occurs at $r$=5.9, 4.6, and
5.4\,fm, respectively. The differences in the overall depths of
the average potentials reflect the associated effective masses
(i.e., the non-local contributions of the two-body
interactions).

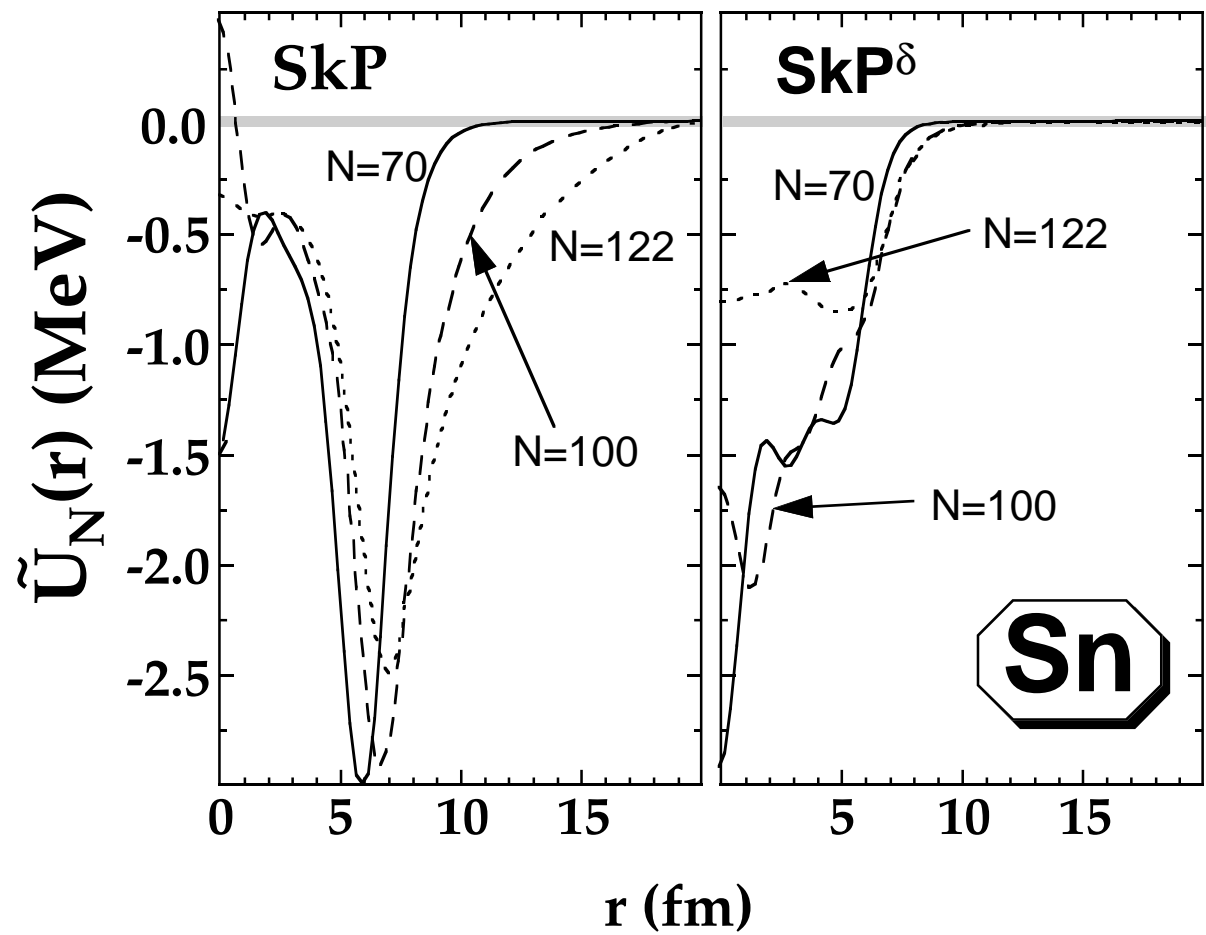
\begin{figure}[ht]
\caption[FF]{
Self-consistent spherical local neutron pairing
potentials $\tilde{U}_N(r)$ calculated with the SkP and
SkP$^\delta$ interactions for selected tin isotopes across the
$\beta$-stability valley.
}
\label{FIG01b}
\end{figure}

The analogous results for the p-p potentials $\tilde{U}(r)$
calculated for the SkP and SkP$^\delta$ interactions are shown
in Fig.{\ }\ref{FIG01b}.  On can see that the different
character of pairing interactions is directly reflected in the
form of the p-p potentials. Particularly  noteworthy  is  the
fact that the density-dependent pairing interaction in SkP
yields a very pronounced surface-peaked potential (the behavior
of $\tilde{U}(r)$ at large distances is further discussed in
Sec.{\ }\ref{sec3e}). One can easily understand its form by
recalling that this potential is equal to the product of the
pairing density $\tilde\rho(r)$ [Fig.{\ }\ref{FIG17}] and the
function which roughly resembles the behavior of DDDI of
Eq.{\ }(\ref{DDDI}); i.e., small in the interior and large in
the outer region. Of course, values of $\tilde\rho(r)$ and
$\tilde{U}(r)$ depend on each other by the fact that they both
result from a self-consistent solution of the complete HFB
equation in which the p-h and p-p channels are coupled together
(see Sec.{\ }\ref{sec2bc}).  Similar results were also
obtained in Refs.{\ }\cite{[Sta91]} (in the HFB+SkP model) and
\cite{[Sta92]} (in the QLM) for the proton-rich rare-earth
nuclei.

Since the p-h channel provides the bulk part of the interaction
energy, the particle densities $\rho(r)$ closely follow the
pattern of the p-h potentials (i.e., the density is large where
the potential is deep). An analogous relation is only partly
true for $\tilde\rho(r)$ and $\tilde{U}(r)$; i.e., even the
dramatic surface character of the SkP p-p potential
(Fig.{\ }\ref{FIG01b}) does not result in the pairing density
being similarly peaked at the surface. Recall that the
contributions to $\tilde\rho(r)$ come mainly from a few wave
functions near the Fermi surface, and that the form of these
wave functions is mainly governed by the p-h channel.  Since
these wave functions must have significant components in the
interior, the resulting pairing densities cannot exactly fit
into the surface-peaked p-p potentials. Nevertheless, a clear
tendency towards surface localization  is evident in
Fig.{\ }\ref{FIG17}.

In the case of the pure contact interaction (SkP$^\delta$
calculations) the p-p potential is exactly proportional to the
pairing density \cite{[Dob84]} with the proportionality constant
$V_0/2$ equal to $-$80\,MeV\,fm$^3$ \cite{[Dob95c]}.  Therefore,
the resulting potential is concentrated at the origin and
increases towards the surface. (Early calculations of p-p
potentials in the QLM with the density-independent delta
interaction can be found in Ref.{\ }\cite{[Smi88]}. The
general behavior of $\tilde{U}(r)$, denoted as $\Delta(r)$
therein, is very similar to our SkP$^\delta$ results.)

\subsection{HFB equations in the coordinate representation}
\label{sec2bap}

The variation
of the HFB energy with respect to independent
parameters $Z(\bbox{r}\sigma,\bbox{r}'\sigma')$ leads to
the HFB equation \cite{[RS80],[Dob84]},
\widetext
   \begin{equation}\label{eq143}
                  \int\text{d}^3\bbox{r}'
                  \sum_{\sigma'}\left(\begin{array}{cc}
                     h(\bbox{r}\sigma,\bbox{r}'\sigma') &
              \tilde h(\bbox{r}\sigma,\bbox{r}'\sigma') \\
              \tilde h(\bbox{r}\sigma,\bbox{r}'\sigma') &
                   - h(\bbox{r}\sigma,\bbox{r}'\sigma') \end{array}\right)
                                \left(\begin{array}{c}
                        \phi_1 (E,\bbox{r}'\sigma') \\
                        \phi_2 (E,\bbox{r}'\sigma')
                                \end{array}\right) =
                                \left(\begin{array}{cc}
                             E+\lambda & 0         \\
                                  0    & E-\lambda \end{array}\right)
                                \left(\begin{array}{c}
                        \phi_1 (E,\bbox{r}\sigma) \\
                        \phi_2 (E,\bbox{r}\sigma)
                                \end{array}\right),
   \end{equation}
where $\phi_1(E,\bbox{r}\sigma)$ and $\phi_2(E,\bbox{r}\sigma)$
are upper and lower components of the two-component
single-quasiparticle HFB wave function, and $\lambda$ is the
Fermi energy.

Properties of the HFB equation in the spatial coordinates,
Eq.{\ }(\ref{eq143}), have been discussed in
Ref.{\ }\cite{[Dob84]}.  In particular, it has been shown that
the spectrum of eigenenergies $E$ is continuous for
$|E|$$>$$-\lambda$ and discrete for $|E|$$<$$-\lambda$.  Since
for $E$$>$0 and $\lambda$$<$0 the lower components
$\phi_2(E,\bbox{r}\sigma)$ are localized functions of
$\bbox{r}$, the density matrices,
   \begin{mathletters}\label{eq144}\begin{eqnarray}
       \rho(\bbox{r}\sigma,\bbox{r}'\sigma') &=&
          \sum_{0<E_n<-\lambda}  \phi_2  (E_n,\bbox{r} \sigma )
                                 \phi^*_2(E_n,\bbox{r}'\sigma')
        + \int_{-\lambda}^\infty \text{d}n(E)
                                 \phi_2  (E  ,\bbox{r} \sigma )
                                 \phi^*_2(E  ,\bbox{r}'\sigma')
                             ,   \label{eq144a} \\
 \tilde\rho(\bbox{r}\sigma,\bbox{r}'\sigma') &=&
        - \sum_{0<E_n<-\lambda}  \phi_2  (E_n,\bbox{r} \sigma )
                                 \phi^*_1(E_n,\bbox{r}'\sigma')
        - \int_{-\lambda}^\infty \text{d}n(E)
                                 \phi_2  (E  ,\bbox{r} \sigma )
                                 \phi^*_1(E  ,\bbox{r}'\sigma')
                             ,   \label{eq144b}
   \end{eqnarray}\end{mathletters}%
are always localized.

{}For the case of a discretized continuum,
Sec.{\ }\ref{sec3a}, the integral over the energy reduces to a
discrete sum \cite{[Dob84]} but one should still carefully
distinguish between contributions coming from the discrete
($E_n$$<$$-\lambda$) and discretized ($E_n$$>$$-\lambda$)
states. The orthogonality relation for the single-quasiparticle
HFB wave functions reads
   \begin{equation}\label{orthog}
                  \int\text{d}^3\bbox{r}
          \sum_{\sigma} \left[ \phi^*_1(E_n,\bbox{r} \sigma )
                               \phi_1  (E_{n'},\bbox{r}\sigma)
                             + \phi^*_2(E_n,\bbox{r} \sigma )
                               \phi_2  (E_{n'},\bbox{r}\sigma)
                  \right] = \delta_{n,n'}.
   \end{equation}
It is seen from Eq.{\ }(\ref{orthog}) that the lower components are
not normalized.
Their norms,
   \begin{equation}\label{norms}
   N_n  =  \int\text{d}^3\bbox{r}\sum_{\sigma}
                      |\phi_2(E_n,\bbox{r} \sigma )|^2,
   \end{equation}
define the total number of particles
   \begin{equation}\label{Ntot}
    N=\int\text{d}^3\bbox{r}\rho(\bbox{r}) = \sum_{n}N_n.
   \end{equation}

In the HFB theory, the localization condition (\ref{eq106})
discussed in Sec.{\ }\ref{sec2a} is automatically guaranteed
for any system with negative Fermi energy $\lambda$. This allows
studying nuclei which are near the particle drip lines where the
Fermi energy approaches zero through negative values.

{}For the Skyrme interaction, the HFB equation (\ref{eq143}) is
a differential equation in spatial coordinates \cite{[Dob84]}.
If the spherical symmetry is imposed, which is assumed in the
following, this equation reads
   \begin{equation}\label{eq152}
   \left[-\frac{\text{d}}{\text{d}r}
      \left(\begin{array}{cc}       M & \tilde M \\
                   \tilde M &       -M \end{array}\right)
          \frac{\text{d}}{\text{d}r}+
      \left(\begin{array}{cc}      {U}-\lambda & \tilde{U} \\
                   \tilde{U}         &      -{U}+\lambda
            \end{array}\right)\right]
    \left(\begin{array}{c} r\phi_1(E,r) \\ r\phi_2(E,r) \end{array}\right) =
   E\left(\begin{array}{c} r\phi_1(E,r) \\ r\phi_2(E,r) \end{array}\right) ,
   \end{equation}
\narrowtext\noindent%
where $M$ and $\tilde M$ are p-h and p-p mass parameters,
respectively, and ${U}$ and $\tilde{U}$ are defined in
Sec.{\ }\ref{sec2ba}.  Due to the spherical symmetry,
Eq.{\ }(\ref{eq152}) is solved separately for each partial
wave ($j,\ell$).  The potentials include also the centrifugal
and spin-orbit terms, and the p-h mass parameter $M$ is
expressed in terms of the effective mass $m^*$; i.e.,
$M$=$\hbar^2/2m^*$, see Ref.{\ }\cite{[Dob84]} for details.

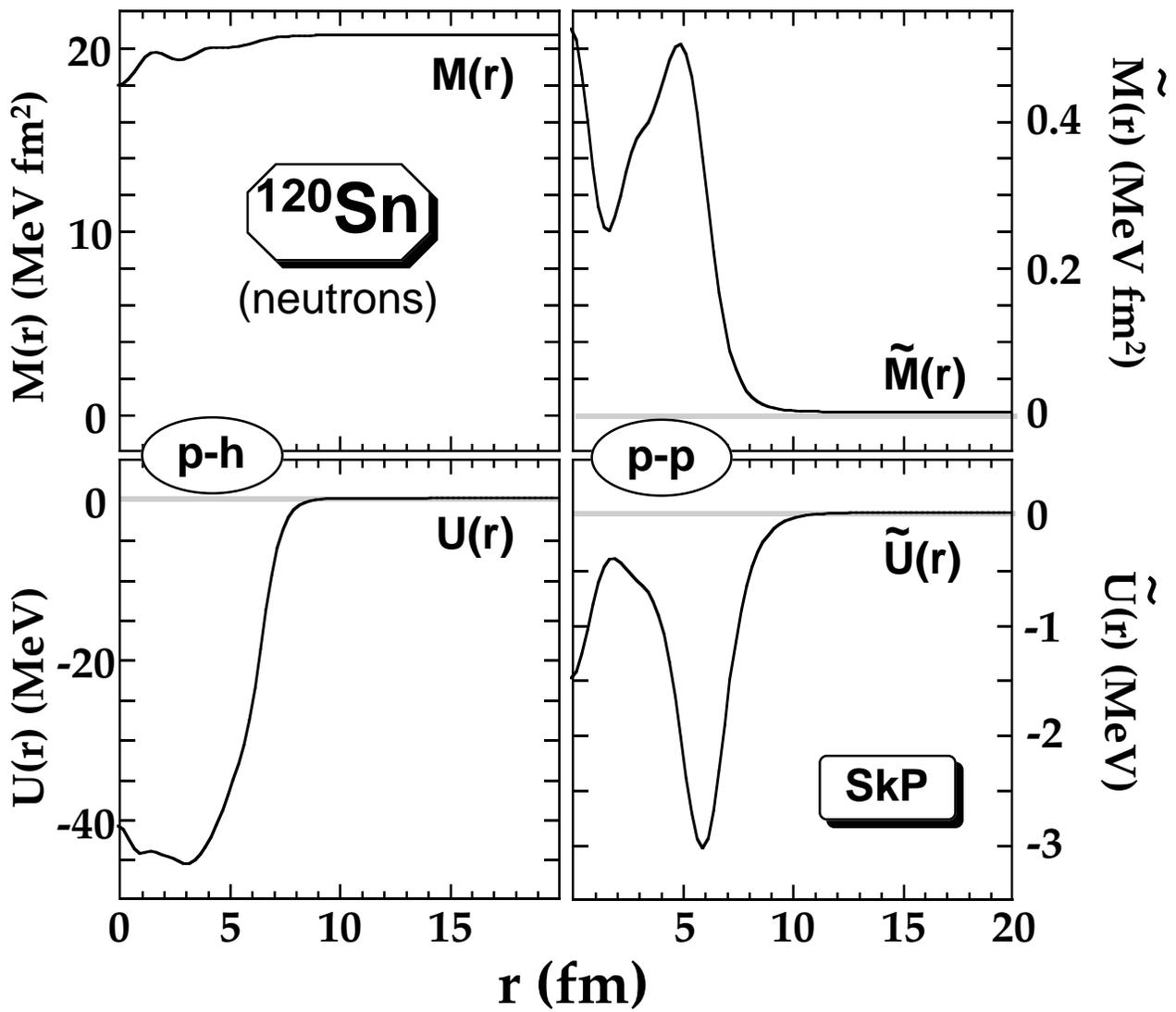
\begin{figure}[ht]
\caption[FF]{
The self-consistent HFB+SkP mass parameters $M$ and
$\tilde M$, and potentials ${U}$ and $\tilde{U}$ (central parts
only), for neutrons in $^{120}$Sn.
}
\label{FIG20}
\end{figure}

Before discussing the properties of the HFB wave functions, we
analyze the structure of the spherical HFB Hamiltonian of
Eq.{\ }(\ref{eq152}).  Figure {\ }\ref{FIG20} shows the
behavior of $M(r)$ and $\tilde M(r)$, and ${U}(r)$ and
$\tilde{U}(r)$ (central parts only) obtained  for neutrons in
$^{120}$Sn in the HFB+SkP model.  The p-h functions, $M(r)$ and
${U}(r)$, are similar to those obtained in other mean-field
theories. $M(r)$ has values close to
$\hbar^2/2m$$\simeq$20\,MeV\,fm$^2$, which corresponds to the
value of the free nucleon mass $m$. In the nuclear interior,
this function has slightly smaller values, because the effective
mass $m^*$ is here slightly larger than $m$. This effect is due
to the non-zero isovector effective mass of the Skyrme SkP
interaction; recall that for this interaction the nuclear-matter
value of the isoscalar effective mass is $m^*$=$m$. The central
potential ${U}(r)$ has the standard depth of about 40\,MeV and
disappears around $r$=7.5\,fm.

The form of the p-p functions, $\tilde M(r)$ and $\tilde{U}(r)$,
characterizes the pairing properties of the system.  One may
note that both these functions are essentially peaked at the
nuclear surface.  In $^{120}$Sn they also exhibit central bumps
resulting from the fact that in this nucleus the neutron 3$s_{1/2}$
orbital is located near the Fermi surface. Values of $\tilde
M(r)$ are (in the chosen units) an order of magnitude smaller
than those of $\tilde{U}(r)$.  This should be compared with the
results obtained for the p-h channel, where the values of $M(r)$
are only about a factor of 2 smaller than those of ${U}(r)$. It
means that,  for the SkP parametrization, the kinetic term in
the p-p channel (which simulates the finite-range effects) is
relatively less important than the kinetic energy term in the
p-h channel.

\subsection{Single-quasiparticle wave functions}
\label{sec2bc}

This section contains the discussion of HFB wave functions
$\phi_1(E,r)$ and $\phi_2(E,r)$ (Sec.{\ }\ref{sec2bca}),
canonical-basis wave functions $\breve\psi_\mu(r)$
(Sec.{\ }\ref{sec2bcb}), and HF+BCS wave functions
(Sec.{\ }\ref{sec2bcc}).  In the following, the HFB equation
(\ref{eq152}) was solved in the spherical box of the radius
$R_{\text{box}}$=20\,fm for the $j$=1/2 and $\ell$=0 ($s_{1/2}$)
neutron states; i.e., for vanishing centrifugal, Coulomb,  and
spin-orbit potentials.  The calculations were performed for
$^{120}$Sn.

\subsubsection{Examples of the single-quasiparticle wave functions}
\label{sec2bca}

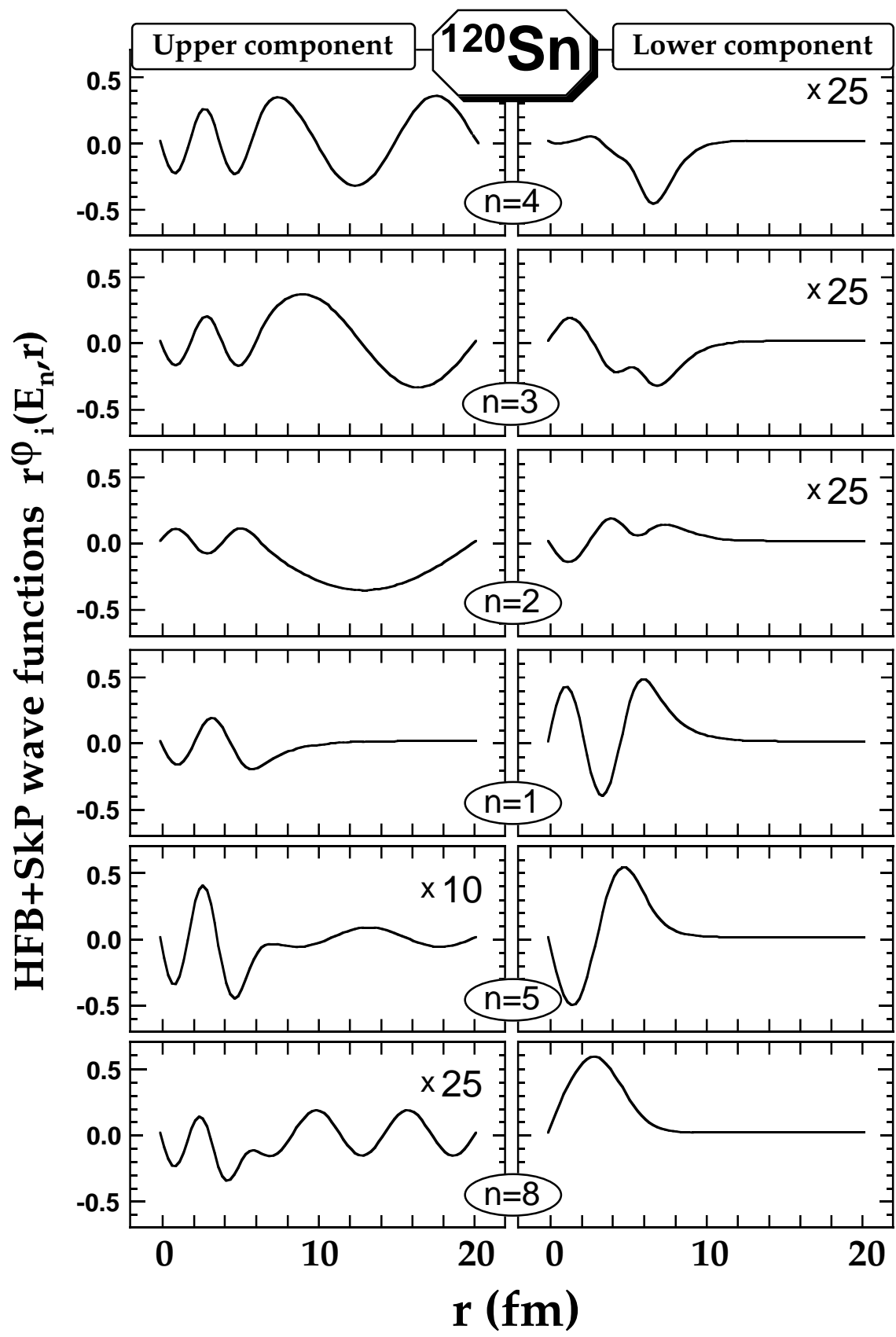
\begin{figure}[ht]
\caption[FF]{
The HFB+SkP radial wave functions $r\phi_i(E_n,r)$ of the
neutron $s_{1/2}$ single-quasiparticle states in $^{120}$Sn.
Upper ($i$=1) and lower ($i$=2) components are plotted in the
left and right columns, respectively. The numbers preceded by a
times ($\times$) sign indicate the scaling factors for the small
wave function components.
}
\label{FIG21}
\end{figure}

The neutron single-quasiparticle wave functions are presented in
Fig.{\ }\ref{FIG21}.  The upper components $r\phi_1(E_n,r)$,
and the lower components $r\phi_2(E_n,r)$, are plotted in the
left and right columns, respectively. Because a box of a finite
radius was used, the particle continuum is discretized. The
positive quasiparticle eigenenergies $E_n$ are in increasing
order numbered by the index $n$, and their values are tabulated
in the left portion of Table \ref{TAB01}, together with the
norms of the lower components (\ref{norms}),
$N_n$=$4\pi\int{r^2}\text{d}r|\phi_2(E_n,r)|^2$. Since  the
lower components define the particle density matrix
[Eq.{\ }(\ref{eq144a})] the numbers $(2j$+1)$N_n$ (i.e.,
2$N_n$ for the $j$=1/2 case considered) constitute contributions
of a given quasiparticle state to the total number of neutrons
(see Eq.{\ }(\ref{Ntot})).

Wave functions in Fig.{\ }\ref{FIG21}, and the entries in
Table \ref{TAB01}, have been ordered from the bottom to the top
not according to the excitation-energy index $n$, but rather
according to numbers of nodes of the {\em large} component.
(The large component is the lower component for hole states and
the upper component for particle states -- see
Fig.{\ }\ref{FIG21}.) The lower component of the $n$=8 state
is large, and it has zero nodes; therefore it is plotted at the
bottom of the figure. Next comes the $n$=5 state, whose lower
component has one node, and the $n$=1 state with two nodes.
Lower components of these three states are larger than their
upper components and they contribute almost 2 particles each to
the total number of neutrons.  Consequently, these quasiparticle
states should be associated with the 1$s_{1/2}$, 2$s_{1/2}$, and
3$s_{1/2}$, single-hole states.

{}For all other calculated $s_{1/2}$ states the upper components
are larger than the lower ones, and these states contribute
small fractions to the particle number, see Table \ref{TAB01}.
Consequently, these quasiparticle states should be associated
with the $s_{1/2}$ single-particle states.  The behavior of
these wave functions  differs in the nuclear interior (i.e., for
$r<R$ where $R$$\sim$7{\ }fm is the nuclear radius) and
outside ($r>R$).  Since the wavelength of the upper component is
roughly proportional to $1/\sqrt{E_n+\lambda-{U}(r)}$, the ratio
of the corresponding wavelengths behaves as
   \begin{equation}\label{wave}
   \frac{\lambda_{\text{out}}}{\lambda_{\text{in}}}
      \approx \sqrt{1+\frac{|{U}(0)|}
    {E_n+\lambda}},
   \end{equation}
where ${U}(0)$ is the depth of the neutron potential well.  For
the $s_{1/2}$ neutron states in $^{120}$Sn the excitation
energy, $E_n+\lambda$,  can be found from Table{\ }\ref{TAB01}
($\lambda$=$-$7.94\,MeV), and ${U}(0)$$\sim$$-$45\,MeV (see
Fig.{\ }\ref{FIG20}).

The upper component of the $n$=2 state has three nodes.
However, for $r>R$ the exterior part of the wave function
corresponds to a half-wave; i.e., it represents the
lowest-energy discretized continuum state.  Since $E_n+\lambda$
is only 0.97\,MeV, the wavelength in the nuclear interior is
$\sim$6.5 times shorter than $\lambda_{\text{out}}$.  The next
two wave functions have four and five nodes in their upper
components.  Compared to the $n$=2 state, they exhibit shorter
wavelengths both outside and inside the nucleus (the
corresponding excitation energies are larger), and the ratio
${\lambda_{\text{out}}}/{\lambda_{\text{in}}}$   decreases
according to Eq.{\ }(\ref{wave}).

The quasiparticle states with $n$=2, 3, and 4 should be
associated with the 4$s_{1/2}$, 5$s_{1/2}$, and 6$s_{1/2}$
states in the particle continuum. Of course, the values of their
quasiparticle energies strongly depend on the size of the box,
because the wavelength of their exterior parts will increase
with increasing $R_{\text{box}}$ (is roughly proportional to
$R_{\text{box}}$).

{}From the above discussion, one can see that the structure of
large components resembles very much that of the HF wave
functions. Moreover, the small components are very small
compared to the large ones; in order to plot both of them in the
same scale (Fig.{\ }\ref{FIG21}) they have to be multiplied by
factors from 10 to 25.  Only the lowest quasiparticle state
($n$=1), which corresponds to the 3$s_{1/2}$ state near the
Fermi surface, has the two components of a similar magnitude.
It is to be noted, however, that the detailed structure of small
components is decisive for a description of the pairing
correlations.  Indeed,  both components are coupled in the HFB
equations by the pairing fields
$\tilde{h}(\bbox{r}\sigma,\bbox{r}'\sigma')$ or $\tilde{U}$.

In agreement with general asymptotic properties of the upper and
lower components \cite{[Bul80],[Dob84]}, one sees in
Fig.{\ }\ref{FIG21} that the lower components vanish at large
distances for all quasiparticle states, regardless of the
excitation energy.  Consequently, the resulting density matrix
is localized.  It is interesting to observe  (Table \ref{TAB01})
that the norms of the lower components $N_n$ do not behave
monotonically with quasiparticle energy. Namely, $N_n$  is about
0.0002 for $n$=2; then it increases to 0.0019 at $n$=6, and only
then it decreases to about 0.0001 at $n$=11.  This  means that
the pairing correlations couple states with very high
quasiparticle excitations and short-wavelength upper components;
i.e., located high up in the particle continuum.  In the
considered example, only by going to the energy region of as
high as 50\,MeV is the pairing coupling to the continuum states
exhausted.

Apart from the $n$=1 state which has the quasiparticle energy
$E$ smaller than $-\lambda$, for all other quasiparticle states
the upper components oscillate at large distances; i.e., these
states belong to the HFB continuum. This seems natural for the
4$s_{1/2}$, 5$s_{1/2}$, and 6$s_{1/2}$ states discussed above,
but it also holds for the deep-hole states 1$s_{1/2}$ and
2$s_{1/2}$. This illustrates the physical property of the
deep-hole states that once such a state  is excited, it is
coupled to the particle continuum and acquires some particle
width.  Of course, before such a hole is created (e.g.,
one-quasiparticle excitation in the neighboring nucleus) the
nucleus (i.e., quasiparticle vacuum) is perfectly particle-bound
and the contributions from the deep-hole-like quasiparticle
states to the density matrix are localized in space.

\subsubsection{Examples of the canonical-basis wave functions}
\label{sec2bcb}

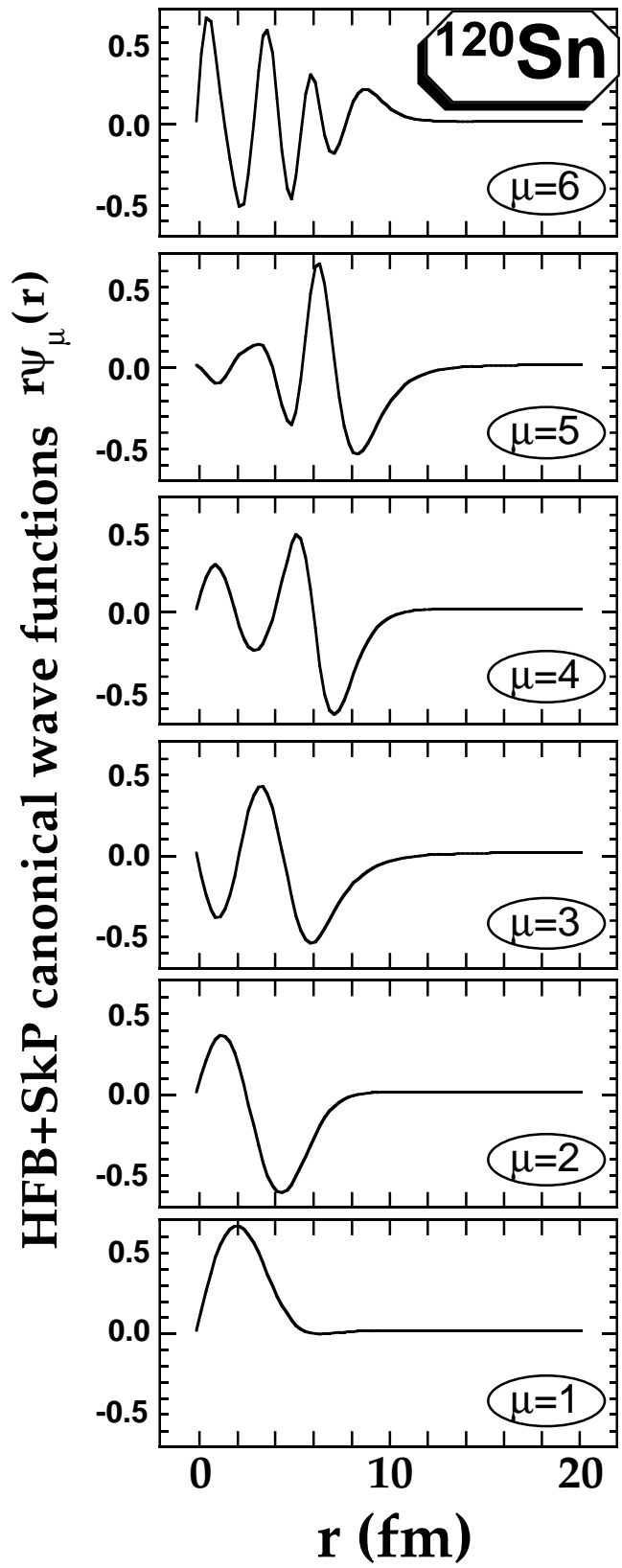
\begin{figure}[ht]
\caption[FF]{
The HFB+SkP  radial canonical-basis wave functions
$r\breve\phi_\mu(r)$ of the neutron $s_{1/2}$ single-particle
states in $^{120}$Sn.
}
\label{FIG22}
\end{figure}

By solving the integral eigenequation for the density matrix
(\ref{eq122a}), one obtains the canonical-basis wave functions
$\breve\psi_\mu(r)$. Actually, when the HFB equation
(\ref{eq152}) is solved by a discretization method  on a spatial
mesh, as is done here, the density matrix is represented by a
matrix  and the integral eigenequation becomes the usual matrix
eigenproblem. In the present application to $^{120}$Sn, the mesh
of equally spaced points with $\Delta r$=0.25\,fm was used and
then the canonical-basis wave functions were obtained on the
same mesh of points. These wave functions are plotted in
Fig.{\ }\ref{FIG22}, while other characteristics of the
canonical states are listed on the right-hand side portion of
Table \ref{TAB01}. Here the states are ordered from bottom to
top according to their occupation probabilities $v_\mu^2$.

When $\mu$ increases from 1 to 5, the number of nodes of the
canonical-basis wave functions increases from zero to four.
Therefore, these states represent the 1$s_{1/2}$ to 5$s_{1/2}$
single-particle states. The first three of them have large
occupation probabilities $v_\mu^2$, negative average values
$\epsilon_\mu$ of the p-h Hamiltonian, and positive pairing gaps
$\Delta_\mu$ [see Eq.{\ }(\ref{eq148})]. These states have all
the characteristics of bound single-particle states, and their
wave functions strongly resemble the large components of the
$n$=8, 5, and 1 quasiparticle states shown in
{}Fig.{\ }\ref{FIG21}. It is interesting to note that the two
states $\mu$=4 and 5 follow exactly the same pattern of
localized wave functions, despite the {\em positive} values of
$\epsilon_\mu$.  Therefore, these two states can be understood
as the representatives of the positive-energy spectrum in the
ground-state of $^{120}$Sn. We purposely avoid using the term
``particle continuum'', because these orbitals represent
discrete and  localized eigenstates of the density matrix.

Table \ref{TAB01} shows that the occupation probabilities of the
canonical-basis states with $\mu$=4,$\ldots$,7 decrease very
rapidly. In fact only states with $\mu$=4 and 5 have tangible
occupation probabilities; one can say that the remaining
orbitals are entirely empty. This feature has to be compared
with the sequence of norms of the lower HFB components, $N_n$,
which do not fall down to zero at even a nearly similar pace.
This demonstrates that even if the convergence of the HFB
eigenproblem requires high quasiparticle energies, the number of
physically important single-particle states is very restrained.
Unfortunately, as discussed below in Sec.{\ }\ref{sec2bb}, one
cannot obtain the canonical-basis states without actually
solving the HFB equations up to high energies.  For $\mu$=6 and
higher, the occupation probabilities are so small that the
numerical procedure used to diagonalize the density matrix
returns accidental mixtures of almost degenerate eigenfunctions.
This is seen in Fig.{\ }\ref{FIG22}, where the wave function
with $\mu$=6  has six nodes instead of five, expected from the
regular sequence. Also the energies $\epsilon_\mu$ are for these
nearly empty  states randomly scattered between 40 and 70\,MeV,
while the pairing gaps $\Delta_\mu$ are scattered around zero.

\subsubsection{Examples of the BCS quasiparticle wave functions}
\label{sec2bcc}

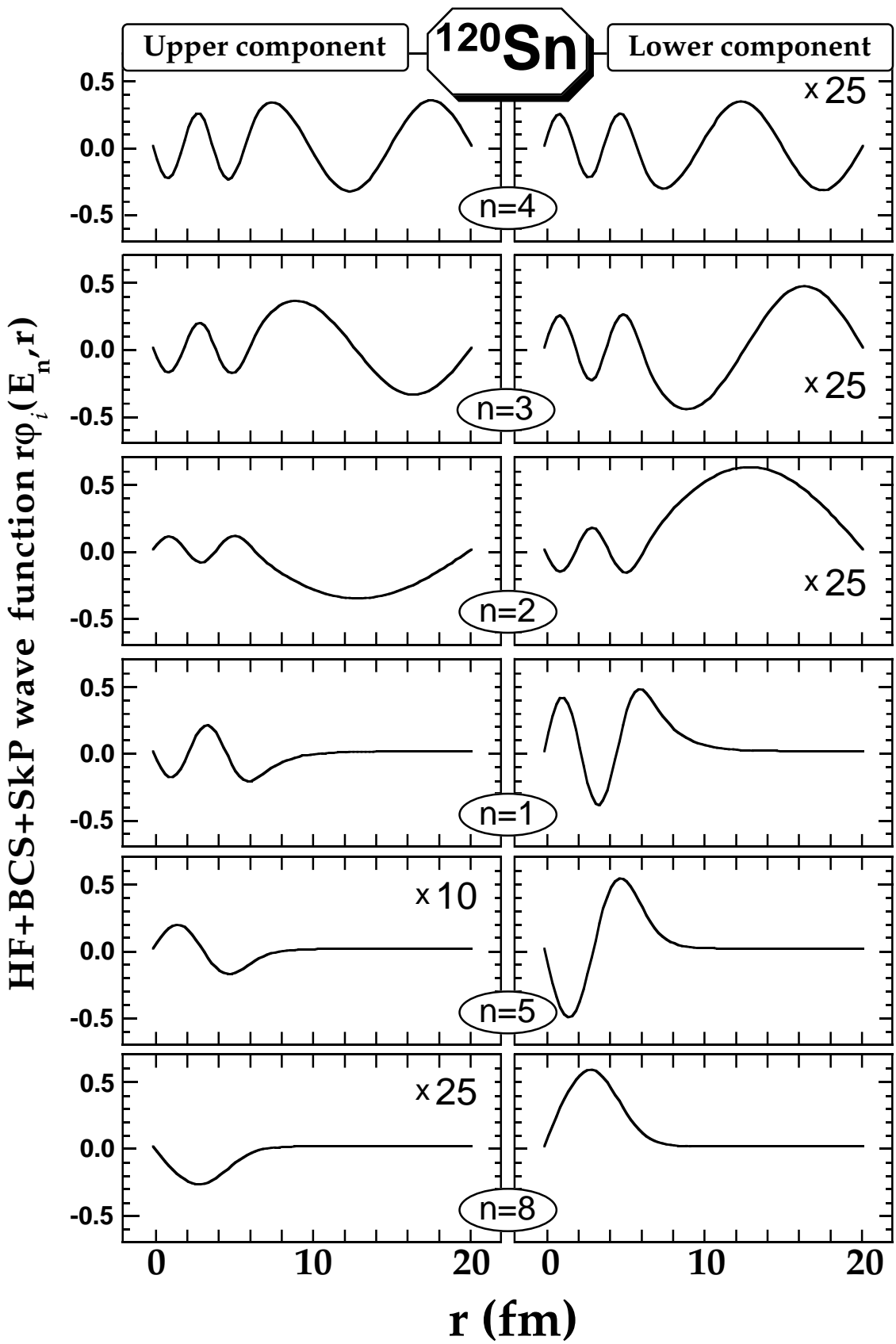
\begin{figure}[ht]
\caption[FF]{
Same as in Fig.{\ }\protect\ref{FIG21}, but for the HF+BCS approach.
}
\label{FIG23}
\end{figure}

The BCS quasiparticle wave functions can be obtained by
enforcing the BCS approximation on the HFB equations.  This is
done by setting the pairing Hamiltonian $\tilde h$ to a
constant; i.e., by using $\tilde M(r)$=0 and
$\tilde{U}(r)$=$-$1.232\,MeV. This value of $\tilde{{U}}$ is
equal to minus the HFB average neutron pairing gap, as defined
in Eq.{\ }(\ref{eq157}).  As seen in Fig.{\ }\ref{FIG23},
the pattern of large components follows closely that obtained in
the HFB method, while the shapes of small components are
entirely different. Indeed, since in the BCS approximation lower
and upper components are simply proportional, small and large
components have the same asymptotic properties.  This leads to
serious inconsistencies, because the small lower components {\em
are not} localized any more, and introduce an unphysical
particle gas in the density matrix, while the small upper
components {\em are} localized and the corresponding deep-hole
states have no particle width.

\subsection{HFB equations in the canonical basis}
\label{sec2bb}

It is seen in Eqs.{\ }(\ref{eq136}) and (\ref{eq141}) that
the two-body interaction enters the p-h and p-p channels in a
different way. This is particularly conspicuous when the
canonical basis (\ref{eq122}) is used; i.e.,
   \begin{eqnarray}
       E_{\text{HFB}} = \sum_\nu \breve T_{\mu\mu}v^2_\mu
         &+&\frac{1}{2}\sum_{\mu\nu}
            \breve F_{\mu\nu}v^2_\mu v^2_\nu \nonumber \\
         &-&\frac{1}{4}\sum_{\mu\nu}
            \breve G_{\mu\nu}u_\mu v_\mu u_\nu v_\nu , \label{eq137}
   \end{eqnarray}
where
   \begin{mathletters}\label{eq138}\begin{eqnarray}
       \breve F_{\mu\nu} &=& \frac{1}{2}\left(
           \breve V_{\mu\nu\mu\nu} + \breve V_{\mu\bar\nu\mu\bar\nu}\right)
                             ,   \label{eq138a} \\
       \breve G_{\mu\nu} &=& -s^*_\mu s_\nu
           \breve V_{\mu\bar\mu\nu\bar\nu}
                             .   \label{eq138b}
   \end{eqnarray}\end{mathletters}%
The two-body matrix elements in the canonical basis are defined
as usual:
\widetext
   \begin{equation}\label{eq139}
       \breve{V}_{\mu\nu\mu'\nu'} =
                  \int\text{d}^3\bbox{r}_1 \text{d}^3\bbox{r}_2
                      \text{d}^3\bbox{r}_1'\text{d}^3\bbox{r}_2'
                  \sum_{\sigma_1\sigma_2\sigma_1'\sigma_2'}
                      V(\bbox{r}_1 \sigma_1 ,\bbox{r}_2 \sigma_2 ;
                        \bbox{r}_1'\sigma_1',\bbox{r}_2'\sigma_2')
                        \breve\psi^*_\mu (\bbox{r}_1 \sigma_1 )
                        \breve\psi^*_\nu (\bbox{r}_2 \sigma_2 )
                        \breve\psi_{\mu'}(\bbox{r}_1'\sigma_1')
                        \breve\psi_{\nu'}(\bbox{r}_2'\sigma_2').
   \end{equation}
\narrowtext\noindent%
Since we include
in $V(\bbox{r}_1\sigma_1,\bbox{r}_2\sigma_2;
\bbox{r}_1'\sigma_1',\bbox{r}_2'\sigma_2')$ the exchange term,
the matrix $V_{\mu\nu\mu'\nu'}$ is antisymmetric in $\mu\nu$ and
in $\mu'\nu'$.  Due to the hermiticity and the time-reversal
symmetry of the interaction, matrices $\breve{F}_{\mu\nu}$ and
$\breve{G}_{\mu\nu}$ obey the following symmetry relations,
   \begin{mathletters}\begin{eqnarray}
       \breve F_{\mu\nu} =  \breve F^*_{\mu\nu}
                         =  \breve F_{\nu\mu}
                        &=& \breve F_{\mu\bar\nu}
                         =  \breve F_{\bar\mu\nu}
                             ,   \label{eq140a} \\
       \breve G_{\mu\nu} =  \breve G^*_{\mu\nu}
                         =  \breve G_{\nu\mu}
                        &=& \breve G_{\mu\bar\nu}
                         =  \breve G_{\bar\mu\nu}
                             .   \label{eq140b}
   \end{eqnarray}\end{mathletters}%
The matrix $\breve{F}$ is defined by different matrix elements
of the interaction than the matrix $\breve{G}$. Namely, the
matrix element $\breve{F}_{\mu\nu}$  represents a ``diagonal''
scattering of pairs of states $\mu\nu$$\rightarrow$$\mu\nu$ (or
$\mu\bar\nu$$\rightarrow$$\mu\bar\nu$).  This type of scattering
concerns {\em all} pairs of states.  The resulting contributions
to the energy, Eq.{\ }(\ref{eq137}), involve the occupation
probabilities of the single-particle states constituting each
pair.  On the other hand, the matrix elements
$\breve{G}_{\mu\nu}$ represent a ``non-diagonal'' scattering of
pairs of {\em time-reversed} states
$\nu\bar\nu$$\rightarrow$$\mu\bar\mu$.  This scattering concerns
only a very special subset of all pairs.

In principle, an effective interaction should describe both
channels of interaction at the same time. This is, for example,
the case for the Gogny interaction \cite{[Gog73]} and for the
Skyrme SkP interaction \cite{[Dob84]}. However, the fact that
both channels of interaction play a different role in the HFB
theory allows the use of different forms of interaction to model
the p-h and p-p channels.  Such an approach is additionally
motivated by the fact that the interaction in the p-h channel,
which defines, e.g.,  the saturation properties,  is much better
known than the p-p interaction. Moreover, the p-h channel
provides a two-orders-of-magnitude larger interaction energy.

Since the canonical-basis wave functions
$\breve\psi(\bbox{r}\sigma)$ are all localized, it is
instructive to consider the HFB equations in this particular
basis. They read:
   \begin{mathletters}\label{eq145}\begin{eqnarray}
       (\breve h - \lambda)_{\mu\nu}\eta_{\mu\nu}
      + \breve{\tilde    h}_{\mu\nu} \xi_{\mu\nu} &=& 0
                             ,   \label{eq145a} \\
       (\breve h - \lambda)_{\mu\nu} \xi_{\mu\nu}
      - \breve{\tilde    h}_{\mu\nu}\eta_{\mu\nu} &=& \breve E_{\mu\nu}
                             ,   \label{eq145b}
   \end{eqnarray}\end{mathletters}%
where
   \begin{mathletters}\begin{eqnarray}
      \eta_{\mu\nu} &:=& u_\mu v_\nu + u_\nu v_\mu
                             ,   \label{eq146a} \\
       \xi_{\mu\nu} &:=& u_\mu u_\nu - v_\nu v_\mu
                             .   \label{eq146b}
   \end{eqnarray}\end{mathletters}%
Equation (\ref{eq145a}) is equivalent to the variational
condition that the HFB energy is minimized, while
Eq.{\ }(\ref{eq145b}) defines the energy matrix
$\breve{E}_{\mu\nu}$.  (The matrix $\breve{E}_{\mu\nu}$
represents the HFB Hamiltonian in the canonical basis.) Since
for every pair of indices $\mu\nu$ it holds that
$(\eta_{\mu\nu})^2$+$(\xi_{\mu\nu})^2$=1,
Eqs.{\ }(\ref{eq145}) can be written as
   \begin{mathletters}\begin{eqnarray}
       (\breve h - \lambda)_{\mu\nu} &=& \breve E_{\mu\nu} \xi_{\mu\nu}
                             ,   \label{eq147a} \\
     - \breve{\tilde    h}_{\mu\nu}  &=& \breve E_{\mu\nu}\eta_{\mu\nu}
                             .   \label{eq147b}
   \end{eqnarray}\end{mathletters}%
The occupation probabilities $v_\mu$ are  solely determined by
the diagonal matrix elements of the p-h and p-p Hamiltonians,
   \begin{mathletters}\label{eq148}\begin{eqnarray}
       \epsilon_\mu &:=&   \breve h_{\mu\mu}
                             ,   \label{eq148a} \\
       \Delta_\mu   &:=& - \breve{\tilde    h}_{\mu\mu}
                             ,   \label{eq148b}
   \end{eqnarray}\end{mathletters}%
and the result is
   \begin{mathletters}\label{eq149}\begin{eqnarray}
       v_\mu =& \text{sign}(\Delta_\mu)
           &\sqrt{\frac{1}{2}-\frac{\epsilon_\mu-\lambda}{2E_\mu}}
                             ,   \label{eq149a} \\
       u_\mu =&
           &\sqrt{\frac{1}{2}+\frac{\epsilon_\mu-\lambda}{2E_\mu}}
                             ,   \label{eq149b}
   \end{eqnarray}\end{mathletters}%
where $E_\mu$ are the diagonal matrix elements of the matrix
$\breve{E}_{\mu\nu}$:
   \begin{equation}\label{eq150}
      E_\mu := \breve E_{\mu\mu} =
                  \sqrt{(\epsilon_\mu-\lambda)^2 + \Delta_\mu^2} .
   \end{equation}
In this representation, the average pairing gap (\ref{eq157})
is given by the average value of $\Delta_\mu$ in the occupied states,
   \begin{equation}\label{eq158}
      \langle\Delta\rangle =           \frac{\sum_\mu\Delta_\mu v^2_\mu}
                                            {\sum_\mu           v^2_\mu}
                           = \frac{1}{N^\tau}\sum_\mu\Delta_\mu v^2_\mu  .
   \end{equation}

Equations (\ref{eq149}) and (\ref{eq150}) misleadingly resemble
those of the simple BCS theory \cite{[RS80]}.  However, in the
HFB theory, $\epsilon_\mu$ is not the single-particle energy
(i.e., the eigenvalue of $h$) but the diagonal matrix element of
$h$ in the canonical basis. Similarly, $\Delta_\mu$ does not
represent the pairing gap in the state $\breve\psi_\mu$, and
$E_\mu$ is not the quasiparticle energy $E$.  However, since
these quantities define the occupation probabilities, they play
a very important role in an interpretation of the HFB results,
and many intuitive, quantitative, and useful  features of the
BCS theory can be reinterpreted in terms of the canonical
representation (cf.{\ }Sec.{\ }\ref{sec3f}).

In particular, the  average values of single-particle p-h and
p-p Hamiltonians fulfill the following self-consistency
equations:
   \begin{mathletters}\label{eq151}\begin{eqnarray}
       \epsilon_\mu =   T_{\mu\mu} +&& \frac{1}{2}\sum_\nu \breve F_{\mu\nu}
                       \left(1-\frac{\epsilon_\nu-\lambda}{E_\nu}\right)
                             ,   \label{eq151a} \\
       \Delta_\mu   =               &&\frac{1}{4}\sum_\nu \breve G_{\mu\nu}
                                \frac{\Delta_\nu}{E_\nu}
                             .   \label{eq151b}
   \end{eqnarray}\end{mathletters}%
{}For a given interaction $\breve{F}_{\mu\nu}$ and
$\breve{G}_{\mu\nu}$, Eqs.{\ }(\ref{eq151}) represent a set of
nonlinear equations which determine $\epsilon_\mu$ and
$\Delta_\mu$.  Equations for $\epsilon_\mu$ (\ref{eq151a}) and
for $\Delta_\mu$ (\ref{eq151b}) are coupled by the values of
$E_\nu$ (\ref{eq150}), which depend on both  $\epsilon_\mu$ and
$\Delta_\mu$. However, it is clear that the interaction in the
p-h channel mainly influences the values of $\epsilon_\mu$,
while that in the p-p channel -- $\Delta_\mu$.

Unfortunately, Eqs.{\ }(\ref{eq151}) cannot replace the
original HFB equations, because they require the knowledge of
the canonical basis to determine the $\breve{F}_{\mu\nu}$ and
$\breve{G}_{\mu\nu}$ matrices.  The only way to determine the
canonical basis is to solve the original  HFB equation
(\ref{eq143}), and then to diagonalize the density matrix
(\ref{eq144a}).  Moreover, solving Eqs.{\ }(\ref{eq151})
ensures that only the $\mu$=$\nu$ subset of variational
equations (\ref{eq145a}) is  met, the minimum of energy being
obtained by solving the whole set (i.e., for all indices $\mu$
and $\nu$).

The diagonalization of the energy matrix $\breve{E}_{\mu\nu}$
gives the spectrum of HFB eigenenergies, $E_n$:
   \begin{equation}\label{Ediag}
   \sum_{\nu}\breve{E}_{\mu\nu}{\cal{U}}_{n\nu}=E_n {\cal{U}}_{n\mu}.
   \end{equation}
The matrix ${\cal{U}}_{n\mu}$ represents the unitary
transformation from the canonical to the quasiparticle basis
\cite{[RS80]}.  Its matrix elements provide the link between the
quasiparticle energies $E_n$ and the diagonal matrix elements
$E_\mu$ which define the occupation probabilities, i.e.,
   \begin{equation}\label{Esumrule}
   E_\mu=\sum_n E_n |{\cal{U}}_{n\mu}|^2.
   \end{equation}

\section{Coupling to the positive-energy states}
\label{sec3}

{}For weakly bound nuclei one may expect that the particle
continuum influences the ground-state properties in a
significant way.  As discussed in Sec.{\ }\ref{sec2bcb}, the
phase space corresponding to positive single-particle energies
should not be confused with the continuum of scattering states
which asymptotically behave as plane waves, and are significant
for genuine scattering phenomena.

\subsection{Boundary conditions}
\label{sec3a}

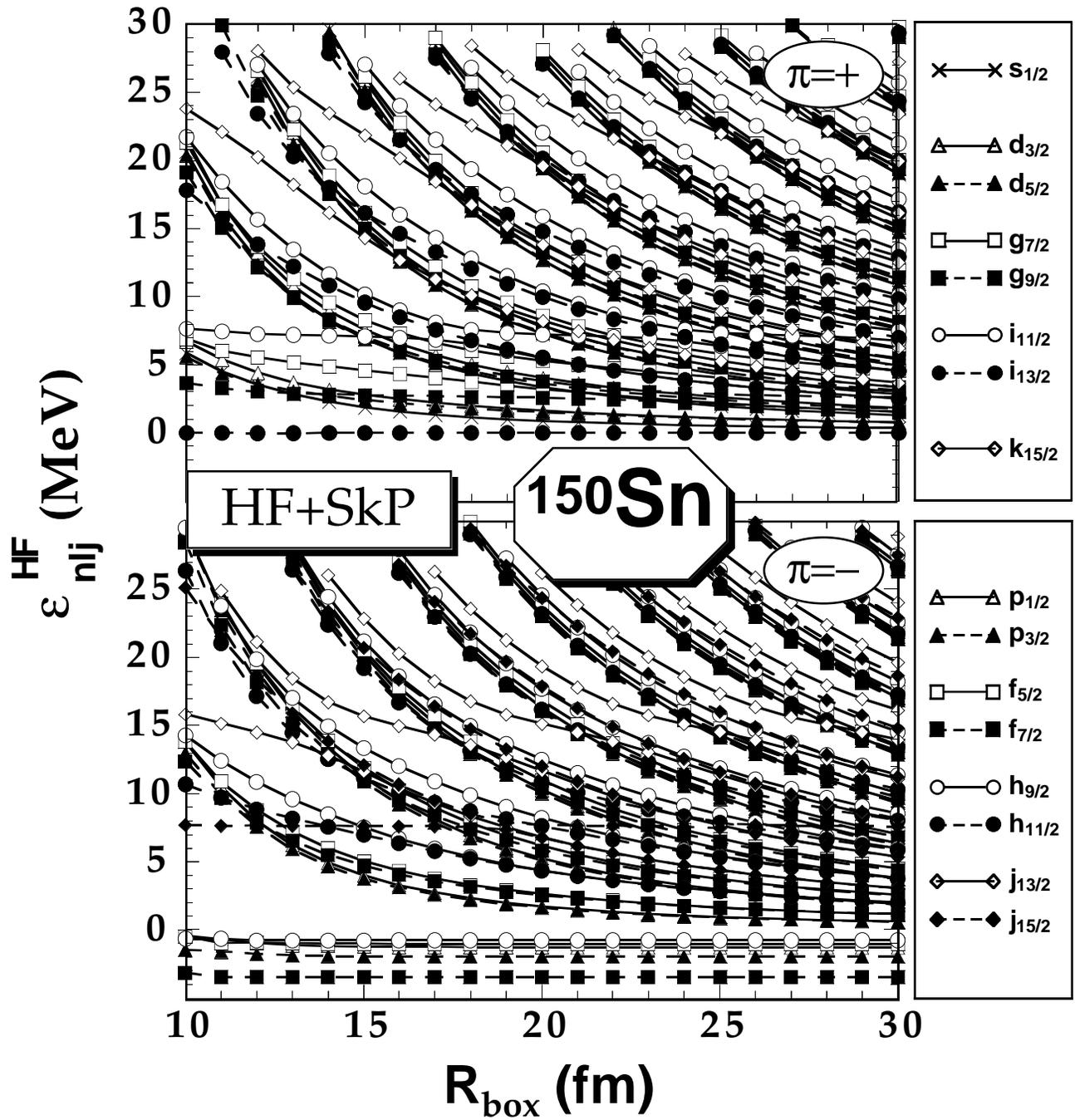
\begin{figure}[ht]
\caption[FF]{
Weakly bound and unbound self-consistent single-neutron HF+SkP
energies $\epsilon^{\text{HF}}_{n{\ell}j}$ for $^{150}$Sn as
function of $R_{\text{box}}$. Top and bottom panels show states
of positive and negative parity, respectively.
}
\label{FIG12}
\end{figure}

Properties of the continuum scattering states are intuitively
well understood in terms of unpaired single-particle orbits.
Shown in Fig.{\ }\ref{FIG12} are  the self-consistent HF+SkP
neutron single-particle energies in $^{150}$Sn,
$\epsilon^{\text{HF}}_{nlj}$, as functions of the radius
$R_{\text{box}}$ of the spherical box in which the HF equations
are solved. It is assumed that the following boundary condition
holds for all single-particle wave functions:
   \begin{equation}\label{eq153}
       \psi_\mu(R_{\text{box}}) = 0.
   \end{equation}
{}For bound single-particle states,
$\epsilon^{\text{HF}}_{nlj}$$<$0, the effect of increasing
$R_{\text{box}}$ beyond 10\,fm is insignificant.  As seen in
Fig.{\ }\ref{FIG12}, the energies of the least bound 3$p$,
2$f$, 1$h_{9/2}$, and 1$i_{13/2}$ states, which form the
82$\leq$$N$$\leq$126 shell, are independent of $R_{\text{box}}$.

The boundary condition (\ref{eq153}) leads to a discretization
of the continuum by selecting only those states which have a
node at $r$=$R_{\text{box}}$. When $R_{\text{box}}$ increases,
the density of the low-energy continuum states increases as
$R_{\text{box}}^3$. This effect is very well visible in
Fig.{\ }\ref{FIG12}.  Among those states whose energies
decrease with $R_{\text{box}}$, one may easily distinguish some
quasi-bound states, which have energies fairly independent of
$R_{\text{box}}$. In Fig.{\ }\ref{FIG12} these are the
high-$\ell$ states $i_{11/2}$, $j_{13/2}$, $j_{15/2}$, and
$k_{15/2}$.  However, at some values of $R_{\text{box}}$ they
are crossed by, and they interact with, the real continuum
states (plane waves) of the same quantum numbers, and their
precise determination is, in practice, very difficult.

A solution of the HFB equation (\ref{eq152}) in the spherical
box amounts to using the analogous boundary conditions,
   \begin{equation}\label{eq154}
       \phi_1(E,R_{\text{box}}) = \phi_2(E,R_{\text{box}}) = 0,
   \end{equation}
for both components of the HFB wave function.  As a result, the
quasiparticle continuum of states with $|E|$$>$$-\lambda$ is
discretized and becomes more and more dense with increasing
$R_{\text{box}}$. However, as discussed in
Sec.{\ }\ref{sec2bc}, the density matrix depends only on the
localized (lower) components of the quasiparticle wave functions
and, therefore, is very well stable with increasing
$R_{\text{box}}$. By the same token, the properties of the
canonical-basis states, which are the eigenstates of the density
matrix, are also asymptotically stable. Of course, the bigger
the value of $R_{\text{box}}$, the larger is the numerical
effort required to solve the HFB equations. Consequently, it is
important to optimize the value of $R_{\text{box}}$; i.e., to
use the smallest box sizes which reproduce all interesting
physical properties of the system.

Apart from ours, there are also other possible approaches to
solving the HFB eigenproblem; in particular: (i) the
diagonalization in the large harmonic oscillator basis, and (ii)
the two-step diagonalization. Scheme (i) has been used, e.g., in
the HFB+Gogny calculations or in the deformed HFB+SkP
calculations of Ref.{\ }\cite{[Sta91]}. Its limitations, due
to the incorrect asymptotics,  are discussed in
Sec.{\ }\ref{sec3e} below. In method (ii) one first solves the
HF  problem and then diagonalizes the full HFB Hamiltonian in
the HF basis.  Such a strategy has been suggested in
Ref.{\ }\cite{[Zve85]} and recently adopted in
Ref.{\ }\cite{[Ter95a]}.

\subsection{Canonical single-particle spectrum}
\label{sec3d}

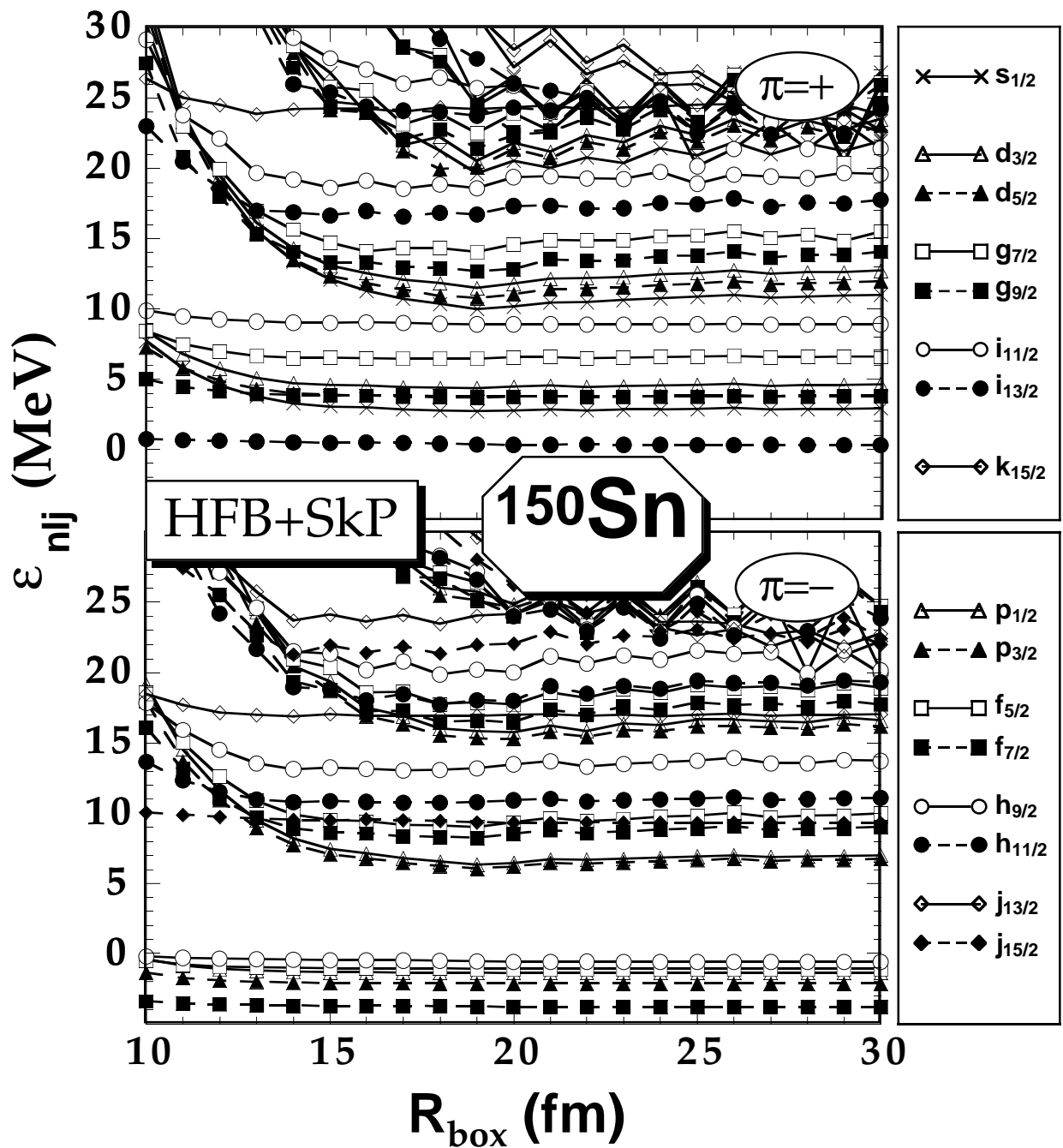
\begin{figure}[ht]
\caption[FF]{
Same as in Fig.~\protect\ref{FIG12}, but for the canonical energies
$\epsilon_{n{\ell}j}$, Eq.{\ }(\protect\ref{eq148a}).
}
\label{FIG10}
\end{figure}

As discussed in Sec.{\ }\ref{sec2bb}, quantities which
determine the p-h properties of the system are the canonical
energies $\epsilon_\mu$ [Eq.{\ }(\ref{eq148a})]. The neutron
canonical energies in $^{150}$Sn are shown in
Fig.{\ }\ref{FIG10} as functions of the box size
$R_{\text{box}}$. In this figure, the single-particle index
$\mu$ is represented by the spherical quantum numbers
$n{\ell}j$; only the states with occupation probabilities
$v^2_{n{\ell}j}$$>$0.0001 are presented. The canonical states
belonging to the shell 82$\leq$$N$$\leq$126 have negative
$\epsilon_{n{\ell}j}$'s, and they are very close to the HF
single-particle energies displayed in Fig.{\ }\ref{FIG12}.
They do not depend on the values of $R_{\text{box}}$ for
$R_{\text{box}}$$>$10\,fm.

At positive values of $\epsilon_{nlj}$, there are several
orbitals which do not depend on the box size even at
$R_{\text{box}}$$<$15\,fm.  These states correspond to the
high-$\ell$ quasibound states $i_{11/2}$, $j_{13/2}$,
$j_{15/2}$, and $k_{15/2}$,  already identified in the HF
spectrum of Fig.{\ }\ref{FIG12}.  The values of
$\epsilon_{n{\ell}j}$ for these states are only slightly higher
than the corresponding values of
$\epsilon^{\text{HF}}_{n{\ell}j}$.  However, these quasibound
canonical-basis states are not accompanied by the sea of
plane-wave  scattering states  (cf.  the $j_{13/2}$, and the
$k_{15/2}$ states in Figs.{\ }\ref{FIG12} and \ref{FIG10}).
One can thus say that the canonical-basis states represent the
quasibound states well decoupled from the scattering continuum.

Many other canonical-basis states, especially those with low
orbital angular momenta $\ell$, significantly depend on the box
size up to about $R_{\text{box}}$=18\,fm, and then stabilize.
Therefore, in all subsequent calculations we use a ``safe"
value of $R_{\text{box}}$=20\,fm, unless stated otherwise.

Above 20\,MeV there appear states with canonical energies
fluctuating  with $R_{\text{box}}$.  These states have very
small occupation probabilities close to the limiting value of
$v^2_{n{\ell}j}$=0.0001, and their determination as eigenstates
of the density matrix is prone to large numerical uncertainties
(see Sec.{\ }\ref{sec2bcb}). One should note that the physical
observables are calculated directly by using the HFB density
matrices, and the above numerical uncertainties do not affect
the results obtained within the HFB theory.

As pointed out in Ref.{\ }\cite{[Dob94]}, the canonical
spectrum presented in Fig.{\ }\ref{FIG10} can be used to
analyze the shell effects far from stability. In particular, the
size of the $N$=126 gap is very small (a 2\,MeV gap between the
1$i_{13/2}$ and 4$s_{1/2}$ states), and hence it cannot yield
any pronounced shell effect (seen, e.g., in the behavior of the
two-neutron separation energies, Sec.{\ }\ref{sec4c}). This
shell-gap quenching is not a result of a too small value of the
spin-orbit splitting. Indeed, a larger spin-orbit strength would
push  the 1$i_{13/2}$ level down in energy, without affecting
the size of the $N$=126 shell gap (several negative-parity
states  are nearby).  The $N$=126 gap, which is equal to about
4\,MeV at $R_{\text{box}}$=10\,fm, closes up with increasing
$R_{\text{box}}$ due to the several low-$\ell$ states whose
energies steadily decrease.  This effect can be attributed to
the pairing-induced  coupling with the positive-energy states
(see Sec.{\ }\ref{sec3e}).

In the energy window between 0 and 20\,MeV, the density of
single-particle canonical energies is fairly uniform and no
pronounced shell effects are visible.  Since the Fermi energy
must stay at negative values, this region of the phase space
cannot be reached.  However, one may say that the influence of
the positive-energy spectrum on the bound states (had we
analyzed it in terms of, e.g., the Strutinsky averaging) is
characterized by a rather structureless distribution of states.
Above 20\,MeV, the occupation probabilities rapidly decrease
(cf. Table \ref{TAB01}), and this part of the phase space can
safely be disregarded, provided one stays in the canonical
basis.

\subsection{Single-quasiparticle spectrum}
\label{sec3f}

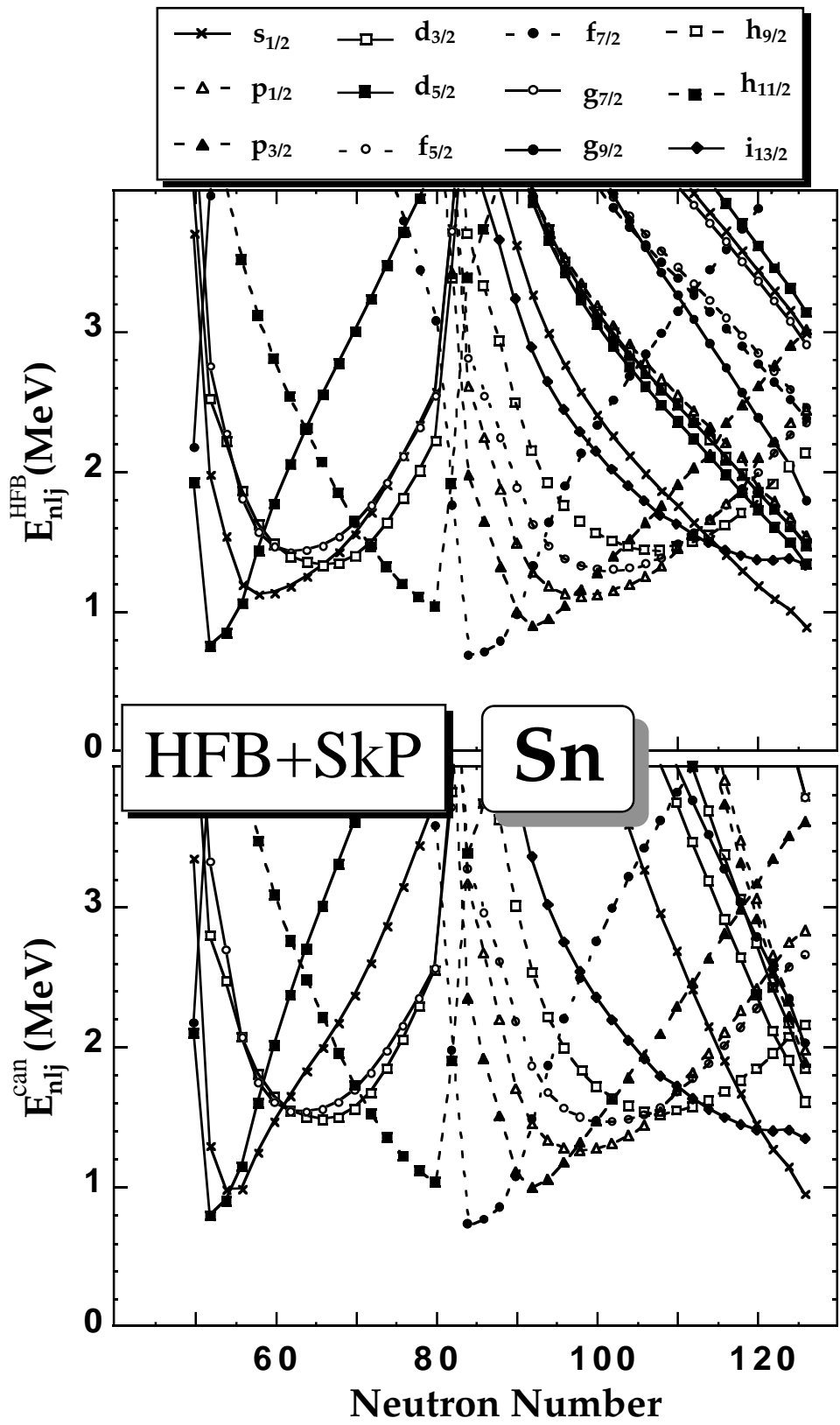
\begin{figure}[ht]
\caption[FF]{
Self-consistent single-quasineutron HFB+SkP energies
$E^{\text{HFB}}_{n{\ell}j}$ (top panel) compared with the
BCS-like canonical single-quasineutron energies
$E^{\text{can}}_{n{\ell}j}$ (Eq.{\ }(\ref{eq150})) (bottom
panel) for the tin isotopes.
}
\label{FIG14}
\end{figure}

The eigenvalues of the HFB equation (\ref{eq152})
(single-quasiparticle energies) carry information on the
elementary modes of the system.  The lowest single-quasineutron
energies $E^{\text{HFB}}_{n{\ell}j}$ in tin isotopes between
$N$=50 and $N$=126 are shown in Fig.{\ }\ref{FIG14} (top
panel).  Apart from the magic shell gaps at $N$=50 and $N$=82,
where the single-quasiparticle energies exhibit sudden jumps,
they depend rather  smoothly  on neutron number. For a given
orbital $n{\ell}j$, the minimum of $E^{\text{HFB}}_{n{\ell}j}$
is attained in the isotope where the corresponding
single-particle state is closest to the Fermi energy. Hence,
from Fig.{\ }\ref{FIG14} one can infer the order of
single-particle energies in the beginning of the
50$\leq$$N$$\leq$82 shell as $2d_{5/2}$, $3s_{1/2}$, $2d_{3/2}$,
$1g_{7/2}$, and $1h_{11/2}$. Similarly, the predicted order at
the bottom of the next major shell is $2f_{7/2}$, $3p_{3/2}$,
$3p_{1/2}$, $2f_{5/2}$, $1h_{9/2}$, and $1i_{13/2}$.  The order
of spherical single-particle states does vary with $N$.  For
instance, according to the HFB+SkP calculations of
Fig.{\ }\ref{FIG14}, the $1g_{7/2}$ shell never becomes lowest
in energy, as it should have done, had the single-particle
energies been $N$-independent.

Noteworthy is the fact that, due to the strong interaction with
the low-$\ell$ continuum (cf.{\ }Sec.{\ }\ref{sec3d}), the
$4s_{1/2}$ excitation becomes lowest at $N$$>$114.  Above the
$4s_{1/2}$ state there appear several quasiparticle states with
excitation energies rapidly decreasing with $N$.  These orbitals
represent the low-energy continuum states.  They are very close
in energy, exhibit small spin-orbit splitting, and the lowest of
them are the low-$\ell$ states: $4p_{1/2}$, $4p_{3/2}$,
$3d_{3/2}$, and $3d_{5/2}$. All these features are
characteristic of the continuum states \cite{[Dob95d]}.  Still
higher in energy, one may distinguish a similar doublet of the
$3f_{5/2}$ and $3f_{7/2}$ states, as well as the $2g_{9/2}$
state which represents a high-$\ell$ resonance.

The bottom panel of Fig.{\ }\ref{FIG14} shows similar results
for the BCS-like canonical energies $E_\mu$ defined in
Eq.{\ }(\ref{eq150}), and denoted here by
$E^{\text{can}}_{n{\ell}j}$.  A comparison between
$E^{\text{HFB}}_{n{\ell}j}$ and $E^{\text{can}}_{n{\ell}j}$
illustrates the fact that the {\em lowest} elementary
excitations of the nucleus are equally well described by both
these quantities.  Indeed, a general pattern and, in most cases,
also the values of $E^{\text{HFB}}_{n{\ell}j}$ and
$E^{\text{can}}_{n{\ell}j}$ are very similar. The differences
mainly concern the $s_{1/2}$ states, and also the low-$\ell$
states in the continuum, which in the canonical representation
appear higher in energy (see Table{\ }\ref{TAB01} for the
direct comparison for $s_{1/2}$ states).  On the other hand, the
position of the high-$\ell$ $2g_{9/2}$ resonance is almost
identical in both representations. Such a similarity supports
the supposition (Ref.{\ }\cite{[Dob94]} and
Sec.{\ }\ref{sec3d}) that the canonical single-particle
energies, which are the main ingredients of
$E^{\text{can}}_{n{\ell}j}$, constitute a fair representation of
single-particle and single-quasiparticle properties of the
system.

\subsection{Relation between canonical and single-quasiparticle
            wave functions}
\label{canqua}

As discussed in Sec.{\ }\ref{sec2ab}, the canonical states
constitute a basis in which the independent-quasiparticle state
$|\Psi\rangle$ has the form of a product of correlated pairs
[Eq.{\ }(\ref{eq116})]. Therefore, these states can be
considered as fundamental building blocks describing the pairing
correlations in a many-fermion system. On the other hand, the
canonical states are determined by a solution of the HFB
equation -- the single-quasiparticle states.

Since the canonical states constitute an orthonormal ensemble,
the lower and upper HFB components can be expanded as
   \begin{mathletters}\label{eqqcan}\begin{eqnarray}
    \phi_1  (E_n,\bbox{r} \sigma )&=&\sum_{\mu} {\cal{A}}^{(1)}_{n\mu}
   \breve\psi_{\mu}(\bbox{r}\sigma)
                             ,   \label{eqqcanb} \\
   \phi_2  (E_n,\bbox{r} \sigma )&=&\sum_{\mu} {\cal{A}}^{(2)}_{n\mu}
   \breve\psi_{\mu}(\bbox{r}\sigma)
                             ,   \label{eqqcana}
   \end{eqnarray}\end{mathletters}%
where
   \begin{equation}\label{overlaps}
       {\cal{A}}^{(i)}_{n\mu}\equiv
                      \int\text{d}^3\bbox{r}\sum_{\sigma}
                      \breve\psi^*_{\mu}(\bbox{r}\sigma)
                      \phi_i(E_n,\bbox{r} \sigma ) ~~~(i=1,2)
   \end{equation}
are the associated overlaps.  In order to find the relation
between ${\cal{A}}^{(1)}_{n\mu}$ and ${\cal{A}}^{(2)}_{n\mu}$
one can employ Eqs.{\ }(\ref{eq122}) and (\ref{eq144}) for the
HFB densities.  This gives the canonical wave functions
expressed as linear combinations of the {\em lower} HFB
components:
   \begin{mathletters}\label{eqqcan1}\begin{eqnarray}
   v^2_\mu \breve\psi_{\mu}(\bbox{r}\sigma) &=&
              \sum_{n}{\cal{A}}^{(2)}_{n\mu}\phi_2  (E_n,\bbox{r} \sigma )
                             ,   \label{eqqcan1a} \\
   -u_\mu v_\mu \breve\psi_{\mu}(\bbox{r}\sigma) &=&
              \sum_{n}{\cal{A}}^{(1)}_{n\mu}\phi_2  (E_n,\bbox{r} \sigma )
                             . \label{eqqcan1b}
   \end{eqnarray}\end{mathletters}%
One should note that the expansions (\ref{eqqcan1}) are valid
regardless of the fact that the lower components
$\phi_2(E_n,\bbox{r}\sigma)$ {\em do not constitute} an
orthogonal ensemble of wave functions. By multiplying both sides
of Eqs.{\ }(\ref{eqqcan1a}) and (\ref{eqqcan1b}) with
$\breve\psi^*_{\nu}(\bbox{r}\sigma)$ and taking the scalar
product, one arrives at the orthogonality relations:
   \begin{mathletters}\label{eqqcan2}\begin{eqnarray}
      \sum_{n}{\cal{A}}^{(2)}_{n\mu}{\cal{A}}^{(2)*}_{n\nu}
         &=& v^2_\mu \delta_{\mu\nu}
                             ,   \label{eqqcan2a} \\
      \sum_{n}{\cal{A}}^{(1)}_{n\mu}{\cal{A}}^{(2)*}_{n\nu}
         &=& -u_\mu v_\mu \delta_{\mu\nu}
                             . \label{eqqcan2b}
   \end{eqnarray}\end{mathletters}%
The above identities express the fact, that both
${\cal{A}}^{(2)}_{n\mu}$ and ${\cal{A}}^{(1)}_{n\mu}$ are
related to the transformation matrix ${\cal{U}}_{n\nu}$ defined
in Eq.{\ }(\ref{Ediag}):
   \begin{equation}\label{transl}
   {\cal{A}}^{(2)}_{n\mu}=v_\mu  {\cal{U}}_{n\mu} \quad,\quad
   {\cal{A}}^{(1)}_{n\mu}=-u_\mu  {\cal{U}}_{n\mu},
   \end{equation}
and Eqs.{\ }(\ref{eqqcan2}) reflect the unitarity of
${\cal{U}}_{n\nu}$.  Equations (\ref{transl}) can be easily
derived by inserting expansions (\ref{eqqcan}) into the HFB
equation (\ref{eq143}), and then expressing the matrix
$\breve{E}_{\mu\nu}$ (\ref{eq145b}) in its eigensystem
(\ref{Ediag}).

It is instructive to express the upper HFB component in a form
similar to that of Eq.{\ }(\ref{eqqcana}):
   \begin{equation}\label{phiB}
     \phi_1  (E_n,\bbox{r} \sigma )=-\sum_{\mu} \frac{u_\mu}{v_\mu}
     {\cal{A}}^{(2)}_{n\mu}
      \breve\psi_{\mu}(\bbox{r}\sigma).
   \end{equation}
For $E_n>-\lambda$, the upper component
$\phi_1(E_n,\bbox{r}\sigma)$ is the scattering wave function. It
can be formally expanded in the {\em localized} canonical wave
functions according to Eq.{\ }(\ref{phiB}), but the main
contribution comes from the particle-like states with very small
values of ${v^2_\mu}$. Hence, this relation is not too useful in
practical applications.

\subsection{Spectral distribution for the canonical-basis wave functions}
\label{sec2bcd}

In order to discuss the importance of the particle continuum on
the structure of canonical states, it is interesting to see how
a given canonical state is distributed among the
single-quasiparticle states.  For this, it is convenient to
rewrite Eq.{\ }(\ref{eqqcan1a}) in the following way:
   \begin{equation}\label{spectral2}
   \breve\psi_{\mu}(\bbox{r}\sigma)=
          \sum_{0<E_n<E_{\text{max}}}  \frac{{\cal{S}}_{n\mu}}{\sqrt{N_n}}
                                 \phi_2  (E_n,\bbox{r} \sigma ).
   \end{equation}
The spectral amplitudes ${\cal{S}}_{n\mu}$ define the
distribution of the canonical states among the
single-quasiparticle states. It is important to recall at this
point that the sum in Eq.{\ }(\ref{spectral2}) represents in
fact the discrete ($E_n$$<$$-\lambda$) states and the
discretized ($E_n$$>$$-\lambda$) continuum states, i.e.,
   \begin{eqnarray}
   \breve\psi_{\mu}(\bbox{r}\sigma) &=&
          {\displaystyle\sum_{0<E_n<-\lambda}
                                 \frac{{\cal{S}}_{n\mu}}{\sqrt{N_n}}}
                                 \phi_2  (E_n,\bbox{r} \sigma ) +
                                                         \nonumber \\
        && {\displaystyle\int_{-\lambda}^\infty \text{d}n(E)
                                 \frac{{\cal{S}}_{E,\mu}}{\sqrt{N_E}}}
                                 \phi_2  (E  ,\bbox{r} \sigma ) ,
                                                         \label{spectral}
   \end{eqnarray}
with the spectral amplitudes ${\cal{S}}_{n\mu}$ and
${\cal{S}}_{E,\mu}$ pertaining to the discrete and continuous HFB
spectrum, respectively.

The spectral amplitudes can be expressed in terms of matrices
${\cal{A}}^{(2)}_{n\mu}$ or ${\cal{U}}_{n\mu}$ introduced in
Sec.{\ }\ref{canqua}:
   \begin{equation}\label{ampli}
   {\cal{S}}_{n\mu}  =  \frac{\sqrt{N_n}}{v_\mu^2}
   {\cal{A}}^{(2)}_{n\mu} = \frac{\sqrt{N_n}}{v_\mu}{\cal{U}}_{n\mu}.
   \end{equation}
We have included in ${\cal{S}}_{n\mu}$ the norms $N_n$ of the
lower components, Eq.{\ }(\ref{norms}).  In this way, the
values of spectral amplitudes measure the influence of
quasiparticle states irrespective of the overall magnitude of
their lower components.

Before discussing the properties of the spectral amplitudes,
let us  write down the two sum rules:
   \begin{equation}\label{eqqcan4a}
 1   =   \sum_{\mu}|{\cal{S}}_{n\mu}|^2\frac{v_\mu^2}{N_n}
     =   \sum_{n}  |{\cal{S}}_{n\mu}|^2\frac{v_\mu^2}{N_n},
   \end{equation}
   \begin{equation}\label{eqqcan4b}
 1   =   \sum_{\mu}|{\cal{S}}_{n\mu}|^2\frac{v_\mu^4}{N_n^2}.
   \end{equation}
The first two sum rules, Eq.{\ }(\ref{eqqcan4a}), come from
the the unitarity of ${\cal{U}}_{n\mu}$. The last one,
Eq.{\ }(\ref{eqqcan4b}), expresses the condition defining the
norm of the lower HFB component.

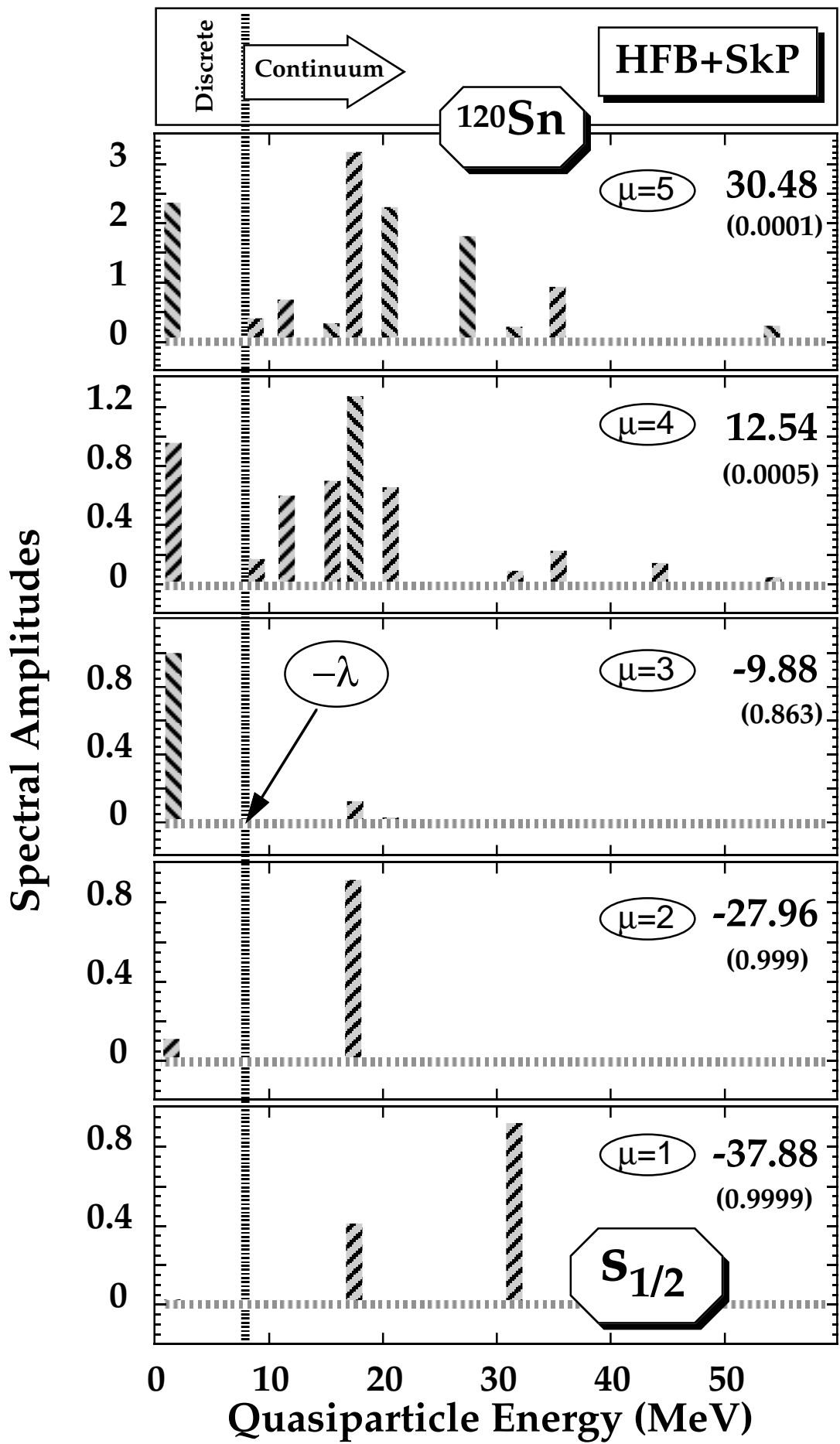
\begin{figure}[ht]
\caption[FF]{
The HFB+SkP spectral amplitudes ${\cal{S}}_{n\mu}$
(\protect\ref{spectral2}) of the canonical $s_{1/2}$ states in
$^{120}$Sn with $\mu$=1--5.  The corresponding canonical energy
$\epsilon_\mu$ is given in MeV and the occupation probability
$v_\mu^2$ is displayed  in parentheses.  All the amplitudes
${\cal{S}}_{n\mu}$ for $\mu$=1 have been assumed to be positive.
 This defines  the relative phases of the spectral amplitudes
for $\mu$$>$1 (shown by bars hashed in opposite directions).
For $E<-\lambda$ the quasiparticle spectrum is discrete, while
for $E>-\lambda$ it is represented by the discretized continuum.
}
\label{FIG33}
\end{figure}

In Fig.{\ }\ref{FIG33} are shown the spectral amplitudes for
the $s_{1/2}$ canonical states in $^{120}$Sn
(cf.{\ }Secs.{\ }\ref{sec2bca} and \ref{sec2bcb}).  The
phases of the single-quasiparticle wave functions have been
fixed in such a way that all the amplitudes ${\cal{S}}_{n\mu}$
for $\mu$=1 are positive (some of these amplitudes  are too
small to be displayed  in the figure). This defines  the
relative phases of the spectral amplitudes for $\mu$$>$1. Then,
the positive and negative amplitudes are in
Fig.{\ }\ref{FIG33} shown by bars hashed in opposite
directions.  Results shown in this figure pertain to the same
single-quasiparticle and canonical states as those shown in
Figs.{\ }\ref{FIG21} and \ref{FIG22}, respectively, and in
Table \ref{TAB01}.

The lowest panel in Fig.{\ }\ref{FIG33} shows that the
$1s_{1/2}$ canonical state ($\mu$=1) is composed mainly of two
components corresponding to the two deep-hole quasiparticles at
$E_8$=31.64\,MeV and $E_5$=17.60\,MeV.  Similarly, the $\mu$=2
and $\mu$=3 canonical states are mixtures of the
$E_5$=17.60\,MeV and $E_1$=1.54\,MeV quasiparticles. For all
three of these canonical states, the diagonal amplitudes
dominate.

Another pattern appears for the positive-energy canonical
states; i.e., for $\mu$=4 and $\mu$=5.  These two canonical
states contain large components of the hole-like quasiparticles
at $E_5$=17.60\,MeV and $E_1$=1.54\,MeV, but in addition, they
also acquire large components of  the particle-type
quasiparticles belonging to the continuum.  These continuum
components are centered around 15 and 20\,MeV for $\mu$=4 and
$\mu$=5, respectively. This illustrates the fact that a correct
description of the positive-energy canonical states requires
solving the HFB equation to rather high energies.  The widths of
the corresponding distributions are rather large, which
indicates that there is not a single resonance in the particle
continuum which would alone describe the high-energy $s_{1/2}$
canonical states. This can be well understood by recalling that
the $\ell$=0  resonances have usually very large widths.

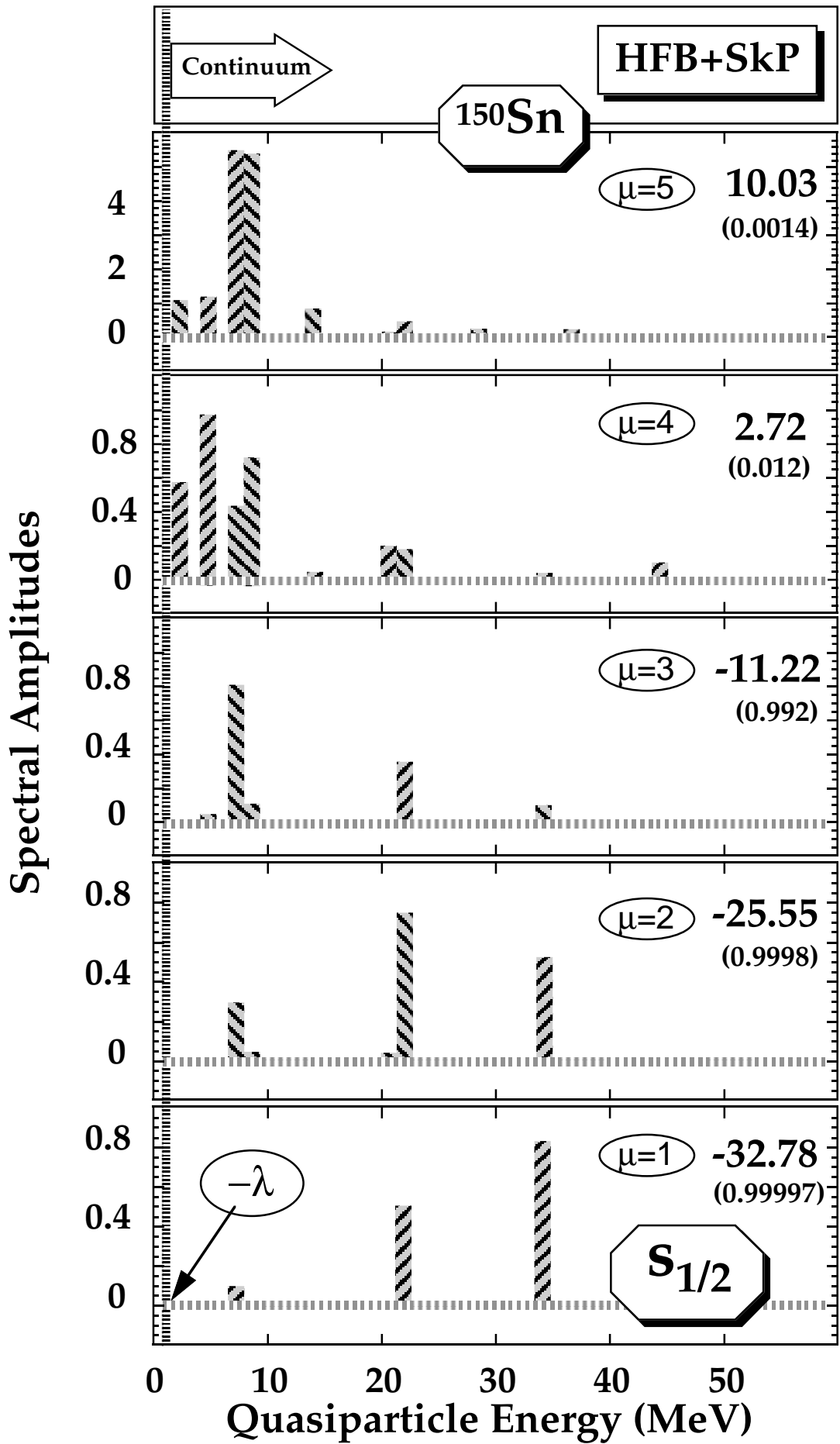
\begin{figure}[ht]
\caption[FF]{
Same as in Fig.~\protect\ref{FIG33}, but for $^{150}$Sn.  No
discrete states appear for $E<-\lambda$.
}
\label{FIG35}
\end{figure}

{}For the drip-line nucleus $^{150}$Sn, the spectral $s_{1/2}$
amplitudes are shown in Fig.{\ }\ref{FIG35}. Similarly to the
case of $^{120}$Sn, the three lowest canonical states for
$\mu$=1, 2, and 3 are mainly composed of the three hole-like
quasiparticles at $E_9$=34.27, $E_7$=22.12, and $E_3$=7.24\,MeV
with dominating diagonal amplitudes. On the other hand, the
low-lying positive-energy canonical $\mu$=4 state has large and
almost equal components coming from the particle-like
quasiparticles at $E_1$=2.40, $E_2$=4.84, and $E_4$=8.93\,MeV.
The following $\mu$=5 canonical state has dominant amplitudes
from the hole-like and particle-like quasiparticles at
$E_3$=7.24 and 8.93\,MeV, respectively. One should note that the
$\mu$=4 and $\mu$=5 canonical $s_{1/2}$ states in $^{150}$Sn
have rather large occupation factors as compared to those in
$^{120}$Sn. Both of them require including the
single-quasiparticle states {\em at least}  up to 10\,MeV. The
following $\mu$=6 state (not shown in the figure) has the
occupation probability of $v^2_6$=0.0003 and the spectral
amplitudes  extending up to 25\,MeV.

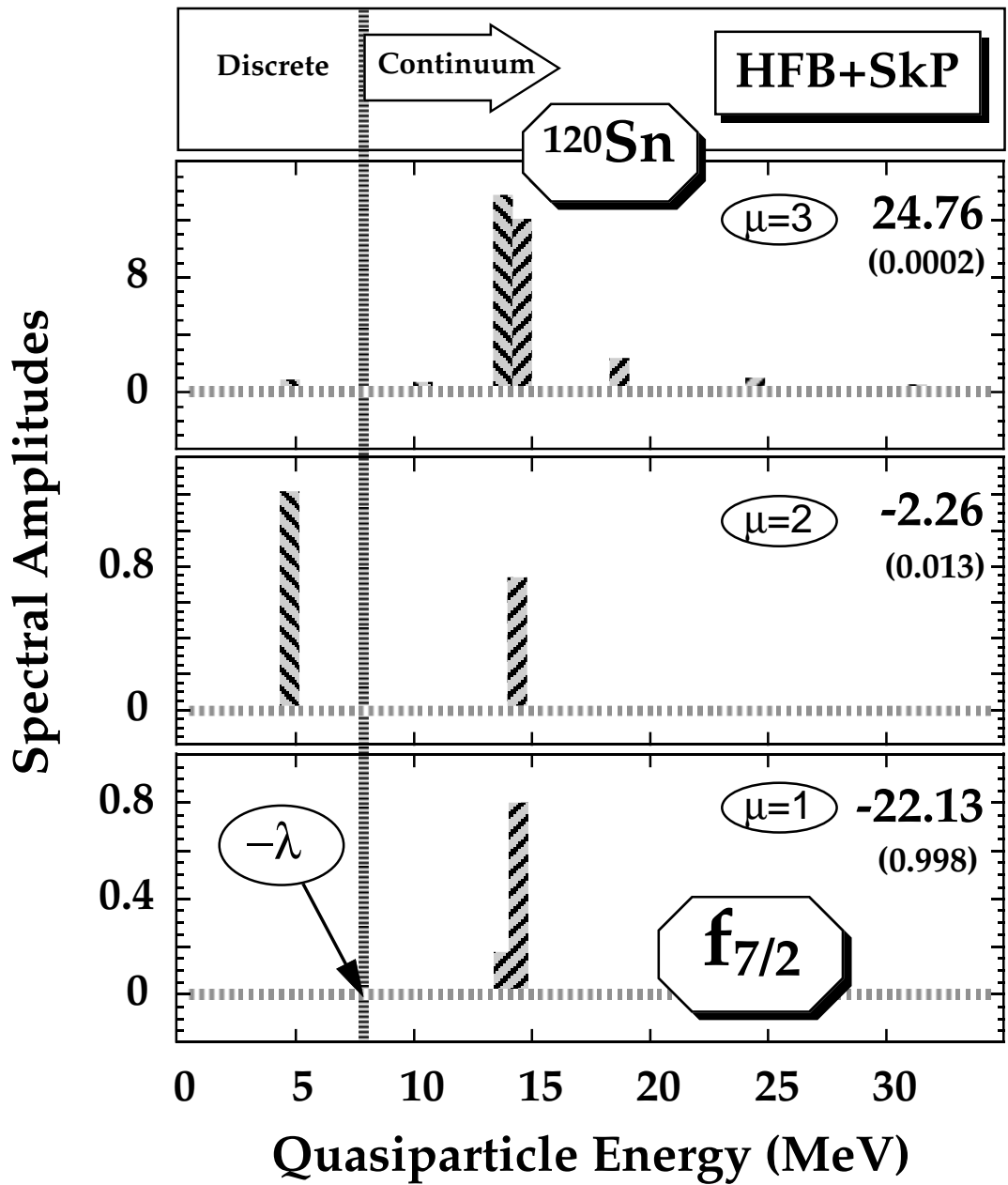
\begin{figure}[ht]
\caption[FF]{
Same as in Fig.~\protect\ref{FIG33}, but for the $\mu$=1--3
canonical $f_{7/2}$ states.
}
\label{FIG34}
\end{figure}

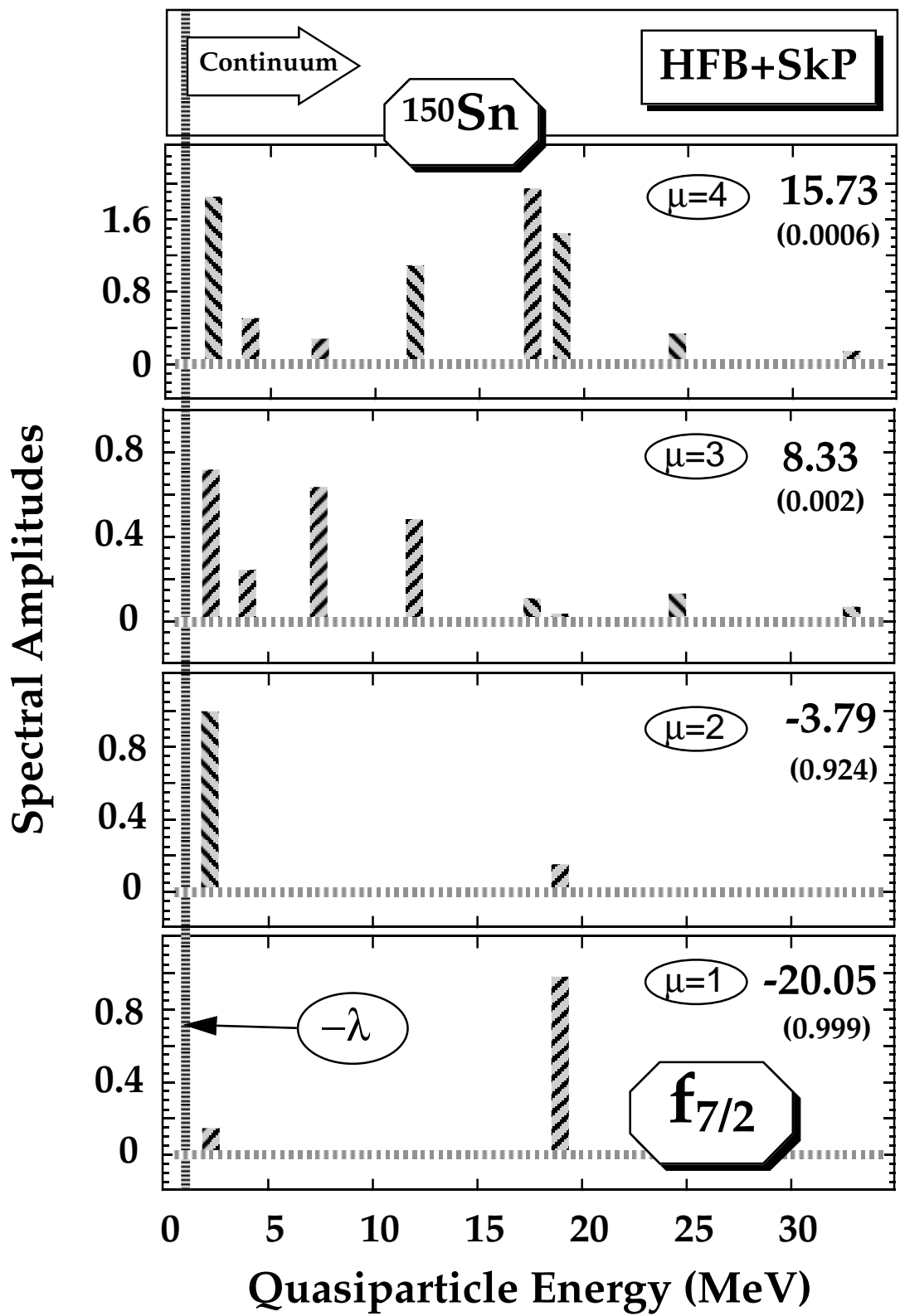
\begin{figure}[ht]
\caption[FF]{
Same as in Fig.~\protect\ref{FIG34}, but for $^{150}$Sn.
}
\label{FIG36}
\end{figure}

The spectral amplitudes for the $f_{7/2}$ states in $^{120}$Sn
and $^{150}$Sn are shown in Figs.{\ }\ref{FIG34} and
\ref{FIG36}, respectively. An interesting situation appears in
$^{120}$Sn where two quasiparticles, one of the particle type
and another one of the hole type, have rather similar
single-quasiparticle energies of 17.63 and 18.97\,MeV.  As a
result, the lowest canonical state ($\mu$=1) acquires a
substantial particle-type quasiparticle component, while both
quasiparticles contribute almost equally to the $\mu$=3
canonical state. In $^{150}$Sn, the positive-energy $f_{7/2}$
canonical states ($\mu$=3 and 4) have large amplitudes from the
hole-like quasiparticles (contributing almost exclusively to the
structure of the negative-energy canonical states with $\mu$=1
(1$f_{7/2}$) and $\mu$=2 (2$f_{7/2}$)), as well as from a wide
distribution of several particle-type quasiparticles extending
up to 20\,MeV.

The spectral amplitudes allow also for a determination of the
asymptotic properties of canonical states.  (See
Ref.{\ }\cite{[Van93]} for a discussion of the the asymptotic
properties of natural orbits.) The lower components
$\phi_2(E_n,\bbox{r}\sigma)$ behave asymptotically as
$\exp($$-$$r\sqrt{2m(E_n-\lambda)/\hbar^2}$
\cite{[Bul80],[Dob84]}.  Therefore, as seen from
Eq.{\ }(\ref{spectral}), the asymptotic properties of
canonical states are governed by the lowest discrete
quasiparticle, provided the corresponding spectral amplitude,
${\cal{S}}_{1\mu}$,  is not equal to zero.  However, if such a
spectral amplitude is non-zero but very small, the corresponding
asymptotic behavior  will be attained only at very large
distances. In practice, the lowest discrete quasiparticle
dominates the asymptotic behavior only if the corresponding
spectral amplitude has a significantly large value.  For the
$s_{1/2}$ states in $^{120}$Sn (Fig.{\ }\ref{FIG33}) such a
situation occurs for the canonical states with $\mu$=2--5 On the
other hand, since the value of $|{\cal{S}}_{1,1}|$ is very
small, the asymptotic behavior of the $\mu$=1 canonical state is
dominated by the hole-like quasiparticle at $E_5$=17.60\,MeV. A
similar situation occurs for the $f_{7/2}$ states in $^{120}$Sn.
Namely, only for the $\mu$=2 canonical state the asymptotic
behavior is determined by the lowest discrete quasiparticle.

An entirely different property can occur in drip-line nuclei,
where the Fermi energy is close to zero  and there may be no
quasiparticle excitations  in the discrete spectrum between 0
and $-\lambda$.  In such a situation, shown in
Figs.{\ }\ref{FIG35} and \ref{FIG36}, the canonical states are
represented by superpositions of lower quasiparticle components
belonging to the particle continuum. Consequently, it is the
integral over the lowest continuum quasiparticle states just
above the $E>-\lambda$ threshold that determines the asymptotic
properties of the canonical states. In other words, the profile
of the level density, $dn(E)/dE$, around $E=-\lambda$ becomes a
crucial factor.  Good examples of a very strong coupling to the
particle continuum are the $\mu$=4 and 5 canonical $s_{1/2}$ and
$f_{7/2}$ states in $^{150}$Sn, where the quasiparticle strength
is distributed in a very wide energy interval ranging from 1.5
to  20\,MeV. On the other hand, the two lowest canonical
$f_{7/2}$ states in $^{150}$Sn can be associated with the two
quasiparticle excitations well localized in energy (see
Fig.{\ }\ref{FIG36}) and their asymptotics is governed by the
energy of the lowest quasiparticle.

An analysis of the spectral distribution, analogous the one
presented above, has recently been performed \cite{[Pol95]} for
the natural orbits in $^{16}$O determined within the Green's
function method using the $NN$ interaction. This method
accounts for a much more general class of correlations as
compared to the HFB correlations of the pairing type studied
here. However, the general features of the spectral
distributions remain essentially the same. Namely, the
low-occupation-number natural orbits are determined mostly
through high-energy continuum contributions, and large box sizes
(15--20\,fm) and large single-particle bases (20 states per
${\ell}j$-block) have to be used to stabilize the solutions.
This is so even if the studied nucleus ($^{16}$O) is
$\beta$-stable, well-bound, and light; one can expect that for
drip-line nuclei the aforementioned features can only be more
pronounced.

\subsection{Asymptotic properties}
\label{sec3c}

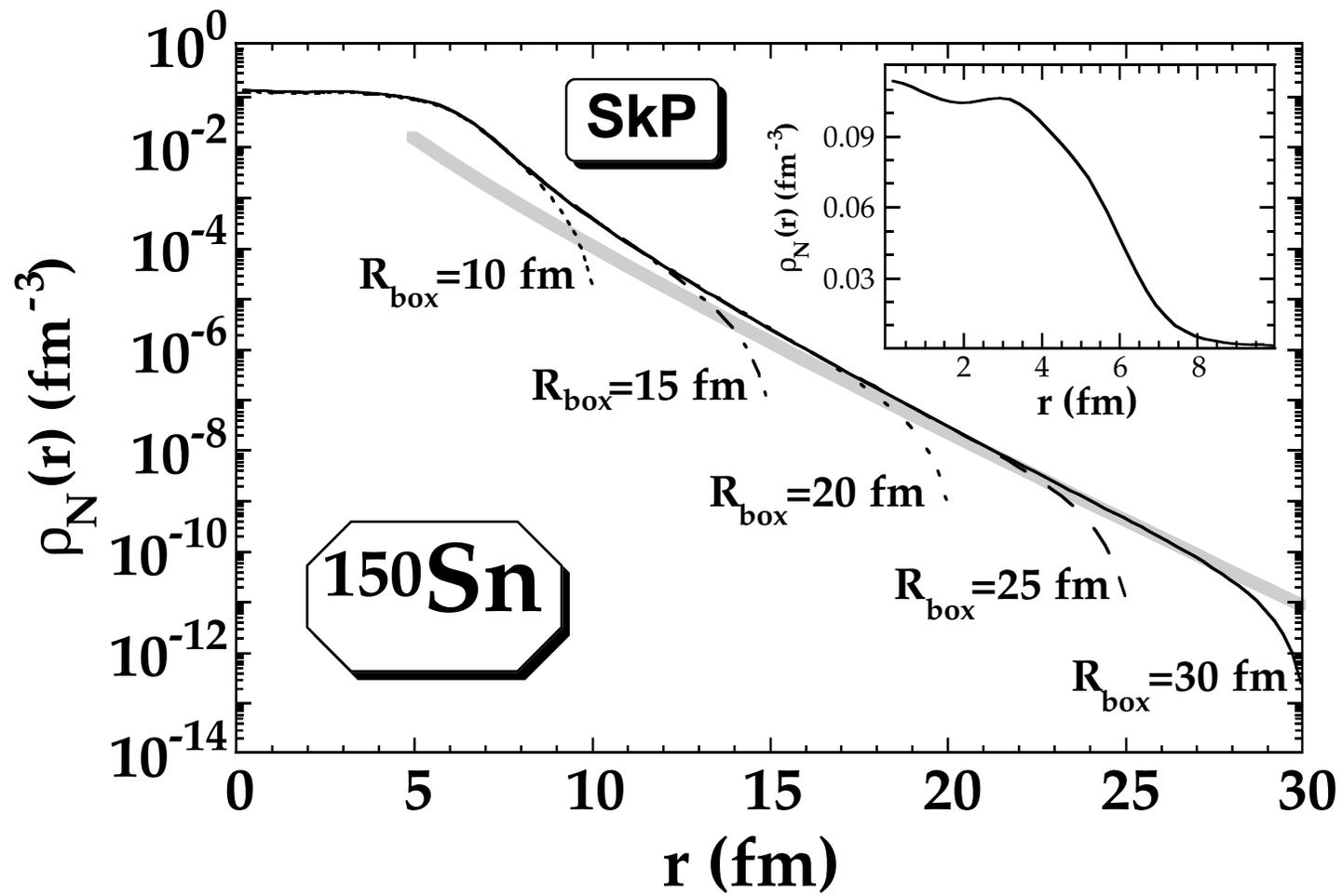
\begin{figure}[ht]
\caption[FF]{
The self-consistent HFB+SkP single-neutron density $\rho_N(r)$
in $^{150}$Sn (in logarithmic scale) calculated with different
values of $R_{\text{box}}$. The inset shows the same data in
linear scale.  The shaded line shows the asymptotic behavior
given by Eq.{\ }(\protect\ref{eq155a}).
}
\label{FIG04}
\end{figure}

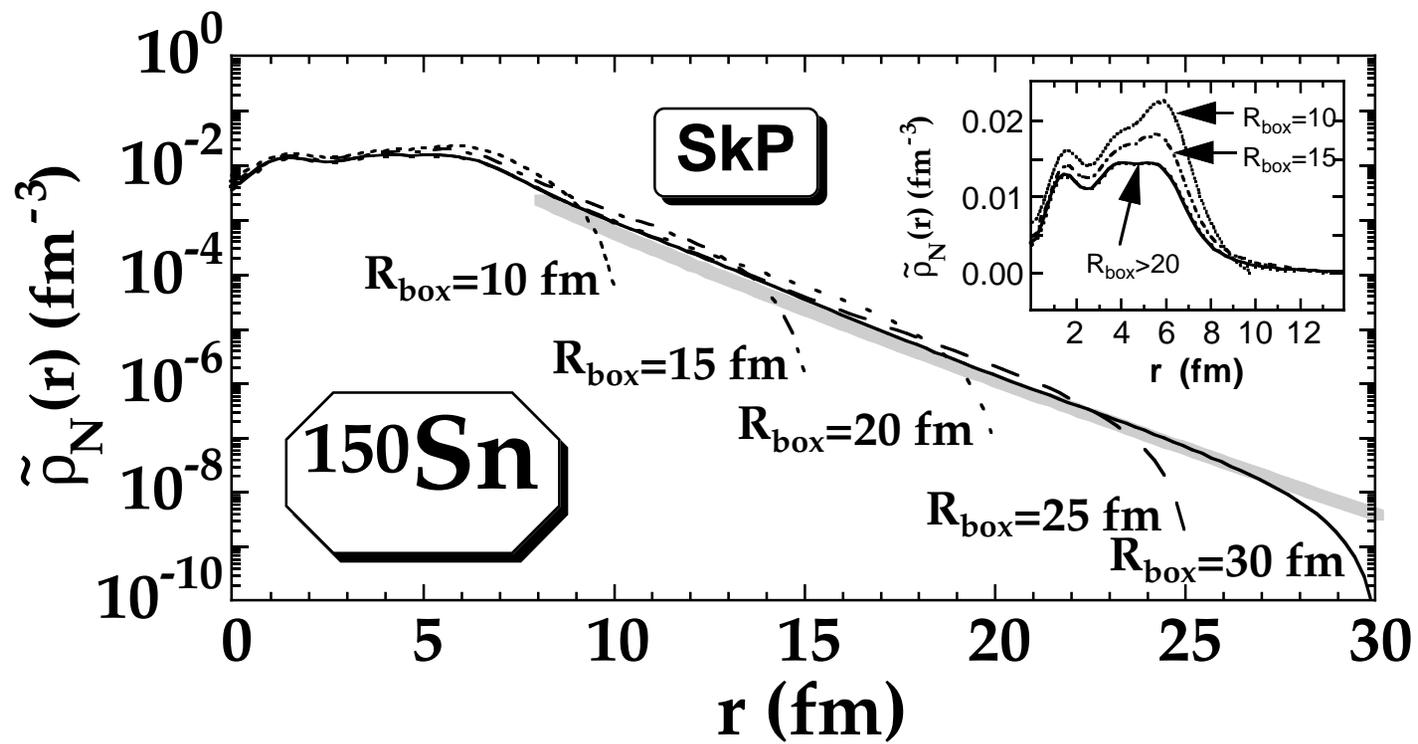
\begin{figure}[ht]
\caption[FF]{
Same as in Fig.{\ }\protect\ref{FIG04}, but for the neutron
pairing density  $\tilde\rho_N(r)$.  The shaded line shows the
asymptotic behavior given by  Eq.{\ }(\protect\ref{eq155b}).
}
\label{FIG04a}
\end{figure}

In the limit of weak binding, radial dimensions of atomic nuclei
increase and it becomes exceedingly important to control the
radial asymptotics of many-body wave functions, not only in
reaction studies but also in nuclear structure applications.
Figure{\ }\ref{FIG04} displays the radial dependence of the
neutron density $\rho(r)$ in $^{150}$Sn calculated with the
values of $R_{\text{box}}$ between 10 and 30\,fm. It is seen
that, for every value of $R_{\text{box}}$, $\rho(r)$ follows its
asymptotic behavior up to about $R_{\text{box}}$$-$3\,fm and
then falls down to zero as a result of the boundary conditions
(\ref{eq154}).  That is, these boundary conditions affect the
density only in a narrow spherical layer of the thickness equal
to about 3\,fm, while inside this layer $\rho(r)$ behaves
independently of the value of $R_{\text{box}}$.  Analogous
results for the pairing density $\tilde\rho(r)$ are shown in
Fig.{\ }\ref{FIG04a}.

At very large distances the asymptotic behavior of the particle
density is governed by the square of the lower component of the
single-quasiparticle wave function corresponding to the lowest
quasiparticle energy $E_{\text{min}}$.  Similarly, the
asymptotic behavior of the pairing density $\tilde\rho(r)$ is
determined by the product of the upper and the lower components
of quasiparticle $E_{\text{min}}$. Using the asymptotic
properties of the HFB wave functions derived in
\cite{[Bul80],[Dob84]}, one obtains:
   \begin{mathletters}\label{eq155}\begin{eqnarray}
           \rho(r) \stackrel{\text{large}~r}{\longrightarrow}
                & \sim {\displaystyle\frac{\exp(-      \chi r)}{r^2}}&
                                                             \quad ; \quad
              \chi=2\kappa_2 , \label{eq155a} \\
     \tilde\rho(r) \stackrel{\text{large}~r}{\longrightarrow}
               & \sim {\displaystyle\frac{\exp(-\tilde\chi r)}{r^2}}&
                                                             \quad ; \quad
        \tilde\chi= \kappa_1 + \kappa_2 , \label{eq155b}
   \end{eqnarray}\end{mathletters}%
where
   \begin{equation}\label{eq156}
     \kappa_1 = \sqrt{\frac{2m(-E_{\text{min}}-\lambda)}{\hbar^2}}\quad,\quad
     \kappa_2 = \sqrt{\frac{2m( E_{\text{min}}-\lambda)}{\hbar^2}}.
   \end{equation}
In the considered example of $^{150}$Sn the calculated values
are $\lambda$=$-$1.46\,MeV and $E_{\text{min}}$= 1.07\,MeV (a
$p_{1/2}$ state). Consequently, $\chi$$\simeq$0.70\,fm$^{-1}$
and $\tilde\chi$$\simeq$0.49\,fm$^{-1}$.  In
Figs.{\ }\ref{FIG04} and \ref{FIG04a} the asymptotic
dependencies given by Eq.{\ }(\ref{eq155}) are shown as shaded
lines.  One can see that for $\rho(r)$ the asymptotic regime is
reached only at distances as large as 25\,fm, which means that
the contributions from other quasiparticle states, and/or from
the next-to-leading-order terms in the Hankel functions, still
influence the particle density at rather large values of $r$.
Interestingly, the pairing density approaches the asymptotic
limit already at $r$$\sim$10\,fm.

A rough estimate of $\chi$ and $\tilde\chi$ can be obtained by
substituting the value of a typical pairing gap
($\Delta$=1\,MeV) for the lowest quasiparticle energy
$E_{\text{min}}$.  {}For stable nuclei
($\lambda$$\simeq$$-$8\,MeV) one obtains
$\chi$$\simeq$1.32\,fm$^{-1}$, while for the one-neutron drip
nuclei, defined by a vanishing separation energy,
$S_n$$\simeq$$\Delta$+$\lambda$$\simeq$0, the result is
$\chi$$\simeq$0.62\,fm$^{-1}$.  This difference illustrates the
increase in the spatial extension of the {\em particle}
densities when going towards the neutron drip line.  On the
other hand, for the {\em pairing} densities the corresponding
numbers are $\tilde\chi$$\simeq$1.24\,fm$^{-1}$ and
$\tilde\chi$=$\chi/2$$\simeq$0.31\,fm$^{-1}$. Therefore, in
stable nuclei both types of densities have rather similar
asymptotic behavior, while in drip-line nuclei the pairing
densities have much longer tails.

In this context, it is instructive to recall the discussion from
Sec.{\ }\ref{sec2ac} regarding  the probabilistic
interpretation of the HFB densities.  The probability
${\cal{P}}_1(x)$ (${\cal{P}}_2(x)$) of finding a particle or a
pair of particles at  $r$=$x$ is proportional to $\rho(x)$ or
$\rho^2(x)+\tilde\rho^2(x)$, respectively. Consequently, in stable nuclei
${\cal{P}}_2(x)$ decays much faster than ${\cal{P}}_1(x)$ at
large distances. This is not true for drip-line nuclei, where
the asymptotics of ${\cal{P}}_1(x)$ and ${\cal{P}}_2(x)$ is the
same.

As discussed above, static pairing correlations can influence
dramatically the asymptotic behavior of density distributions in
drip-line nuclei. In addition, a significant modification of the
density tails comes from the dynamical coupling to collective
modes through the particle continuum. Such a coupling can be
treated in terms of the continuum QRPA and has been shown to be
very important for light systems \cite{[Len93],[Sch95]}.  An
analysis of the asymptotic behavior of the particle density
$\rho(r)$ has recently been performed \cite{[Ben95]} by finding
the {\em exact} solutions for weakly bound two particles
interacting through a contact force.  In that study, the role of
one-particle resonant states on the density asymptotics has been
discussed.

\subsection{Pairing coupling to positive-energy states}
\label{sec3e}

As illustrated in Sec.{\ }\ref{sec3a}, the density of the
scattering continuum states increases with $R_{\text{box}}$. In
the limit of very large values of $R_{\text{box}}$, the set of
discretized continuum states can be considered as a fair
approximation of the real continuum, and the sums over the
positive-energy states can correctly represent integrals over
the continuous energy variable.  Therefore, we may consider this
limit in order to study the dynamical coupling between the bound
single-particle states and the positive-energy states. In the
language of pairing correlations, one may think of this coupling
in terms of a virtual scattering of pairs of fermions from the
bound states to positive-energy states, and back. Such a pair
scattering gives rise to the additional pairing energy to the
ground-state energy.

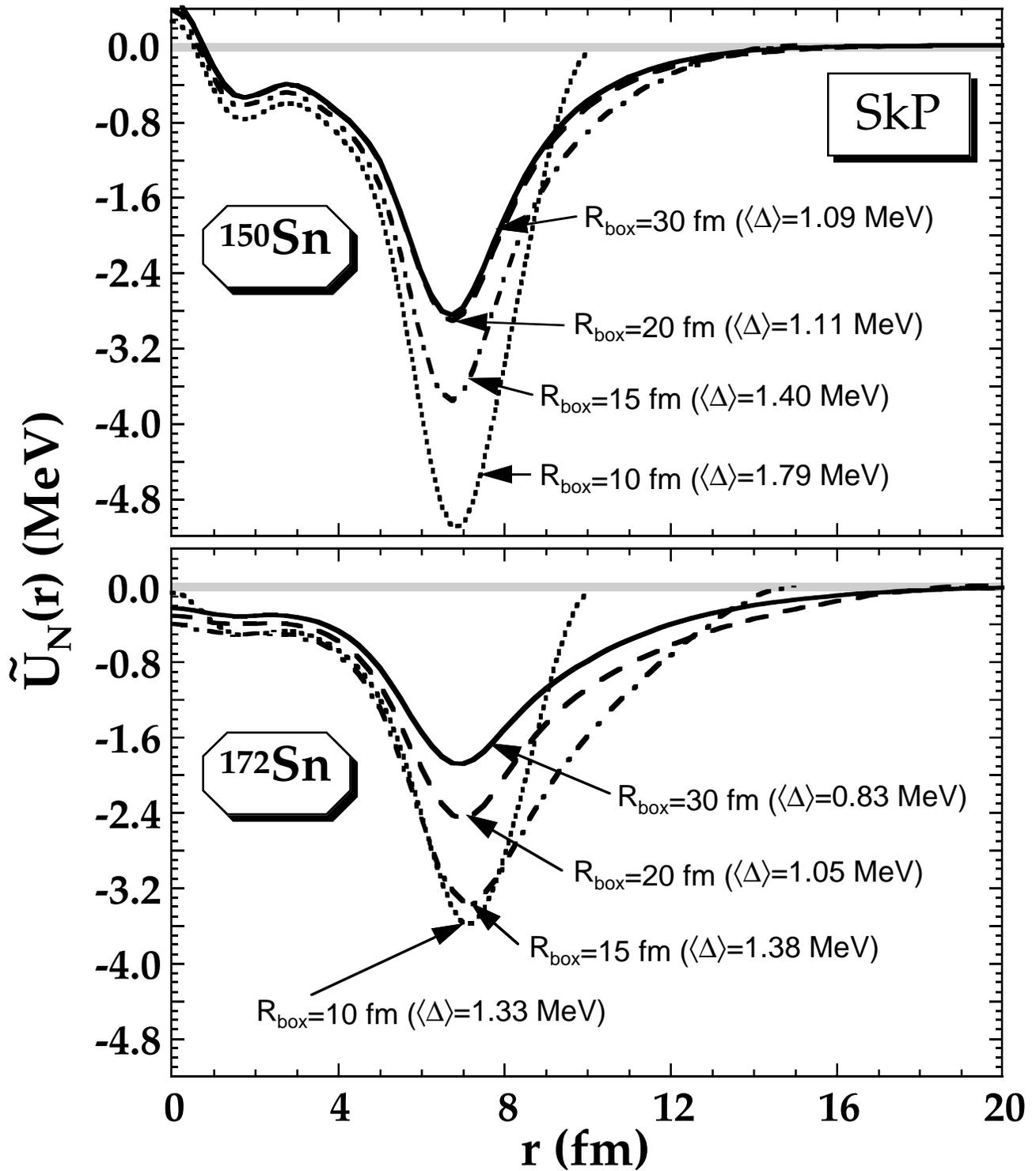
\begin{figure}[ht]
\caption[FF]{
Self-consistent HFB+SkP neutron pairing potentials
$\tilde{U}(r)$ in $^{150}$Sn (top panel) and $^{172}$Sn (bottom
panel) calculated with four values of $R_{\text{box}}$. The
corresponding average gap values $\langle\Delta\rangle$
[Eq.{\ }(\protect\ref{eq157})] are indicated.
}
\label{FIG03}
\end{figure}

To illustrate the stability of results with increasing box size,
in Fig.{\ }\ref{FIG03} we show the neutron p-p potentials
$\tilde{U}(r)$ in $^{150}$Sn and $^{172}$Sn calculated in the
HFB+SkP model for several values of $R_{\text{box}}$.  In these
two nuclei, the values of $\tilde{U}(r)$ do not change when
$R_{\text{box}}$ is larger than 20 and 22\,fm, respectively, but
at smaller values of $R_{\text{box}}$, one observes significant
variations.  A rather unexpected result of this analysis is that
the overall magnitude of pairing correlations, represented by
the average pairing gap $\langle\Delta\rangle$, {\em decreases}
with increasing $R_{\text{box}}$.  This occurs in spite of the
fact that the actual density of scattering states dramatically
{\em increases} with increasing $R_{\text{box}}$.

This effect can be understood by noting that the pairing
correlations produced by a density-dependent p-p interaction
(and hence for the SkP force used here) are concentrated at the
nuclear surface; i.e., at a fixed location in space.  For small
values of $R_{\text{box}}$, the boundary conditions
(\ref{eq154}) have a tendency to push the continuum wave
functions towards smaller distances, and into the surface
region. This increases the magnitude of pairing correlations. On
the other hand, with increasing $R_{\text{box}}$, the scattering
states spread out uniformly outside the nucleus and effectively
leave the surface region.  Hence $\langle\Delta\rangle$
decreases.  As a consequence, with increasing $R_{\text{box}}$
the self-consistent attractive pairing potential $\tilde{U}(r)$
decreases in magnitude and significantly spreads out towards
large distances.

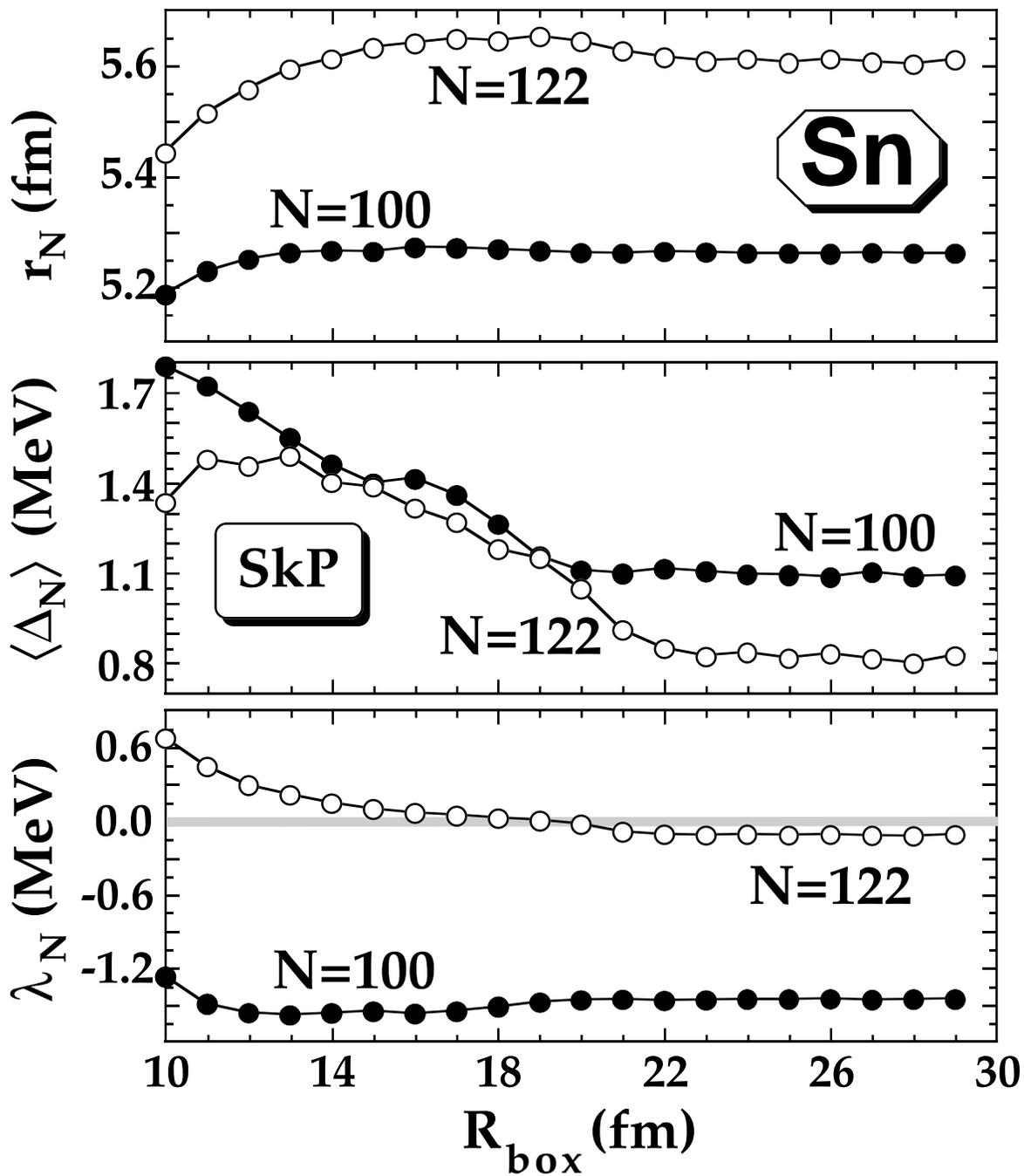
\begin{figure}[ht]
\caption[FF]{
Neutron rms radius $r_N$ (top panel), average pairing gap
$\langle\Delta_N\rangle$ (middle panel), and the Fermi energy
$\lambda_N$ (bottom panel) calculated in the HFB+SkP model for
$^{150}$Sn (full circles) and $^{172}$Sn (open circles) as
functions of $R_{\text{box}}$.
}
\label{FIG05a}
\end{figure}

The importance of allowing the pairing interaction to couple
properly  to the particle continuum is illustrated in
Fig.{\ }\ref{FIG05a}, where the neutron rms radius, the
average pairing gap, and the Fermi energy  are shown as
functions of $R_{\text{box}}$. The two upper plots confirm that
a stability of results is attained beyond 20 or 22\,fm, while
the bottom plot indicates that the pairing coupling to the
positive-energy states can be a decisive factor influencing the
nuclear binding. Indeed, below $R_{\text{box}}$$\simeq$20\,fm
the nucleus $^{172}$Sn is unbound, and it becomes bound only
when its ground state is allowed to gain an additional binding
from the pairing correlations at large distances.  This
indicates that, for the surface-type pairing interaction, one
has to consider a rather dense particle continuum before the
pairing coupling to positive-energy states is exhausted. (For a
similar discussion in a schematic model see
Ref.{\ }\cite{[Bel87]}. There, it has been pointed out that
because of strong coupling to the continuum, $\lambda$ is
significantly lowered in the case of  surface pairing as
compared to the case of  volume pairing.)

Since, for the Gogny interaction, the HFB equations are solved
by expansion in the harmonic oscillator basis, one can test the
coupling to the positive-energy states by increasing the number
$N_{\text{sh}}$ of the oscillator shells used in the basis. In
practice, calculations must be restricted to
$N_{\text{sh}}$$\leq$20, which allows one to describe the wave
functions up to about
$R_{\text{max}}$$\simeq$$\sqrt{2N_{\text{sh}}\hbar/m\omega_0}$,
where $\omega_0$ is the frequency of the harmonic oscillator
\cite{[Naz94]}.  For $N_{\text{sh}}$=20 this corresponds to
about $R_{\text{max}}$=14\,fm.

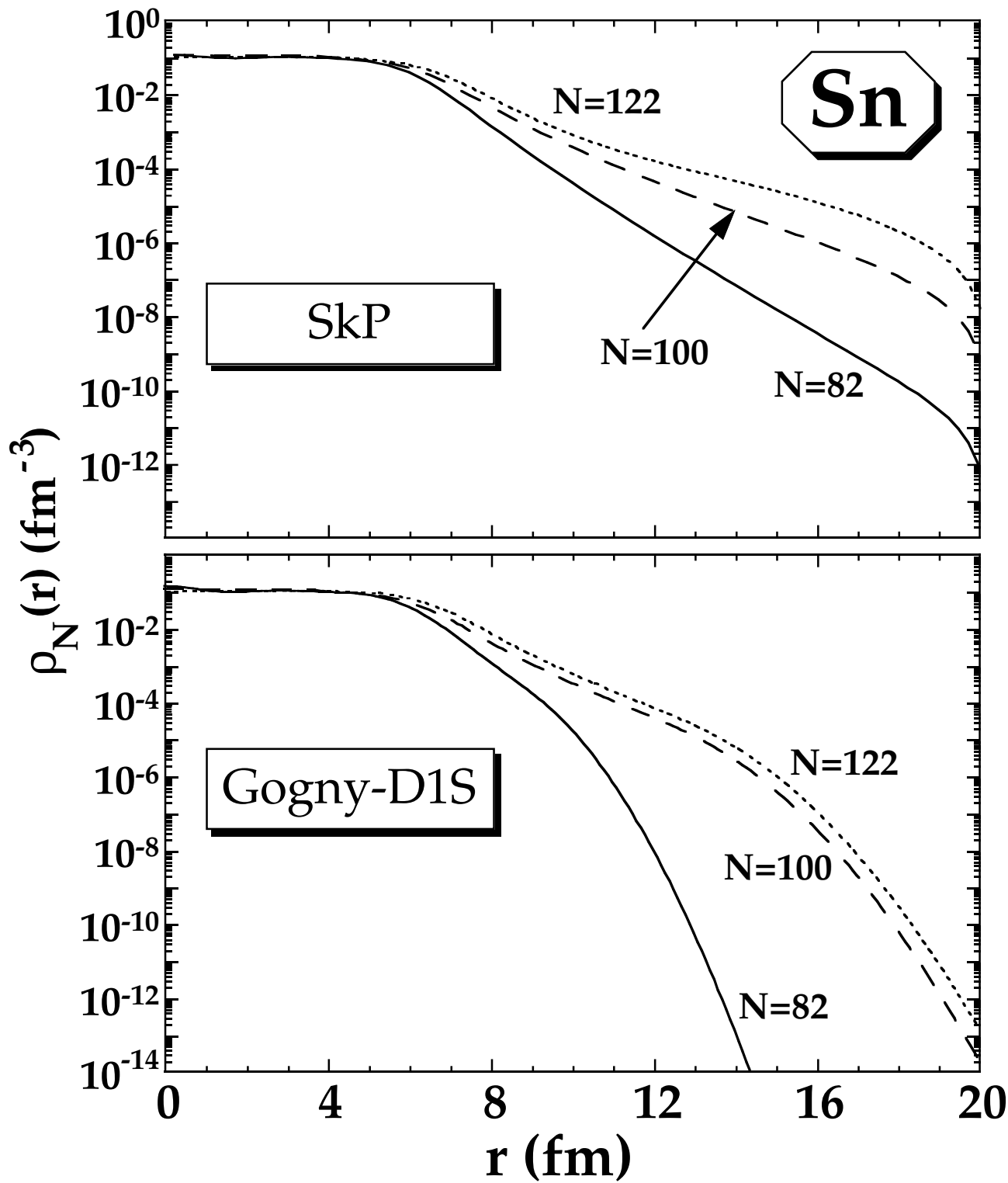
\begin{figure}[ht]
\caption[FF]{
Neutron densities $\rho_N(r)$ (in logarithmic scale) calculated
in the HFB+SkP and HFB+D1S models for $^{132}$Sn, $^{150}$Sn,
and $^{172}$Sn.
}
\label{FIG06}
\end{figure}

{}Figure{\ }\ref{FIG06} compares the asymptotic behavior of
the neutron particle densities in three neutron-rich tin
isotopes calculated in the spatial coordinates (SkP) or in the
harmonic-oscillator basis (D1S). In the former case one obtains
a clean region of the asymptotic dependence governed by
Eq.{\ }(\ref{eq155a}), which around $r$=18\,fm is perturbed by
the box boundary conditions (\ref{eq154}) at
$R_{\text{box}}$=20\,fm. In the latter case, the region of
proper asymptotic behavior becomes perturbed by the
$\exp(-m{\omega_0}r^2/\hbar)$ dependence characteristic of the
harmonic-oscillator-basis wave functions.  The $\omega_0$
values, obtained by minimizing the total energy for the
$N_{\text{sh}}$=17 basis, are equal to 13.4, 6.6, and 6.3\,MeV
in $^{132}$Sn, $^{150}$Sn, and $^{172}$Sn, respectively. Due to
this, a study of the continuum influence using such a basis can
be performed only up to densities of scattering states
corresponding to about $R_{\text{box}}$=14\,fm in the heavier
isotopes and only $R_{\text{box}}$=10\,fm in $^{132}$Sn, as can
be seen in Fig.{\ }\ref{FIG12}. Let us note, however, that the
neutron densities beyond $r$=10\,fm are typically smaller than
10$^{-4}$\,fm$^{-3}$, which explains the stability of the HFB
calculations with increasing size of the basis.

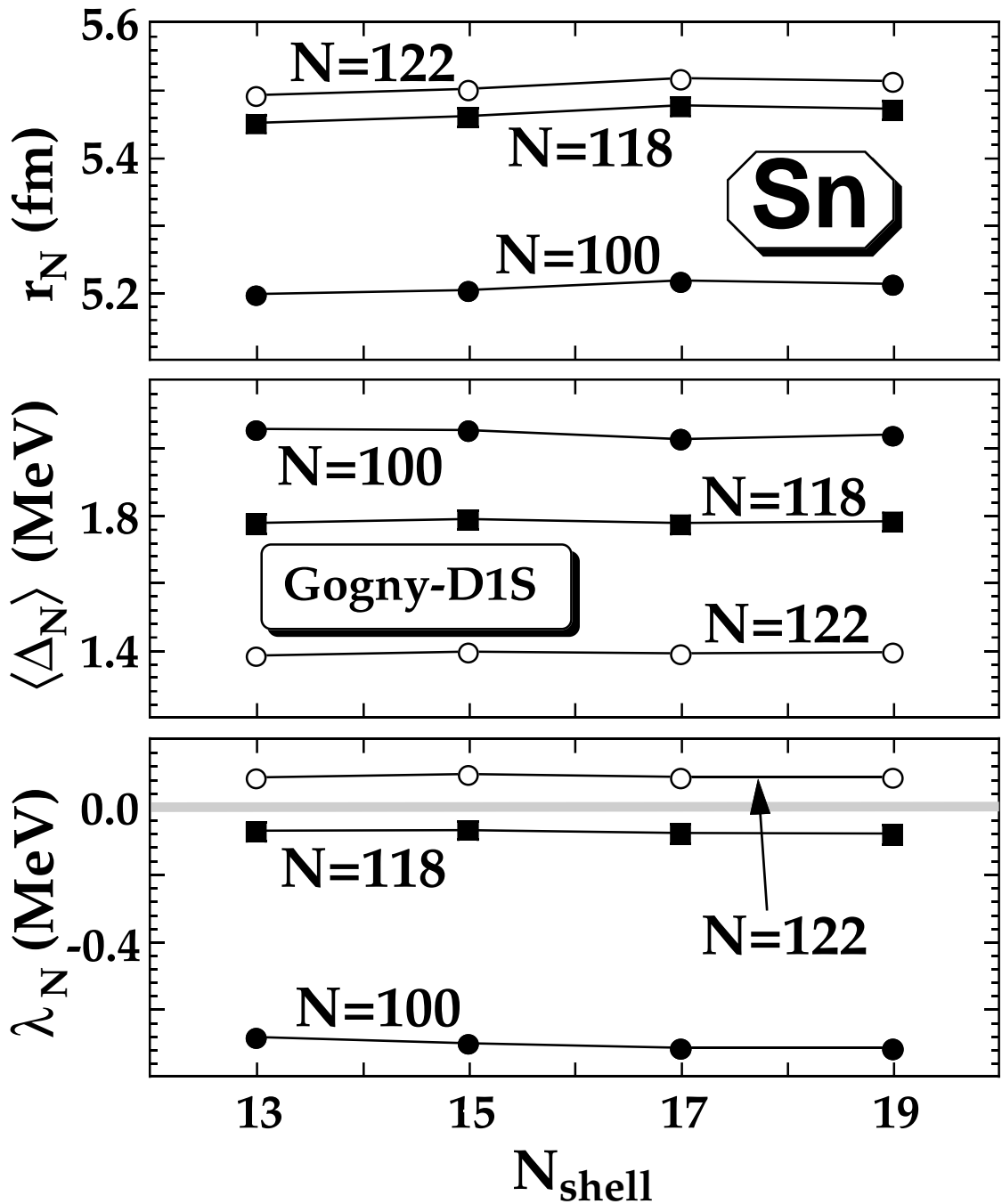
\begin{figure}[ht]
\caption[FF]{
Same as Fig.~\protect\ref{FIG05a} for the D1S interaction, but
as a function of the number of oscillator shells
$N_{\text{sh}}$.  {}For every $N_{\text{sh}}$ the
oscillator-basis frequency $\omega_0$ is adjusted as described
in the text.
}
\label{FIG05b}
\end{figure}

This is illustrated in Fig.{\ }\ref{FIG05b} which is analogous
to the similar study presented for the SkP interaction in
Fig.{\ }\ref{FIG05a}.  Here, for each value of $N_{\text{sh}}$
and for each nucleus, the value of $\omega_0$ was optimized so
as to minimize the total energy.  As can be seen, one obtains a
nice stability of results by using $N_{\text{sh}}$=17.  This
test corresponds to testing the coordinate-representation
solutions (Fig.{\ }\ref{FIG05a}) in the range of box sizes
between 12\,fm$\leq$$R_{\text{box}}$$\leq$14\,fm.  In this
rather narrow region, the SkP results are not stable because of
the dominant surface-type character of its pairing interaction.
Since the p-p Gogny interaction is more of the volume type
(Sec.{\ }\ref{sec2aca}) it requires much smaller distances to
saturate.

\subsection{BCS approximation}
\label{sec3b}

When inspecting Fig.{\ }\ref{FIG12}, it is obvious that by
applying the BCS approximation to the state-independent pairing
force and by allowing the BCS-type pairing correlations to
develop in such a dense spectrum, the result can be disastrous.
The seniority force gives rise to the {\em non-localized pairing
field} \cite{[Dob84]},
   \begin{equation}\label{VBCS}
   \tilde{h}_{\text{BCS}}
   (\bbox{r}\sigma,\bbox{r'}\sigma')=-\Delta_{\text{BCS}}
   \delta(\bbox{r}-\bbox{r'})
   \delta_{\sigma\sigma'},
   \end{equation}
i.e., to a constant pairing gap, identical for all states.  The
high density of single-particle states in the particle continuum
immediately results in an unrealistic increase of BCS pairing
correlations \cite{[Naz94]}.  One may, in principle,
artificially readjust the pairing strength constant to avoid
such an increase, but then the predictive power of the approach
is lost  and,  moreover, the spatial asymptotic properties of
the solutions are still going to be incorrect.

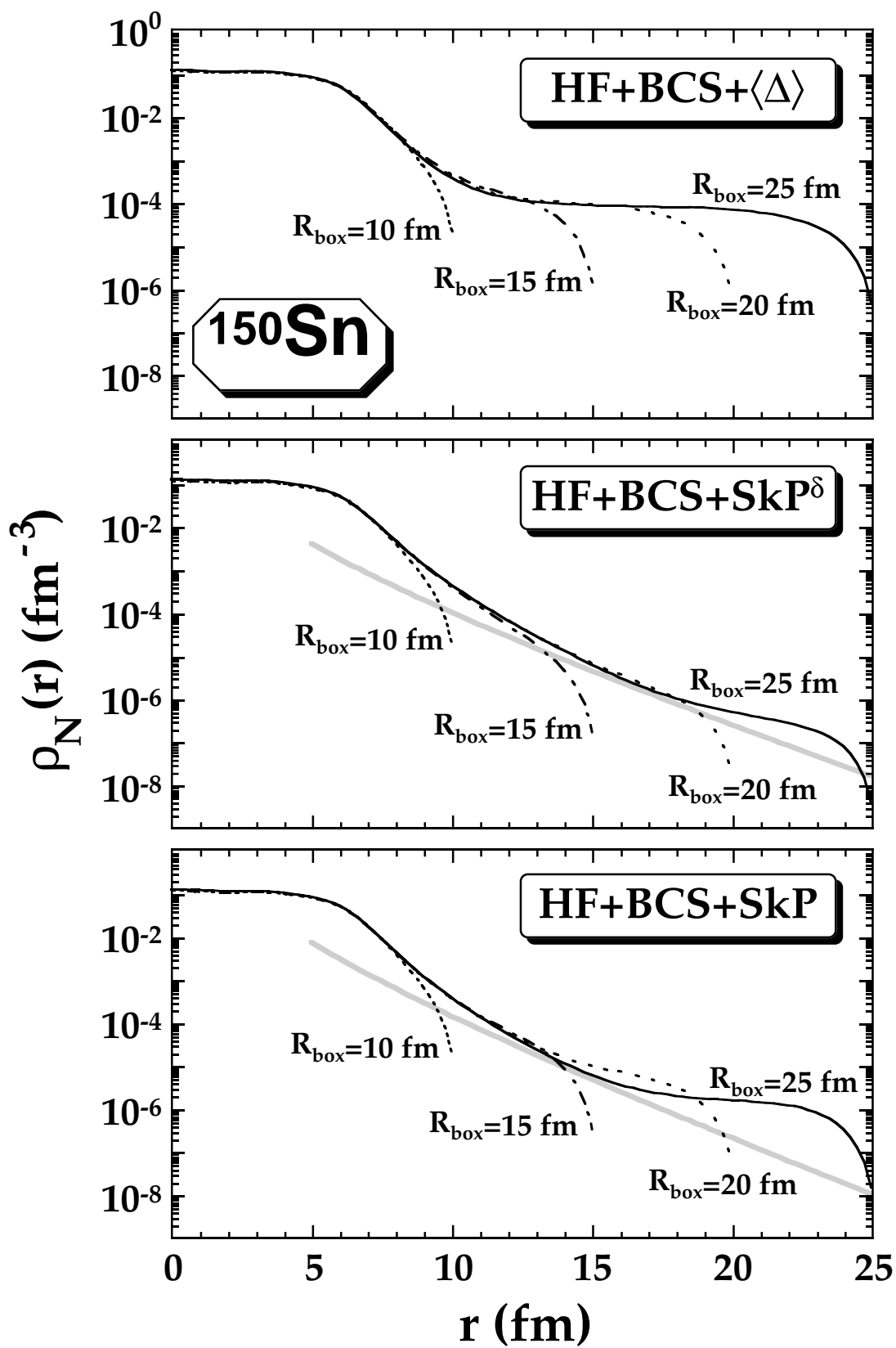
\begin{figure}[ht]
\caption[FF]{
Same as in Fig.{\ }\protect\ref{FIG04}, but for the
HF+BCS+$\langle\Delta\rangle$ approach with the constant pairing
gaps as listed in Fig.{\ }\protect\ref{FIG03} (top panel), the
HF+BCS+SkP$^\delta$ model (middle panel), and the HF+BCS+SkP
model (bottom panel). In all these  calculations  the same
pairing space (i.e., energy cut-off) was used as in
Fig.{\ }\protect\ref{FIG04}.
}
\label{FIG25}
\end{figure}

To illustrate the latter point, Fig.{\ }\ref{FIG25} (top
panel)  shows the neutron densities in $^{150}$Sn calculated for
several values of $R_{\text{box}}$ within the HF+BCS
approximation.  In order to avoid the increase of pairing
correlations with increasing density of states, the calculations
have been performed by fixing the values of the pairing gap. For
every box size $R_{\text{box}}$, the value of
$\Delta_{\text{BCS}}$ has been set equal to the average pairing
gap $\langle\Delta\rangle$ obtained within the HFB method. The
corresponding $\langle\Delta\rangle$ values are quoted in
Fig.{\ }\ref{FIG03}.

It is not too surprising to see that the asymptotic behavior of
the density calculated in the HF+BCS+$\langle\Delta\rangle$
method (top panel) is entirely different than that shown in
Fig.{\ }\ref{FIG04}.  Due to a nonzero occupation probability
of quasibound  states, there appears an unphysical gas of
neutrons surrounding the nucleus.  In Fig.{\ }\ref{FIG25} this
gas has a constant density of
$\rho$$\simeq$6$\times$10$^{-5}$\,fm$^{-3}$, independent of
$R_{\text{box}}$. This result means that an external pressure
would have been necessary to keep the neutrons inside the box.
Namely, had the box boundary condition been released, one would
have observed a stream of neutrons escaping the nucleus. This is
a completely artificial (and unwanted) feature of the BCS
approximation, because for a negative value of the Fermi energy,
neutrons cannot be emitted.

In the above  example the density of the neutron gas at
$R_{\text{box}}$=25\,fm corresponds to about 4 neutrons
uniformly distributed in the sphere of $R$=$R_{\text{box}}$.
Needless to say, by increasing the box radius, the number
neutrons in the gas grows at the expense of the number of
neutrons constituting the nucleus in the center of the box.
Since the total average number of neutrons is conserved, by
changing $R_{\text{box}}$ one actually performs an unphysical
study of {\em different} nuclei, surrounded by a neutron gas of
a fixed density.  Another consequence of the presence of a gas
of particles is that the rms nuclear radius cannot be calculated
in the BCS theory, because the results strongly depend on the
box size (see discussion in Refs.{\ }\cite{[Dob84],[Dob95a]}).

It has been suggested in the literature \cite{[Ton79]} that the
above deficiencies of the BCS approximation can be cured by
applying to them the state-dependent-pairing-gap version, where
the pairing gap is calculated for every single-particle state
using an interaction which is not of the seniority type.  (The
corresponding BCS equations resemble the canonical-basis
relations (\ref{eq151}).) In such an approach one hopes that the
majority of continuum states would neither contribute to the
pairing field (e.g., because of their very different spatial
character) nor result in the appearance of the unphysical gas.
This conjecture is tested in Fig.{\ }\ref{FIG25} (middle and
bottom panel) where the neutron densities obtained within the
state-dependent version of the BCS approximation using the
SkP$^\delta$ and the SkP interactions are presented.  It is seen
that a reduced coupling of some continuum states to the pairing
field does indeed decrease the gas density, however, the
asymptotic behavior of the density is still incorrect.

In the above  plots, the shaded lines represent the asymptotic
behavior given by Eq.{\ }(\ref{eq155a}) assuming
$E_{\text{min}}$=0, i.e., that of a single-particle state at the
Fermi energy. It is seen that a surplus density above this
asymptotic limit appears at large distances.  However, the
deficiencies of the state-dependent BCS approximation, as used
for example in Refs.{\ }\cite{[Ton79],[Len91],[Nay95]}, are
certainly less acute than those of the seniority-pairing BCS.
For example, in this type of approach one may probably calculate
radii of nuclei much nearer to the drip line.

It is clear that the neutron gas appears in the BCS solutions
because of the nonzero occupation probabilities of scattering
states. Therefore, one may think that excluding the scattering
states from the pairing phase space could be a decisive solution
to the problem. However, for drip-line nuclei, where the Fermi
energy is by definition close to zero, the remaining phase space
would then be small, and this would lead to an artificial
quenching of pairing correlations. Moreover, even if the density
obtained in such method would vanish asymptotically, the
corresponding factor $\chi$ would not be governed by
$\Delta$$-$$\lambda$$\simeq$2\,MeV, as discussed in
Sec.{\ }\ref{sec3c}, but by the single-particle energy,
$\epsilon$$\simeq$0, of the highest-energy single-particle
state considered in BCS calculations.  This again would lead to
densities vanishing at much slower pace than it is required by
the HFB theory.

\section{Physical observables far from stability}
\label{sec4}

In this section discussed are some experimental consequences
of the HFB theory, particularly important for weakly bound nuclei.

\subsection{Pairing gaps}
\label{sec4a}

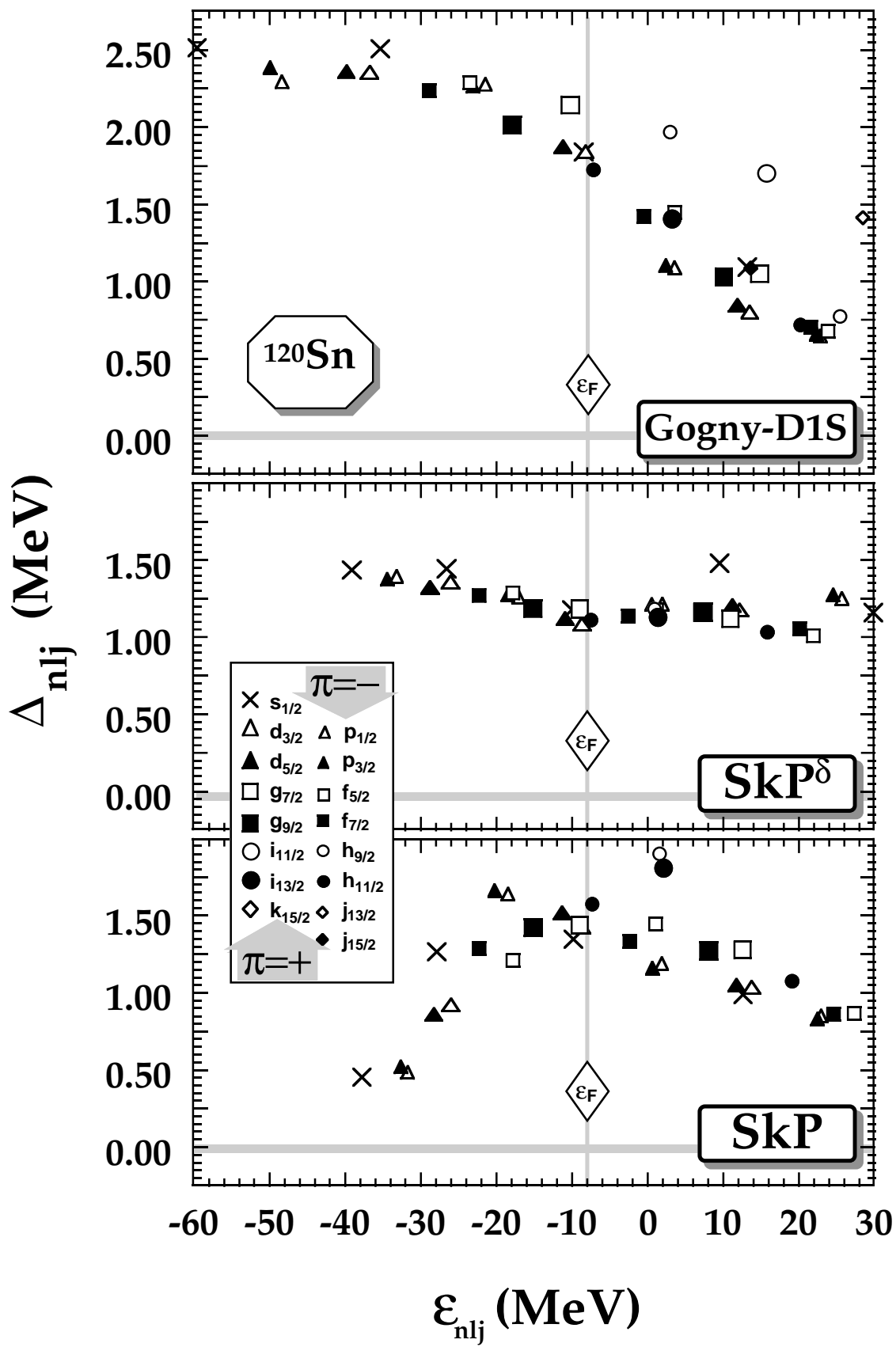
\begin{figure}[ht]
\caption[FF]{
Average values of the neutron p-h and p-p potentials,
$\epsilon_{n{\ell}j}$ and $\Delta_{n{\ell}j}$
[Eqs.{\ }(\ref{eq148})] in the canonical states calculated for
$^{120}$Sn in HFB+D1S (top), HFB+SkP$^{\delta}$ (middle), and
HFB+SkP (bottom) model. Only the states with
$v_{n{\ell}j}^2$$>$0.0001 are displayed.
}
\label{FIG24a}
\end{figure}

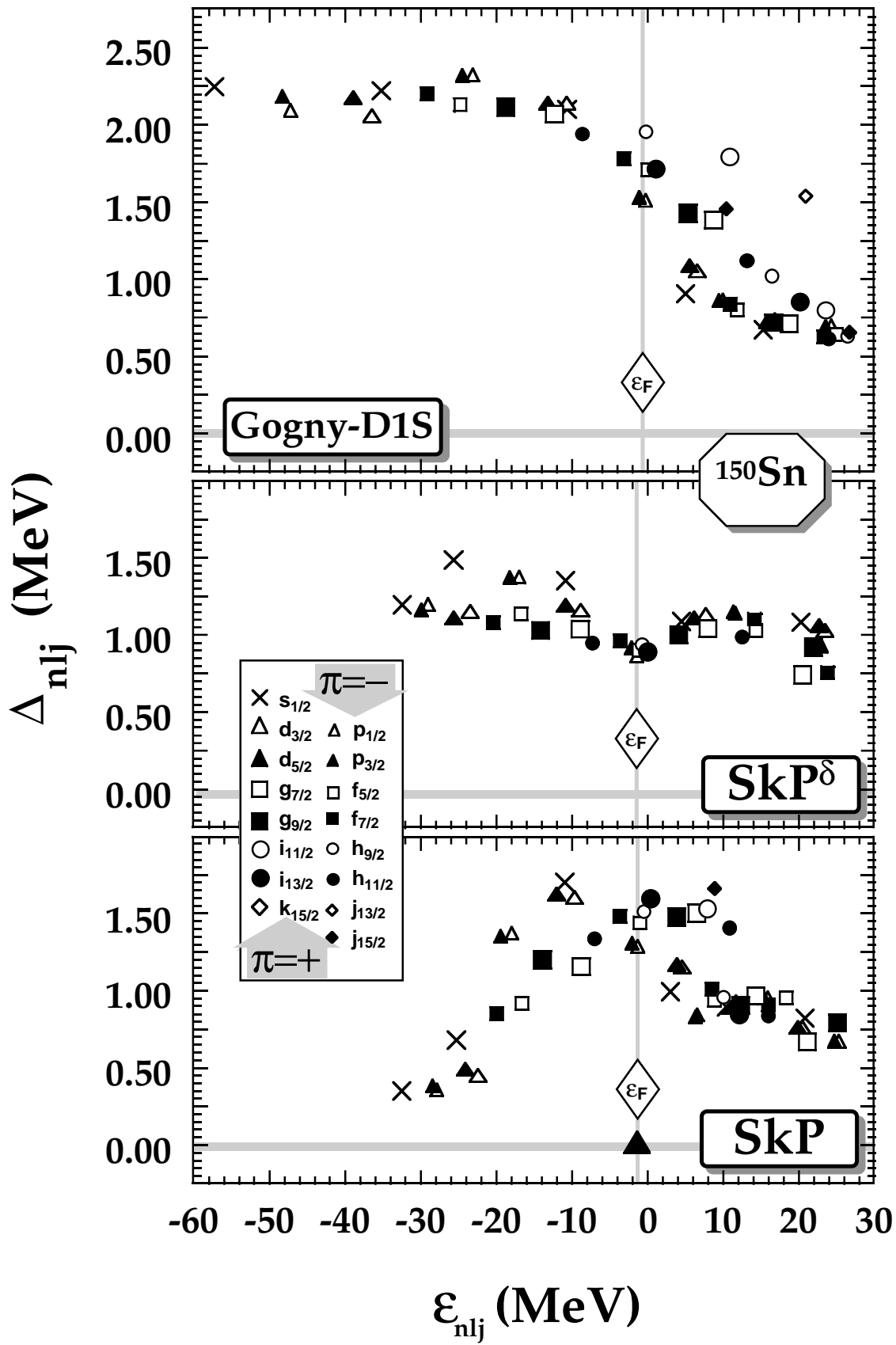
\begin{figure}[ht]
\caption[FF]{
Same as in Fig.{\ }\protect\ref{FIG24a}, but for $^{150}$Sn.
}
\label{FIG24b}
\end{figure}

Pairing gaps are p-p analogs of single-particle energies. They
carry the information about the energies of non-collective
excitations, level occupations, odd-even mass differences, and
other observables.  The average neutron canonical pairing gaps
(\ref{eq148b}) are shown in Figs.{\ }\ref{FIG24a}
($^{120}$Sn) and \ref{FIG24b} ($^{150}$Sn) as functions of the
canonical single-particle energies (\ref{eq148a}).

As seen in the middle part of Fig.{\ }\ref{FIG24a}, pairing
gaps obtained with the volume-type pairing interaction exhibit
very weak configuration dependence.  In $^{120}$Sn they decrease
slightly with $\epsilon_\mu$ but remain confined between 1.0 and
1.5\,MeV. In general, the values of $\Delta_\mu$  for the
$s_{1/2}$ states are slightly larger than for other orbitals,
which is again related to the volume character of volume delta
interaction.

The results presented in the bottom part of
Fig.{\ }\ref{FIG24a} nicely illustrate the surface character
of the SkP pairing interaction. Indeed, here the pairing gaps
increase from 0.5\,MeV (deep-hole states) to about
1.25--1.5\,MeV when the single-particle energies increase
towards the Fermi energy, and then they decrease again to about
1.0\,MeV for positive single-particle energies.  This is related
to the fact that orbitals near the Fermi level are concentrated
in the surface region.

Still another type of behavior is obtained for the finite range
Gogny interaction (top part of Fig.{\ }\ref{FIG24a}).  Here,
the pairing gaps decrease steadily with single-particle energy.
In $^{120}$Sn the values of $\Delta_\mu$ decrease from  about
2.5\,MeV for deep-hole states to about 0.75\,MeV for
positive-energy states.  (A similar energy dependence of pairing
gaps was obtained in the BCS calculations of
Ref.{\ }\cite{[Del95]} with the renormalized Paris potential.)
Interestingly, the values obtained for the high-$\ell$,
$j$=$\ell$$-$$\frac{1}{2}$ orbitals (antiparallel $L$$-$$S$
coupling) are significantly larger than those for other
orbitals.  The different ranges of $\epsilon_\mu$ values for SkP
and D1S in Fig.{\ }\ref{FIG24a} reflects the different
effective masses in both models. A rather low effective mass in
D1S, $m^*/m$=0.70, gives rise a reduced level density and a more
bound 1$s_{1/2}$ ground state as compared with the  SkP model
($m^*/m$=1). In fact, due to the non-local exchange
contributions to the p-h mean field (Appendix \ref{appA}), the
1$s_{1/2}$ state in the Gogny model has the canonical energy
lower than the bottom of the local potential well, shown in
Fig.{\ }\ref{FIG01a}.

In $^{120}$Sn, the HFB+D1S pairing gaps at the Fermi energy are
of the order of 1.75\,MeV, which slightly overestimates the
values corresponding to the odd-even mass staggering in this
region. However, one should bear in mind that the pairing gaps
at the Fermi energy are rather rough approximations to the
odd-even mass difference. A more accurate description can be
obtained by performing blocked HFB calculations for odd-mass
isotopes. In the vicinity of $^{120}$Sn this method yields  the
odd-even mass staggering of 1.6\,MeV \cite{[Dec80]} for the D1S
interaction and of 1.3\,MeV \cite{[Dob84]} for the SkP
interaction.  Another contribution to the odd-even mass
difference comes from the coupling to the low-lying collective
modes.  Therefore, the D1S parameters have been adjusted
\cite{[Dec80]} to give the pairing gap in tin to be 0.3\,MeV
larger than the experimental one. On the other hand, such a
margin has not been taken into account for the SkP and
SkP$^\delta$ forces.  Clearly, a detailed comparison of the
values of pairing gaps for the  interactions discussed in
Fig.{\ }\ref{FIG24a} is delicate. Much more information can
actually be derived from the comparison of their dependence on
the single-particle energies, which is markedly different.

The general pattern of $\Delta_\mu$ remains very similar when
going to  the  neutron-rich nucleus $^{150}$Sn
(Fig.{\ }\ref{FIG24b}).  In particular, the magnitude of the
average pairing gap in deep-hole states depends strongly on the
range and density dependence of pairing interaction.

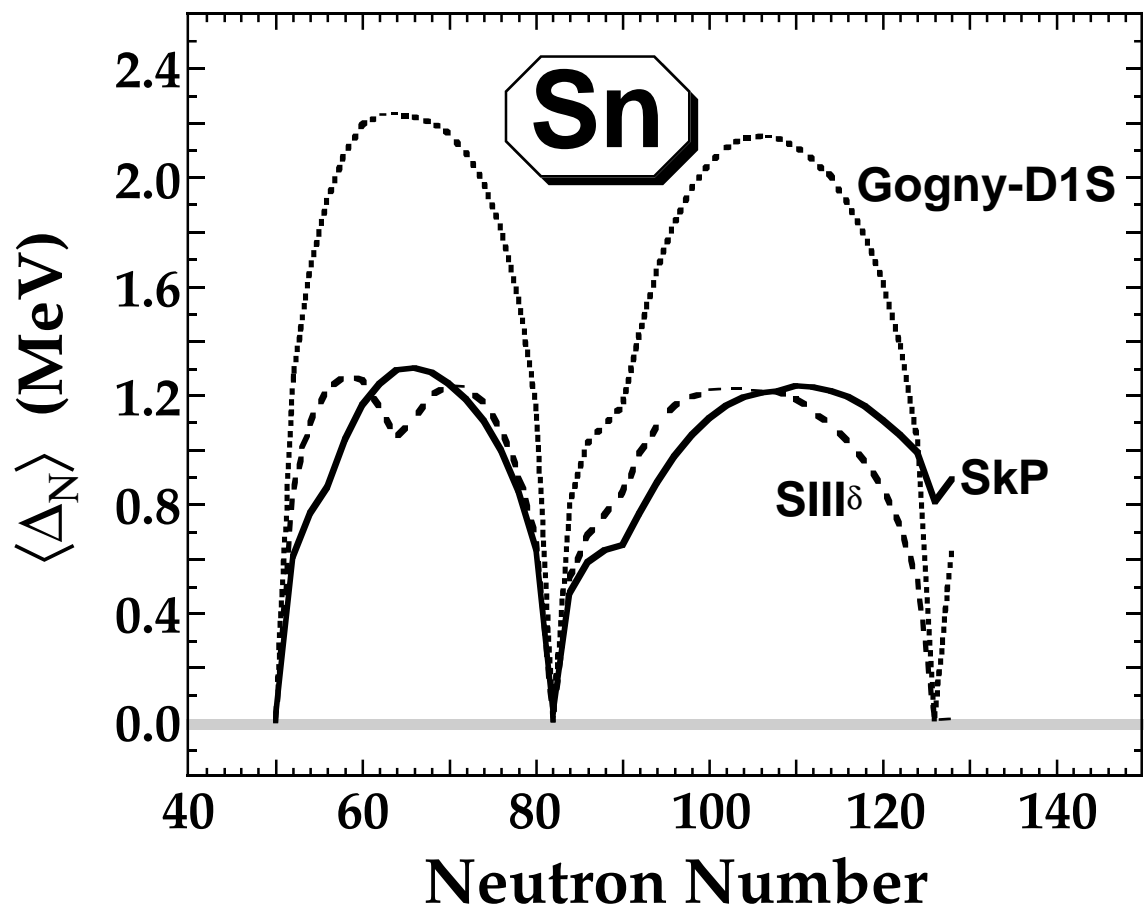
\begin{figure}[ht]
\caption[FF]{
Average neutron pairing gaps $\langle\Delta_N\rangle$
[Eq.{\ }(\protect\ref{eq157})] calculated for SkP (solid),
SIII$^{\delta}$ (dashed), and D1S (dotted) interactions for the
series of tin isotopes.
}
\label{FIG07}
\end{figure}

{}Figures{\ }\ref{FIG07} shows the average neutron pairing
gaps (Eqs.{\ }(\ref{eq157}) and (\ref{eq158})) for SkP,
SIII$^\delta$, and D1S interactions.  The large values of
$\langle\Delta\rangle$ obtained in HFB+D1S can be explained by:
(i) an overall larger magnitude of pairing correlations in tin
nuclei, and (ii) strong pairing correlations in deep-hole states
which strongly contribute to the average,
Eq.{\ }(\ref{eq158}).  It is to be noted, however, that
despite stronger pairing in D1S, the HFB+D1S pairing gaps vanish
at $N$=126 (near the two-neutron drip line), in contrast to the
HFB+SkP result.  This difference may be traced back to a much
larger continuum phase space taken into account in our HFB+SkP
calculations (Sec.{\ }\ref{sec3e}) which are performed in the
coordinate representation, and to a larger $N$=126 shell gap
(4.2 MeV in $^{168}$Sn) obtained with D1S.  (The increase of
proton pairing gaps when approaching the proton drip line has
been calculated previously in  Ref.{\ }\cite{[Sta91]} with the
HFB+SkP model and explained in a similar way.) The disappearance
of the neutron pairing at $N$=126 in the HFB+SIII$^\delta$ model
is partly due to the volume character of $\tilde{h}$ (a weaker
coupling to the particle continuum) and partly  due to a larger
$N$=126 shell gap  \cite{[Dob95c]}.

\subsection{Shell effects}
\label{sec4b}

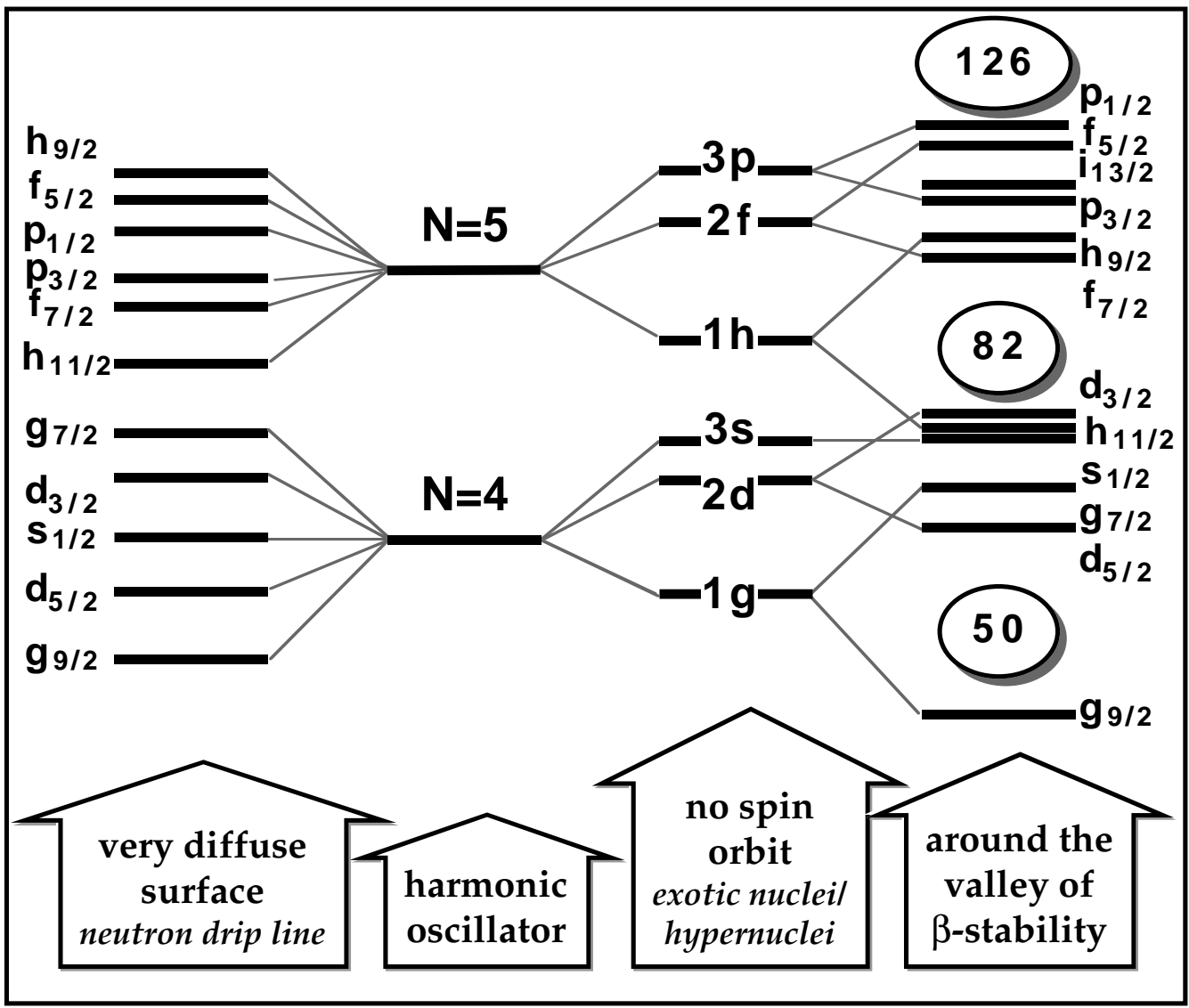
\begin{figure}[ht]
\caption[FF]{
Sequences of nuclear single-particle levels for various
potentials.  Orbitals are labeled by the spherical quantum
numbers. From left to right:  (i) shell structure for a
potential with spin-orbit term but with a very diffuse surface,
(ii) the $N_{\text{osc}}$=4 and  5 shells of the harmonic oscillator
potential, (iii) no spin-orbit term, leading to a degenerate
spin-orbit pattern as observed in, e.g.,  hypernuclei, and (iv)
shell structure characteristic of  nuclei near the stability
valley.
}
\label{FIG27}
\end{figure}

As discussed in Sec.~\ref{sec2ba}, diffused nucleonic densities
and very strong, surface-peaked, pairing fields obtained with
the density-dependent pairing interaction are expected to lead
to very shallow single-particle potentials in drip-line nuclei.
Because of a  very diffuse surface (no flat bottom), the resulting
single-particle spectrum resembles that of a harmonic oscillator
with a spin-orbit term (but with a weakened ${\ell}^2$ term)
\cite{[Dob94]}.  Schematically, this effect is illustrated in
the left panel of Fig.~\ref{FIG27}.  By comparing with the
situation characteristic of stable nuclei (right panel of
Fig.~\ref{FIG27}), new shell structure emerges with a more
 uniform distribution of normal-parity orbits, and the
unique-parity intruder orbit which reverts towards its parent
shell.  Such a new shell structure, with no pronounced shell gaps,
would give rise to different kinds of collective  phenomena
\cite{[Naz94],[Cho95]}.

The effect of the weakening of shell effects in drip-line
nuclei, first mentioned in the astrophysical context
\cite{[Hae89]}, was further investigated in
Refs.~\cite{[Smo93],[Dob94],[Dob95c]}.  First analyses of its
consequences for the nucleosynthesis have also been performed
\cite{[Che95],[Pfe95]}.  Microscopically, it can be explained
by:  (i) the changes in the mean field itself due to weak
binding (see above), and (ii) a strong pairing-induced coupling
between bound orbitals and the low-$\ell$ continuum.

\subsection{Separation energies}
\label{sec4c}

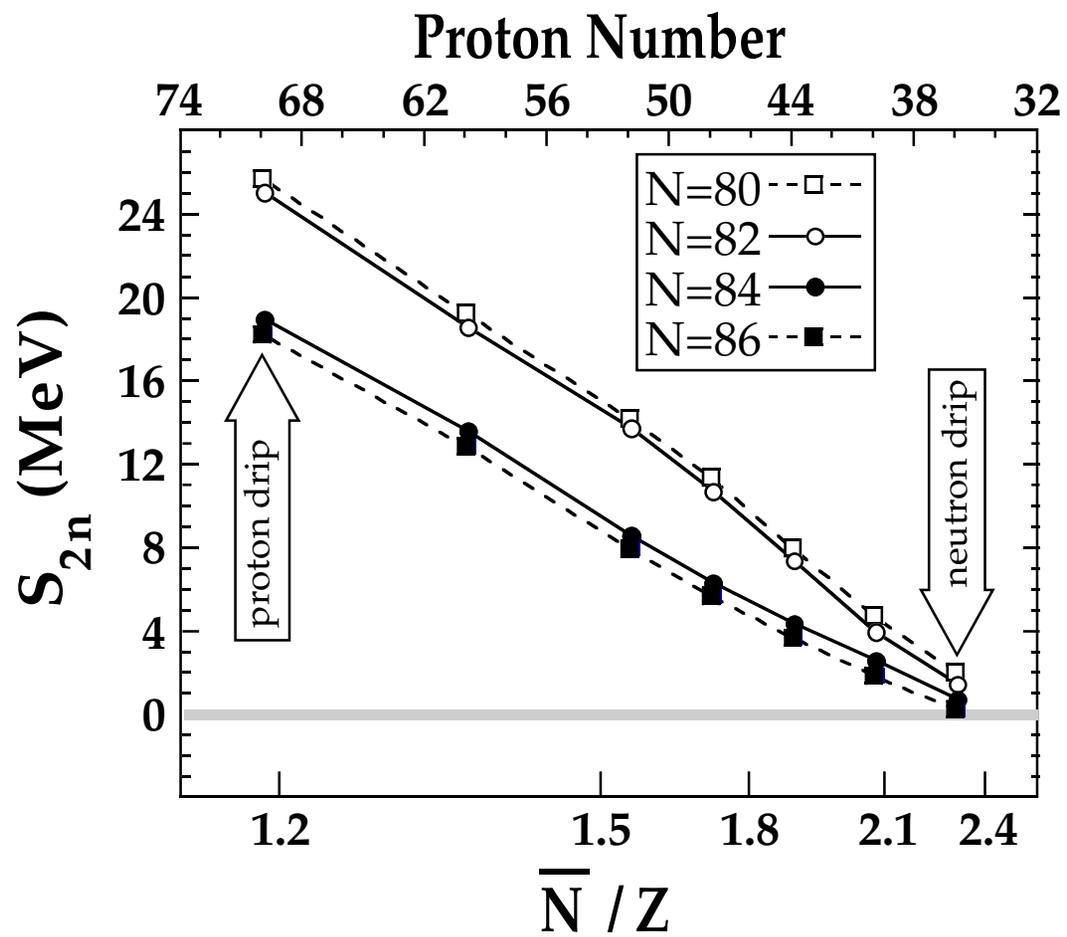
\begin{figure}[ht]
\caption[FF]{
Two-neutron separation energies for the $N$=80, 82, 84, and 86
spherical even-even isotones calculated in the HFB+SkP model as
a function of $\bar{N}/Z$ (lower scale, $\bar{N}$=83) or $Z$
(upper scale). The arrows indicate the proximity of neutron and
proton drip lines -- see Fig.{\ }\protect\ref{FIG26} for
detailed predicted positions of two-particle drip lines.
}
\label{FIG28}
\end{figure}

Weakening of shell effects with neutron number manifests itself
in the behavior of two-neutron separation energies.  This is
illustrated in Fig.~\ref{FIG28} which displays the two-neutron
separation energies for the $N$=80, 82, 84, and 86 spherical
even-even isotones.  The large $N$=82 magic gap, clearly seen in
the nuclei close to the stability valley and to the proton-drip
line, gradually closes down when approaching  the neutron drip
line. A similar effect is seen in the ($Z, N$) map of the
spherical two-neutron separation energies for the particle-bound
even-even nuclei calculated in the HFB+SkP model
(Fig.{\ }\ref{FIG26}). Namely, the neutron magic gaps $N$=20,
28, 50, 82, and 126, clearly seen as cliffs in the $S_{2n}$
surface, disappear for  neutron-rich systems.

The gradual disappearance of the neutron shell structure with
$N$ is not a generic property of all effective interactions.  As
seen in the plot of $S_{2n}$ and $\lambda_N$ for the tin
isotopes (Fig.~\ref{FIG02}) this effect is seen in the  SkP and
SkP$^{\delta\rho}$ models, and, to some degree, also in the
SkP$^{\delta}$ model. (A weak  irregularity at $N$=126 reflects
the weaker coupling to continuum for the volume pairing
\cite{[Bel87]}.) The strong shell effect seen in the SIII and
SkM$^*$ results has been discussed in Ref.~\cite{[Dob95c]}; it
can be attributed to the low effective mass in these forces. The
result of the D1S model, both for $S_{2n}$ and $\lambda_N$, is
close to that of the  SkP$^{\delta}$ model.  It is interesting
to point out that the QLM calculations of Ref.~\cite{[Zve85]}
(with $m^*/m=1$) for the Sn isotopes yield very similar results
to those of HFB+SkP.

The very neutron-rich nuclei, as those shown in
Fig.~\ref{FIG02}, cannot be reached experimentally under present
laboratory conditions.  On the other hand, these systems are the
building blocks of the astrophysical r-process; their separation
energies, decay rates, and cross sections are the basic
quantities determining the results of nuclear reaction network
calculations.  Consequently, one can learn  about properties of
very neutron-rich systems by studying element abundances
\cite{[Kra93]}.  The recent r-process network calculations
\cite{[Che95]}, based on several mass formulae, indicate a
quenching of the  shell effect at $N$=82 in  accordance with the
results of HFB+SkP model.

\subsection{Deep hole states}
\label{sec4d}

Pairing interaction between  bound orbitals and particle
continuum is partly responsible for the appearance of particle
widths of  deep-hole states and the term-repulsion phenomenon
(strong repulsion between single-particle levels)
\cite{[Bul80],[Bel87]}.  In the DWBA and for the local pairing
field $\tilde{U}$ the particle width is given by
   \begin{equation}\label{width}
    \Gamma_i = 2\pi \left|\int\text{d}^3 \bbox{r} \varphi_i(\bbox{r})
               \tilde{U}(\bbox{r}) \varphi_\epsilon(\bbox{r})
   \right|^2.
   \end{equation}
Here, $\varphi_i(r)$ is the HF wave function of the bound
deep-hole state $i$ with the single-particle energy
$\lambda-E_i$ in the absence of pairing, while
$\varphi_\epsilon(r)$ is the HF wave function of the unbound
state with the energy $\lambda+E_i$.

Equation (\ref{width}) is obtained by assuming that the p-p
field of the HFB Hamiltonian can be treated perturbatively.  A
more consistent way would be to estimate $\Gamma_i$ based on
self-consistent HFB solutions containing pairing correlations.
The proper formulation of the nonperturbative HFB-based  theory
of deep hole states and one-particle transfer process still
needs to be developed.

As discussed in Ref.{\ }\cite{[Bel87]}, $\Gamma_i$ is
sensitive to the type of the pairing force. In general, the
widths are larger for surface pairing than for volume pairing.
However, the result for an individual state strongly depends on
its angular momentum and excitation energy.

Experimentally, total widths of deep hole states,
$\Gamma_{\text{tot}}$,  are of the order of MeV's, (see, e.g.,
Refs.{\ }\cite{[Mou76],[Her88],[Gal88],[Van93a]}).  That is,
the partial width (\ref{width}), of the order of 10-100\,keV,
constitutes an extremely small fraction of
$\Gamma_{\text{tot}}$. Consequently, the experimental
determination of $\Gamma_i$ alone is very unlikely.

\subsection{Pair transfer form factors}
\label{sec4f}

There are many interesting aspects of physics of unstable nuclei
which are related to reaction mechanism studies: weak binding,
large spatial dimensions, skins (see, e.g.,
Refs.{\ }\cite{[Mue93],[Das94],[Kim94]}).  Below, we discuss
some consequences of surface-peaked pairing fields for pair
transfer studies.

An experimental observable that may probe the character of the
pairing field is the pair transfer form factor, directly related
to the pairing  density $\tilde\rho$. The difference in the
asymptotic behavior of single-particle density $\rho$ and  pair
density $\tilde\rho$ in a weakly bound system (see
Secs.~\ref{sec2aca} and \ref{sec3c}) can be probed by comparing
the energy dependence of one-particle and pair-transfer cross
sections.  Such measurements, when performed for both stable and
neutron-rich nuclei, can shed some light on the asymptotic
properties of HFB densities; hence on the character of pairing
field.

{}Figure{\ }\ref{FIG32} displays the pair transfer form
factors $r^2\tilde\rho(r)$ calculated in $^{120}$Sn, $^{150}$Sn,
and $^{172}$Sn with the SkP interaction. These microscopic
results are compared with the macroscopic form factors
$r^2\delta\rho(r)$ \cite{[Das85]} which are determined by using
the derivative of the particle density with respect to the
neutron number:
   \begin{equation}\label{eq305}
    \delta\rho(r) = 2\frac{-E_{\text{pair}}}{\langle\Delta\rangle}
                    \frac{\text{d}\rho(r)}{\text{d}N},
   \end{equation}
where $E_{\text{pair}}$ is given by Eq.{\ }(\ref{epair}).
This expression can be motivated by the fact that only the
orbitals near the Fermi surface make significant contributions
to the pair density.  In the BCS theory, the normalization
constant in $\delta\rho(r)$ is usually chosen \cite{[Bes86]} as
$\Delta/G$=$-$$E_{\text{pair}}/\Delta$.  Here, we use neither
the BCS approximation nor the constant pairing strength $G$.
Therefore, the normalization
$-$$E_{\text{pair}}/\langle\Delta\rangle$ is employed.  The
derivative in Eq.{\ }(\ref{eq305}) is calculated from the
finite difference between the self-consistent results for the
HFB vacuum corresponding to particle numbers $N$+1 and $N$$-$1.
In these calculations, in order to explore the smooth dependence
on the particle number $N$, the odd-average-particle-number
vacua have been calculated without using the blocking
approximation.  It should be mentioned at this point that the
further approximation \cite{[Das85],[Das89]} of the derivative
$\text{d}\rho(r)/\text{d}N$ by the spatial derivative
$\text{d}\rho(r)/\text{d}r$ is not justified, because the
volume-conservation condition is not valid for the neutron
density distribution (see Fig.{\ }\ref{FIG16}).

The pair transfer form factors in Fig.{\ }\ref{FIG32} clearly
show that this process has a predominantly surface character.
The macroscopic form factors have smaller widths and higher
maxima than the microscopic ones. On the other hand, they are
smaller in the interior of the nucleus as well as in the
asymptotic region. In $\beta$-stable nuclei the macroscopic
approximation works fairly well, while in the drip-line nuclei
the differences between the two form factors are markedly
larger. In general, the corresponding differences are much
larger than those obtained within the BCS  and the
particle-number-projected BCS approaches for the seniority
interaction \cite{[Civ92]}.

A comparison of the results obtained for different isotopes
conspicuously shows a significant increase in  the pair transfer
form factors in the outer regions of drip-line nuclei.  In
$^{120}$Sn, the form factors vanish around 9\,fm, while in
$^{150}$Sn and $^{172}$Sn they extend to much larger distances.
This effect is particularly pronounced for the microscopic pair
transfer form factors.

\subsection{Other observables}
\label{sec4e}

The importance of the HFB treatment for calculations of nuclear
radii has been discussed in several papers
\cite{[Dob84],[Sta91],[Fay94],[Dob95a]}. As mentioned in
Sec.~\ref{forces}, odd-even staggering of rms charge radii is
one of the best experimental indicators of the density-dependent
pairing.  The proper treatment of the pairing effect on radii is
especially important for weakly bound systems which exhibit halo
or skin effects \cite{[Ber91],[Sta91],[Dob95a]} (cf. discussion
in Sec.~\ref{sec3e}).

Apart from the information on the nuclear rms radii, one may
also gain some experimental insight into the ratios of neutron
and proton densities at large distances from the center of
nucleus \cite{[Lub94],[Wyc95]}.  This is possible due to
experiments on antiproton annihilation from atomic orbits, which
leads to different reaction products depending on whether the
process involves a proton or a neutron.

The role of deformation in neutron drip-line nuclei still needs
to be investigated.  One can anticipate that due to: (i) very
diffused surfaces, and (ii) strong pairing correlations, the
geometric concept of collective deformation (defined as a
deviation of nuclear surface from sphericity)  should be
revisited.  In this context, the symmetry-unrestricted HFB
calculations in coordinate space are called for.

\section{Summary and Conclusions}
\label{sec6}

The advent of radioactive nuclear beams provides many exciting
opportunities to create and study unstable nuclei far from the
$\beta$ stability valley.  One of the unexplored areas far from
stability is physics of nuclear pairing in weakly bound nuclei,
especially near the neutron drip line. Contrary to the situation
characteristic of stable nuclei, the coupling between the p-h
field and the p-p field in nuclei with extreme $N/Z$ ratios is
dramatic; i.e., no longer can pairing be treated as a residual
interaction.

The main objective of this study was to perform a  detailed
analysis of various facets of pairing fields in atomic nuclei.
The first part contains the comprehensive summary of the HFB
formalism, with particular attention on the physical
interpretation of the underlying densities and fields.  Very
little is known about the p-p component of the nuclear effective
interaction; its structure is of considerable importance not
only for nuclear physics but also for nuclear astrophysics and
cosmology. Therefore, the second part of this work  focuses on
the differences between various pairing interactions.  In
particular, the role of density dependence and finite range of
the p-p force has been illuminated, and the importance of the
coupling to the particle continuum has been emphasized.
Finally, the third part of our study relates the theoretical
formalism to experimental observables; i.e.,  energy spectra,
masses, radii, and pair transfer form factors. It is
demonstrated that these observables carry invaluable information
that can pin down many basic questions regarding the effective
$NN$ force, and its pairing component in particular.  It should
be stressed, however, that in order to see clearly some of the
predicted effects, the excursion far from the valley of
$\beta$-stability is necessary.

The analysis presented in this paper should be viewed as a
useful starting point for future investigations. One of them is
the coupling between collective surface modes (e.g.,
deformation) and pairing fields in weakly bound nuclei. Another
interesting avenue of explorations is the role of dynamics;
e.g., the importance of the particle number conservation and the
coupling to pair vibrations.  A fascinating and difficult
research program is the microscopic description of excited
states, especially those lying above the particle emission
threshold, for which the boundary conditions used in this study
(an impenetrable box) have to be modified to account explicitly
for outgoing waves. We are only beginning to explore many
unusual  aspects of the nuclear many-body problem offered by
systems with extreme $N/Z$ ratios.

\acknowledgments

Interesting discussions with H. Flocard, P.-H. Heenen, and H. Lenske
are gratefully acknowledged.
Oak Ridge National Laboratory is managed for the U.S. Department
of Energy by Lockheed Martin Energy Systems under Contract No.
DE-AC05-84OR21400.  The Joint Institute for Heavy Ion Research
has as member institutions the University of Tennessee,
Vanderbilt University, and the Oak Ridge National Laboratory; it
is supported by the members and by the Department of Energy
through Contract No. DE-FG05-87ER40361 with the University of
Tennessee.  We thank the Department of Energy's Institute for
Nuclear Theory at the University of Washington for its
hospitality and partial support during the completion of this
work.  This research was supported in part by the U.S.
Department of Energy through Contract No.  DE-FG05-93ER40770 and
the Polish Committee for Scientific Research under Contract
No.~2~P03B~034~08.

\appendix
\section{The p-h and p-p mean-field Hamiltonians for a local two-body
finite-range Gogny interaction}
\label{appA}

The Gogny force \cite{[Gog73],[Dec80]} is composed of the
central, spin-orbit, density-dependent, and Coulomb
interactions. The spin-orbit and density-dependent terms have
zero-range, and their contributions to the p-h and p-p mean
fields are identical to those of the Skyrme interaction.  The
corresponding expressions can be found in several papers; e.g.,
Refs.{\ }\cite{[Eng75],[Dob84]}, and will not be repeated
here.  In the following we only consider the central
finite-range and Coulomb terms.  The central components read
   \begin{equation}\label{1a}
      \hat V_{\text{cen}}   =  \sum_{j=1}^{2}
           e^{-{{(\bbox{r} - \bbox{r}')^2} \over {\mu_j^{2}}}}
      (W_{j} + B_{j} P_{\sigma} - H_{j} P_{\tau} -M_{j} P_{\sigma}
      P_{\tau}),
   \end{equation}
where $P_{\sigma}$ and $P_{\tau}$ are the exchange operators for
spin and isospin variables, respectively.  This interaction is
local; i.e., it should be multiplied by
$\delta(\bbox{r}_1-\bbox{r}_1')\delta(\bbox{r}_2-\bbox{r}_2')$
before it is inserted in the integrals (\ref{eq141}) defining
the mean fields. Moreover, it should also be multiplied by the
antisymmetrizing operator $(1-P_{r}P_{\sigma}P_{\tau})$, where
$P_{r}$ is the exchange operator for space variables.  One
usually calls the term involving $P_{r}$ the exchange term,
while the term involving no space exchange is called the direct
term.

The space, spin, and isospin variables are denoted by
$\bbox{r}$, $\sigma$=$\pm\frac{1}{2}$, and
$\tau$=$\pm\frac{1}{2}$, respectively.  The parameters $\mu_j$,
$W_j$, $B_j$, $H_j$, and $M_j$, belong to the set called D1S
\cite{[Ber91b]} which has been used in this paper.  Since the
 expressions given by the $j$=1 and 2 components are identical,
in what follows we drop the index $j$ to increase the legibility
of the formulae.

\widetext

\subsection{Contribution of the central direct interaction to the
            p-h mean field}

Since the interaction (\ref{1a}) is local, the direct term gives
the p-h mean field (\ref{eq141a}) which is also local, i.e.,
   \begin{eqnarray}
      \Gamma^{\tau}_{\text{dir}} (\bbox{r} \sigma, \bbox{r}' \sigma')
         & = & \delta(\bbox{r} - \bbox{r}')
               \delta_{\sigma \sigma'} \int\text{d}^{3}\bbox{r}_1
               e^{-{{(\bbox{r} - \bbox{r}_1)^2} \over {\mu^{2}}}}
               \sum_{\tau_1} \Big[(W - H \delta_{\tau \tau_1})
               \rho^{\tau_1} (\bbox{r}_1)
           +    (B -  M \delta_{\tau \tau_1})
               \rho^{\tau_1} (\bbox{r}_1 \sigma, \bbox{r}_1 \sigma)
                                                     \Big]
   \nonumber  \\
         & + & \delta(\bbox{r} - \bbox{r}') \delta_{\sigma -\sigma'}
               \int\text{d}^{3} \bbox{r}_1
               e^{-{{(\bbox{r} - \bbox{r}_1)^2} \over {\mu^{2}}}}
               \sum_{\tau_1} (B - M \delta_{\tau \tau_1}) \rho^{\tau_1}
               (\bbox{r}_1 \sigma, \bbox{r}_1 -\sigma),  \label{6}
    \end{eqnarray}
where $\rho^{\tau}(\bbox{r})$ is the density of nucleons
(\ref{eq320a}) of type $\tau$.

Assuming that we consider only the states which are even with
respect to the time reversal, the density matrix (\ref{eq118a})
obeys the relation (\ref{7}). Consequently, the densities
$\rho^{\tau}(\bbox{r}\sigma,\bbox{r}\sigma)$ for
$\sigma$=$\pm\frac{1}{2}$ are equal to
$\frac{1}{2}\rho^{\tau}(\bbox{r})$, and the densities
$\rho^{\tau}(\bbox{r}\sigma,\bbox{r}-\sigma)$ vanish. Therefore,
the term in (\ref{6}), which is proportional to
$\delta_{\sigma-\sigma'}$, vanishes,  and the contribution of
the direct term to the p-h mean field is the local,
spin-independent potential:
   \begin{equation}\label{10}
   \Gamma^{\tau}_{\text{dir}} (\bbox{r} \sigma, \bbox{r}' \sigma') =
         \delta(\bbox{r} - \bbox{r}')\delta_{\sigma \sigma'}
         {U}(\bbox{r}),
   \end{equation}
where
   \begin{equation}
    {U}(\bbox{r}) =
            \int\text{d}^{3} \bbox{r}_1
              e^{-{{(\bbox{r} - \bbox{r}_1)^2} \over {\mu^{2}}}}
                \Big[ (W  + B /2)\rho(\bbox{r}_1)
                    - (H  + M /2)\rho^{\tau}(\bbox{r}_1) \Big].
   \end{equation}
One should note that due to the locality of the interaction,
the direct term depends only on the local densities.

\subsection{Contribution of the central exchange interaction to the
            p-h mean field}

Due to the locality of the interaction, the contribution of the
exchange term to the p-h mean field involves no integration:
   \begin{eqnarray}
      \Gamma^{\tau}_{\text{exc}} (\bbox{r} \sigma, \bbox{r}' \sigma')  =
          e^{-{{(\bbox{r} - \bbox{r}')^2}\over {{\mu}^{2}}}}
              \sum_{\tau_1} \Big[  && \delta_{\sigma \sigma'}
                    \big( (M - B \delta_{\tau \tau_1})
      \sum_{\sigma_1} \rho^{\tau_1} (\bbox{r} \sigma_1, \bbox{r}' \sigma_1)
                        + (H - W \delta_{\tau \tau_1})
                      \rho^{\tau_1} (\bbox{r} \sigma, \bbox{r}'  \sigma) \big)
                                                             \nonumber \\
                                 + && \delta_{\sigma -\sigma'}
                          (H - W \delta_{\tau \tau_1})
    \rho^{\tau_1} (\bbox{r} \sigma, \bbox{r}' -\sigma) \Big]. \label{12}
   \end{eqnarray}
Here the time-reversal symmetry does not bring any
simplification.  However, a simpler formula is obtained in cases
where $\rho^{\tau_1} (\bbox{r} \sigma, \bbox{r}' \sigma')$ is
real. It follows from Eq.{\ }(\ref{7}) that the densities
$\rho^{\tau_1}(\bbox{r}\sigma,\bbox{r}'\sigma)$ are equal to
$\frac{1}{2}\sum_{\sigma}\rho^{\tau_1}(\bbox{r}\sigma,\bbox{r}'\sigma)$,
which finally leads to
   \begin{eqnarray}
       \Gamma^{\tau}_{\text{exc}} (\bbox{r} \sigma, \bbox{r}' \sigma) &=&
           e^{-{{(\bbox{r} - \bbox{r}')^2} \over {\mu^{2}}}}
               \sum_{\tau_1} \Big[ M + H/2
                                - (B + W/2) \delta_{\tau \tau_1}) \Big]
               \sum_{\sigma_1} \rho^{\tau_1} (\bbox{r} \sigma_1,
                                              \bbox{r}' \sigma_1),
                                                             \label{14a} \\
       \Gamma^{\tau}_{\text{exc}} (\bbox{r} \sigma, \bbox{r}' -\sigma) &=&
           e^{-{{(\bbox{r} - \bbox{r}')^2} \over {\mu^{2}}}}
               \sum_{\tau_1} ( H - W \delta_{\tau \tau_1})
                               \rho^{\tau_1} (\bbox{r} \sigma,
                                              \bbox{r}' -\sigma).
                                                             \label{14b}
   \end{eqnarray}

\subsection{Contribution of the Coulomb interaction to the
            p-h mean field}

Derivation of the direct and exchange Coulomb fields is similar
to the one of the finite range term  (\ref{1a}) with several
additional simplifications.  When the nuclear state is
time-reversal invariant, one obtains the following contributions
to the proton p-h mean field in terms of the proton densities:
 \begin{eqnarray}
     \Gamma^p_{\text{Coul-dir}} (\bbox{r} \sigma, \bbox{r}' \sigma') &=&
        \delta(\bbox{r} - \bbox{r}')\delta_{\sigma \sigma'}
          \int\text{d}^{3} \bbox{r}_1
                   {e^2 \over \vert \bbox{r} - \bbox{r}_1 \vert}
                                 \rho^p(\bbox{r}_1),
                                                  \label{18} \\
     \Gamma^p_{\text{Coul-exc}} (\bbox{r} \sigma, \bbox{r}' \sigma') &=&
                   {e^2 \over \vert \bbox{r} - \bbox{r}'  \vert}
                                 \rho^p (\bbox{r} \sigma, \bbox{r}' \sigma').
                                                  \label{19}
 \end{eqnarray}

\subsection{Contribution of the central interaction to the
            p-p mean field}

The general form of the pairing field is given by
Eq.{\ }(\ref{eq141b}).  In this case the direct and the
exchange contributions are equal.  For the local central force
(\ref{1a}), the total contributions to the p-p mean field have
the form:
   \begin{eqnarray}
       \tilde{h}^{\tau} (\bbox{r} \sigma, \bbox{r}' \sigma)  &=&
           e^{-{{(\bbox{r} - \bbox{r}')^2} \over {\mu^{2}}}}
             \big[ (W - H) \tilde\rho^{\tau}
                         (\bbox{r} \sigma, \bbox{r}' \sigma)
                 - (B - M) \tilde\rho^{\tau}
                         (\bbox{r}' \sigma, \bbox{r} \sigma) \big],
                                                             \label{27} \\
       \tilde{h}^{\tau} (\bbox{r} \sigma, \bbox{r}' -\sigma)  &=&
           e^{-{{(\bbox{r} - \bbox{r}')^2} \over {\mu^{2}}}}
                   (W + B - H - M) \tilde\rho^{\tau}
                         (\bbox{r} \sigma,\bbox{r}' -\sigma).
                                                             \label{28}
   \end{eqnarray}
Again it is to be noted that due to the locality of the
interaction, the corresponding p-p mean fields do not involve
any integration but are proportional to the pairing density
matrices.  In the case considered in this study (time-even
densities), the contribution (\ref{28}) vanishes.

Since the exchange parameter of the zero-range density-dependent
term of the Gogny D1S interaction is fixed at $x_0$=1, this term
does not contribute to the p-p mean field. Moreover, the
spin-orbit and Coulomb terms usually give small contributions as
compared to those of the central force (\ref{1a}).

\narrowtext

\subsection{Numerical methods used for the calculation of the mean fields}

Computation of the exchange p-h mean fields,
Eqs.{\ }(\ref{14a}), (\ref{14b}), and (\ref{19}), and the
pairing fields, Eqs.{\ }(\ref{27}) and (\ref{28}), is
straightforward. It only requires the knowledge of the spatial
spin-dependent non-local particle, $\rho^{\tau} (\bbox{r}
\sigma, \bbox{r}'\sigma')$, and pairing $\tilde\rho^{\tau}
(\bbox{r} \sigma, \bbox{r}'\sigma')$ densities.

Computation of the direct p-h mean field, Eq.{\ }(\ref{10}),
and the direct Coulomb mean field, Eq.{\ }(\ref{18}), is more
complicated since it requires the evaluation of
three-dimensional integrals of the form:
    \begin{eqnarray}
       I_{\mu} (\bbox{r}) & = & \int\text{d}^{3} \bbox{r}'
             e^{-{(\bbox{r} - \bbox{r}')^2 \over \mu^2}}
                   \; \rho (\bbox{r}'),
                                                             \label{30} \\
       I_{C}   (\bbox{r}) & = & \int\text{d}^{3} \bbox{r}'
            { 1 \over { \vert \bbox{r} - \bbox{r}' \vert }}
                   \; \rho (\bbox{r}').
                                                             \label{31}
    \end{eqnarray}

In order to compute $I_{\mu}(\bbox{r})$ of Eq.{\ }(\ref{30})
we note that for the single-particle wave functions expanded in
the harmonic oscillator basis, the local density $\rho
(\bbox{r})$ is the product of a Gaussian factor and of a
polynomial in the spatial coordinates $x_1$,  $x_2$, and  $x_3$
   \begin{equation}
     \rho (\bbox{r}) =
        \exp \left[ - \sum_{k=1}^{3} \left( {x_k \over b_k} \right)^2 \right]
              \; P ( x_1 , x_2 , x_3 )  ,
   \end{equation}
where $b_1$, $b_2$ and $b_3$ are the HO lengths of the basis
states.  Consequently, this integral can be evaluated exactly
using the Gauss-Hermite quadrature.

The computation of the Coulomb integral (\ref{31}) is more
difficult due to the infinite range of the Coulomb force. The
method we have used consists of expressing the Coulomb force as
a sum of Gaussians:
   \begin{equation}\label{31c}
      { 1 \over { \vert \bbox{r} - \bbox{r}' \vert }} =
         { 2 \over \sqrt{\pi}} \int_{0}^{\infty} {\text{d} \mu \over \mu^{2}}
                e^{-{(\bbox{r} - \bbox{r}')^2 \over \mu^2}}.
   \end{equation}
Then one obtains
   \begin{equation}\label{32}
       I_{C} (\bbox{r})  =
          { 2 \over \sqrt{\pi}} \int_{0}^{\infty} {\text{d} \mu \over \mu^{2}}
              I_{\mu} (\bbox{r})
   \end{equation}
where $I_{\mu} (\bbox{r})$ is given by Eq.{\ }(\ref{30}).  In
order to perform the remaining one-dimensional integration, the
variable $\mu$ is changed to
   \begin{equation}\label{32a}
      \xi = b / \sqrt{b^2 + \mu^{2}} ,
   \end{equation}
where $b$ is the largest of the three harmonic-oscillator
lengths $b_1$, $b_2$ and $b_3$.  This change of variable is very
convenient, since then the range of integration becomes [0, 1].
The integral in Eq.{\ }(\ref{32}) can be very accurately
computed using the Gauss-Legendre quadrature.

\section{The energy cut-off}
\label{appB}

Calculations which are based on the schematic pairing
interaction, or on the contact force [Eq.{\ }(\ref{DIDI})]
require a finite space of states in the p-p channel. For such
interactions, when this space is increased, the pairing energy
diverges for any fixed strength of the interaction.  This
divergence is a well-known effect \cite{[Mig67]} related to the
fact that for the contact interactions the matrix elements do
not  (or too slowly) decrease with the excitation energy. This
is not a case for finite-range interactions, such as Gogny, for
which  the pairing energy converges to a finite value.

Since it is considerably easier to use the zero-range
interactions than the finite-range interactions, one applies the
former ones in a limited configuration space determined by a
cut-off in the single-particle energy or in the
single-quasiparticle energy. This can be understood as a
phenomenological introduction of the finite range
\cite{[Ber91]}.  There are two other arguments in favor of such
a procedure.  First, the scattering of particles in the nuclear
medium at very high energies (or at very small distances) is
very little known, and the particular form offered by any
phenomenological finite-range force is very uncertain. Second,
the single-particle wave functions are primarily determined by
the p-h channel of interaction, and they, in general, spread
throughout distances which are much larger than the range of the
p-p interaction. Therefore, physical differences between the
zero- and short-range p-p forces cannot be expected to be very
pronounced.

Within the BCS approximation, and assuming a constant density of
the single-particle states at large energies, one can derive
\cite{[Mig67],[Bra74]} a prescription to renormalize the
strength of the p-p interaction in such a way that the pairing
gap $\Delta$ does not depend on the energy cut-off.  Suppose
that the single-particle states with energies
$-\epsilon_l$$\leq$$\epsilon$$-$$\lambda$$\leq$$\epsilon_u$ are
used to solve the BCS equations for the force of strength $V_0$.
Then, within the specified approximations, the following
relation holds:
   \begin{equation}\label{V0}
    V_0 = -\frac{C_0}{\ln(2\sqrt{\epsilon_l \epsilon_u}/\Delta)},
   \end{equation}
where $C_0$$\simeq$300\,MeV\,fm$^3$ is a constant inversely
proportional to the density of single-particle states near the
Fermi energy.  In other words, for given values of $C_0$ and
$\Delta$, Eq.{\ }(\ref{V0}) gives values of $V_0'$ for any other
choice of the cut-off energies $\epsilon_l'$ and $\epsilon_u'$.

Since in the present study we use the HFB method instead of the
BCS approximation, and since the density of states can hardly be
considered to be independent of energy (actually for fixed
$R_{\text{box}}$ it increases as $\sqrt{\epsilon}$), the formula
(\ref{V0}) cannot be directly used. However, the question as to
what extent the pairing strengths can be renormalized for a
zero-range p-p interaction can be addressed by analyzing the
numerical solutions of the HFB equations.

{}Figure{\ }\ref{FIG29} (top panel) shows the neutron pairing
energies $E_{\text{pair}}$ [Eq.{\ }(\ref{epair})] calculated
for the SkP$^\delta$ interaction which uses the contact p-p
interaction (\ref{DIDI}) with $V_0$=$-$160\,MeV\,fm$^3$.  It
should be recalled at this point that for all coordinate-space
HFB calculations presented in this study, the cut-off energy,
$E_{\text{max}}$, depends on the quantum numbers ${\ell}j$
(cf.{\ }Ref.{\ }\cite{[Dob84]}).  In the tin nuclei,
$E_{\text{max}}$ decreases from about 40\,MeV for the $s_{1/2}$
states to zero for the $k_{17/2}$ states.  In
Figs.{\ }\ref{FIG29}-\ref{FIG31} different curves correspond
to different cut-off energies, $E_{\text{max}}'=E_{\text{max}}
+\Delta E_{\text{max}}$ ($\Delta E_{\text{max}}$ varied between
0 and 40\,MeV).  Hence, $\Delta E_{\text{max}}$=40\,MeV
 corresponds to the cut-off energy $E_{\text{max}}'$ of 80\,MeV
for the $s_{1/2}$ states, and 40\,MeV for the largest values of
$\ell$.

As expected, pairing energies depend significantly on the
cut-off energy. Comparing results for $\Delta
E_{\text{max}}$=40\,MeV with those for $\Delta
E_{\text{max}}$=0, one obtains  differences of $E_{\text{pair}}$
of the order of 10$\div$20\,MeV in the mid-shells.  Due to the
self-consistent readjustment of the p-h and p-p energies, the
corresponding differences in the total energies
(Fig.{\ }\ref{FIG30}) are much smaller, 2$\div$4\,MeV, but still
significant.

In the bottom panels of Figs.{\ }\ref{FIG29} and \ref{FIG30} are
shown similar results for the renormalized strengths of the
contact force (\ref{DIDI}).  The values of $V_0$, quoted in the
caption of Fig.{\ }\ref{FIG29}, have been obtained by requiring
that the average neutron pairing gap in $^{120}$Sn,
$\langle\Delta_N\rangle$=1.245\,MeV, does not depend on $\Delta
E_{\text{max}}$. With such renormalized interactions, one
obtains very small changes of  total energies
(Fig.{\ }\ref{FIG30}, bottom panel).  The largest deviations do
not exceed 200\,keV and 800\,keV in stable and exotic isotopes,
respectively, and can be safely disregarded when compared to all
other uncertainties of methods used to extrapolate to unknown
nuclei, or when studying the separation energies.

{}Figure{\ }\ref{FIG31} shows the effective pairing-interaction
strengths defined schematically as
$G_{\text{eff}}$=$-$$\langle\Delta\rangle^2/E_{\text{pair}}$.
The top panel presents the results obtained for
$V_0$=$-$160\,MeV\,fm$^3$ and for different values of  $\Delta
E_{\text{max}}$.  One can see that the dependence on the cut-off
energy is very weak, and the $\Delta E_{\text{max}}$-dependence
of $\langle\Delta\rangle^2$ and $E_{\text{pair}}$ cancels out in
$G_{\text{eff}}$.  (At $N$=82 the pairing gap and the pairing
energy both vanish, and hence the $G_{\text{eff}}$ values cannot
be calculated.) A fixed value of $V_0$ gives, therefore, a
well-defined, cut-off-independent value of the effective pairing
strength for every isotope. This result, together with the
analysis of pairing gap distributions in Sec.{\ }\ref{sec4a},
demonstrates that calculations employing the volume contact p-p
interaction are, in many respects, similar to those with the
schematic seniority-pairing force (cf., however,
Sec.{\ }\ref{sec3c}).

The values of $G_{\text{eff}}$ monotonically decrease with
increasing neutron number $N$. The obtained dependence can be
very well described by the simple Madland-Nix formula
\cite{[Mad88]}, $G$=11\,MeV/(11+$N$), while the Jensen-Miranda
formula \cite{[Jen86]}, $G$=0.18\,MeV[1$-$1.2$I$$-$2.8$I^2]$
($I$=($N$$-$$Z$)/$A$), gives a much faster decrease of
$G_{\text{eff}}$ with $N$.  (In a recent study \cite{[Kuz94]},
based on a schematic finite-range force, the isospin dependence
of $G$ has been discussed. The authors found no sign of the
$I^2$ term suggested in Ref.{\ }\cite{[Jen86]}.  This probably
explains the disagreement seen in Fig.{\ }\ref{FIG31}.) In both
expressions we have normalized the multiplicative constants to
obtain $G$=0.18\,MeV at $N$=$Z$=50.

The bottom panel of Fig.{\ }\ref{FIG31} shows similar results,
but for the renormalized values of $V_0$, quoted in the caption
of Fig.{\ }\ref{FIG29}.  The insert shows the values of $V_0$
(dots) as function of $\Delta E_{\text{max}}$ compared with the
simple fit by the formula (\ref{V0}) with
$\epsilon_u$=$\epsilon_l$+$\Delta E_{\text{max}}$, for
$\epsilon_l$=40\,MeV, $C_0$=430\,MeV\,fm$^3$, and
$\Delta$=5.58\,MeV (solid line).  One can see that the generic
dependence of the renormalized values of $V_0$ on $\Delta
E_{\text{max}}$ is fairly well reproduced, although the
numerical constants $C_0$ and $\Delta$ obtained from the fit do
not exactly correspond to the values inferred from the BCS
theory with a constant density of single-particle states.


\clearpage

\begin{table}[ht]
\caption[TT]{
Results of the HFB calculations with SkP force for the $s_{1/2}$
neutrons in $^{120}$Sn.  {}For the $n$th quasiparticle state,
$E^{\text{HFB}}_n$ and $N_n$ are the quasiparticle energy and
the norm of the lower component, respectively. For the $\mu$th
canonical-basis state, $v_\mu^2$ is the occupation probability,
$\epsilon_\mu$ and $\Delta_\mu$ are, respectively, the average
values of the p-h and p-p mean-field Hamiltonians
[Eq.{\ }(\protect\ref{eq148})] and $E^{\text{can}}_\mu$ is the
BCS-like quasiparticle energy defined in
Eq.{\ }(\protect\ref{eq150}).  All energies are in MeV.
}
\begin{tabular}{c|rc||c|rl@{}rr}
\multicolumn{3}{c||}{Quasiparticle states} &
\multicolumn{5}{c}  {Canonical-basis states} \\
\hline
  $n$               & $E^{\text{HFB}}_n~~$          &  $N_n$        &
  $\mu$             & $E^{\text{can}}_\mu~~$        &  $~v^2_\mu$   &
  $\epsilon_\mu~~$  & $\Delta_\mu~$                                       \\
\hline
 11 &   54.27   & 0.0001  & 11 & 47.27   & 0.0    &    39.32  & $-$0.37   \\
 10 &   44.38   & 0.0001  & 10 & 78.07   & 0.0    &    70.12  & $-$0.03   \\
  9 &   35.44   & 0.0006  &  9 & 73.14   & 0.0    &    65.20  & $-$0.81   \\
  7 &   27.49   & 0.0008  &  8 & 54.84   & 0.0    &    46.89  &    0.13   \\
  6 &   20.82   & 0.0019  &  7 & 55.22   & 0.000003&   47.27  &    0.07   \\
  4 &   15.58   & 0.0008  &  6 & 62.46   & 0.00003&    54.51  & $-$0.76   \\
  3 &   11.61   & 0.0006  &  5 & 38.44   & 0.0001 &    30.48  &    0.76   \\
  2 &    8.92   & 0.0002  &  4 & 20.50   & 0.0005 &    12.54  &    0.99   \\
  1 &    1.54   & 0.8372  &  3 &  2.36   & 0.8362 &  $-$9.88  &    1.35   \\
  5 &   17.60   & 0.9942  &  2 & 20.06   & 0.9990 & $-$27.96  &    1.27   \\
  8 &   31.64   & 0.9992  &  1 & 29.94   & 0.9999 & $-$37.88  &    0.45   \\
\end{tabular}
\label{TAB01}
\end{table}


\renewcommand{\topfraction}{0.0}
\renewcommand{\bottomfraction}{0.0}
\renewcommand{\floatpagefraction}{0.0}
\setcounter{topnumber}{20}
\setcounter{bottomnumber}{20}
\setcounter{totalnumber}{20}

\clearpage

\widetext
\begin{figure}[ht]
\caption[FF]{
Plot of the two-neutron separation energies $S_{2n}$ for all
particle-bound even-even nuclei with $A$$\geq$16 and
$N$$\leq$208 calculated within the spherical HFB+SkP approach.
Results for 1905 nuclei are shown using the color codes
spaced by 0.4\,MeV for $S_{2n}$$<$4\,MeV and by 2\,MeV
for $S_{2n}$$>$4\,MeV.
}
\label{FIG26}
\end{figure}
\narrowtext

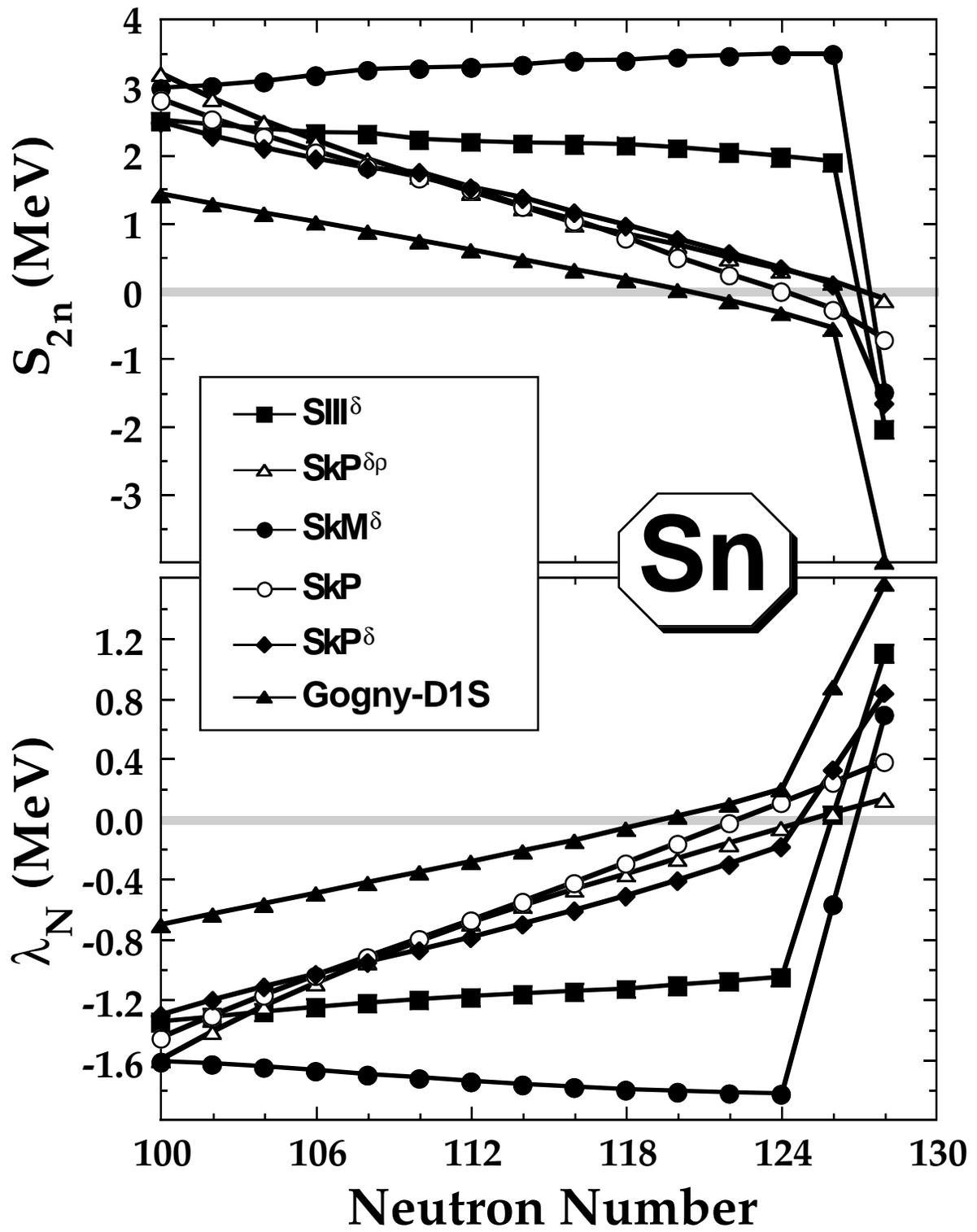
\begin{figure}[ht]
\caption[FF]{
Two-neutron separation energies $S_{2n}$ (top) and Fermi
energies $\lambda_N$ (bottom) for the Sn isotopes,  calculated
in the HFB approach with several Skyrme interactions and the
Gogny-D1S interaction.
}
\label{FIG02}
\end{figure}

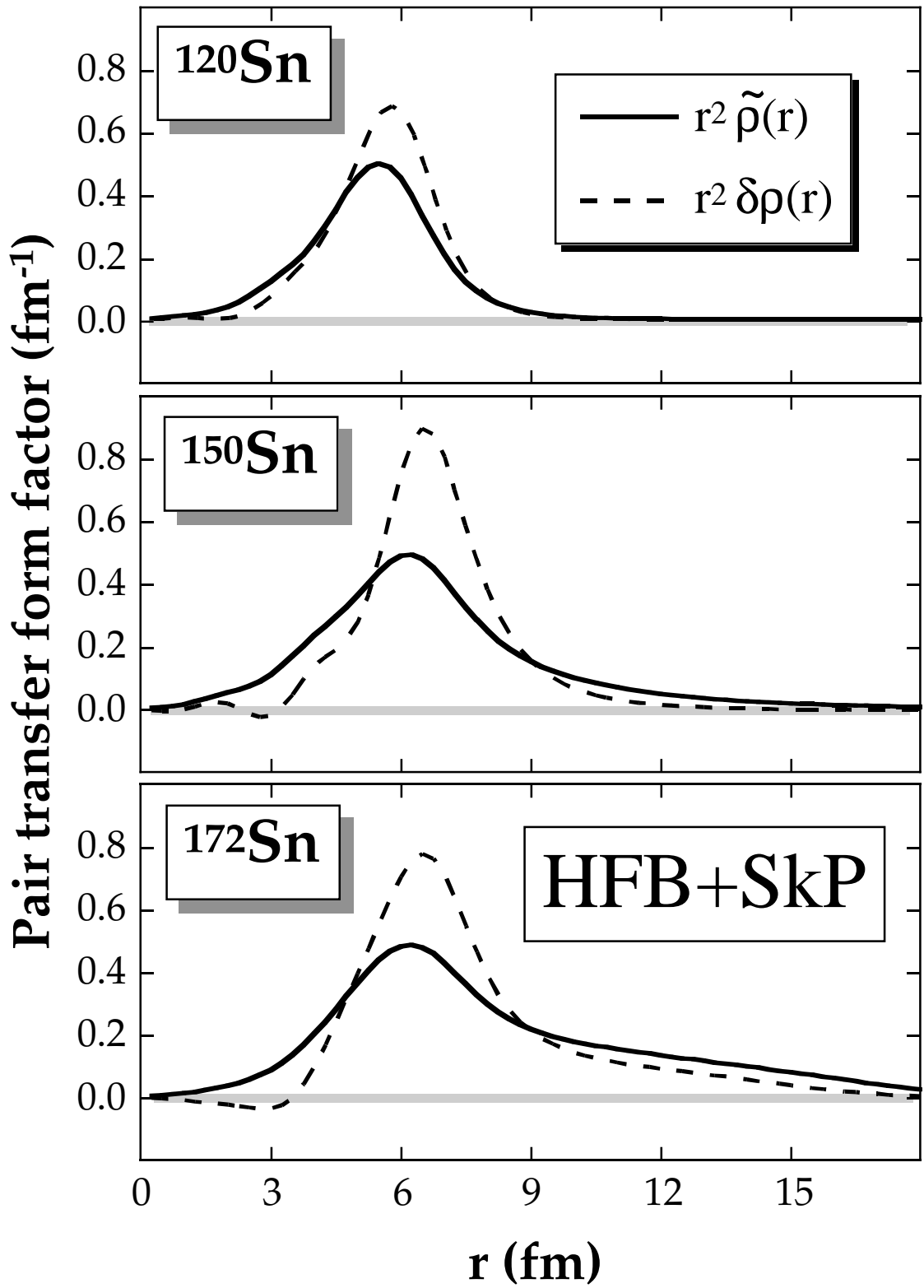
\begin{figure}[ht]
\caption[FF]{
Pair transfer form factor, $r^2\tilde\rho(r)$, calculated
directly from the HFB pairing density $\tilde\rho(r)$ (solid
lines), compared with the macroscopic form factor calculated
from the derivative of the particle density $\delta\rho(r)$
(dashed lines).
}
\label{FIG32}
\end{figure}

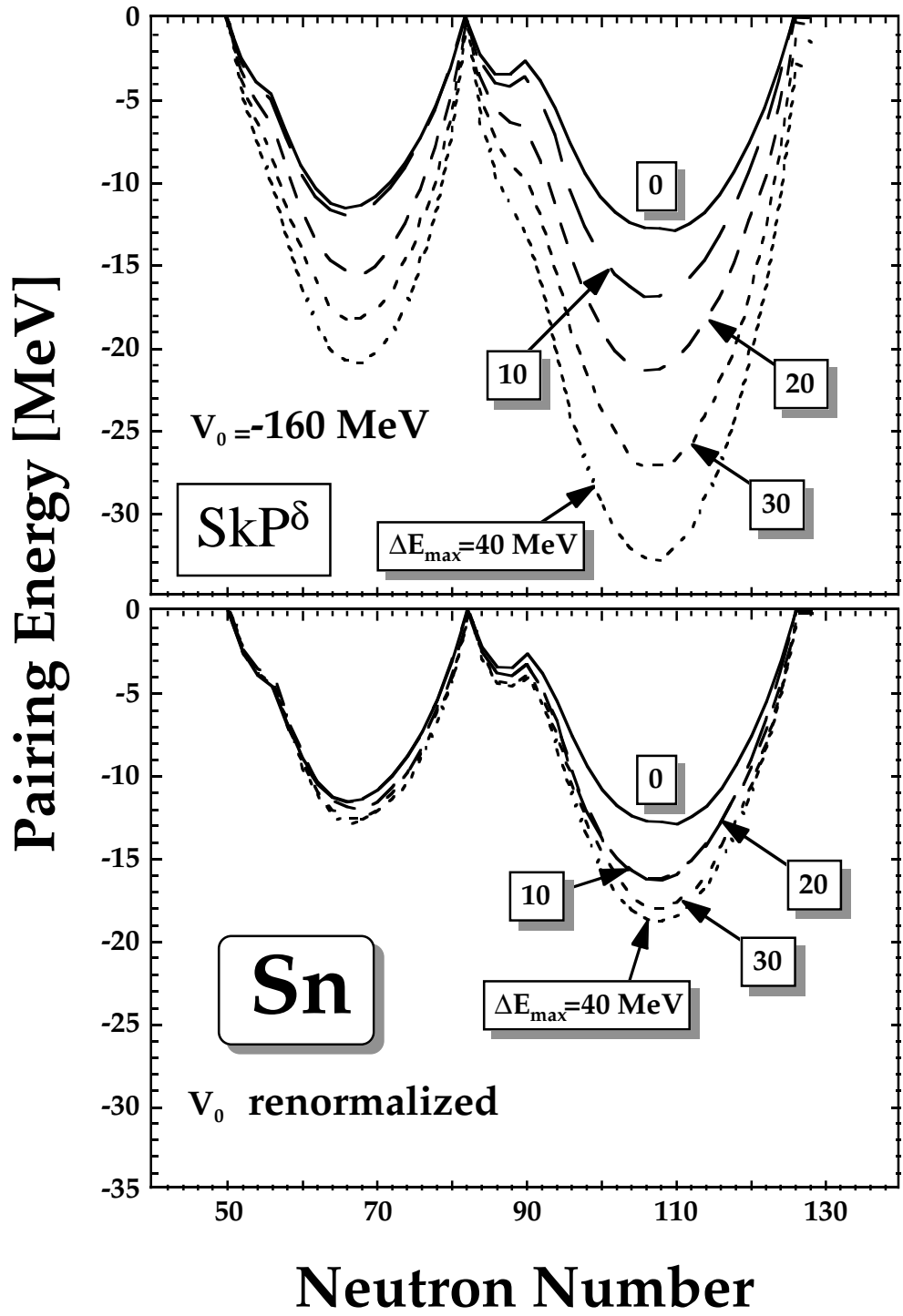
\begin{figure}[ht]
\caption[FF]{
Pairing energies $E_{\text{pair}}$ in the tin isotopes
calculated within the HFB+SkP$^\delta$ model.  Top panel shows
the results for the fixed interaction strength
$V_0$=$-$160\,MeV\,fm$^3$ and for several cut-off energies
$\Delta E_{\text{max}}$ {\em added} to the usual
${\ell}j$-dependent cut-off energy $E_{\text{max}}$
\protect\cite{[Dob84]}. Bottom panel shows similar results when
the values of $V_0$ are renormalized to $-$158.64, $-$149.57,
$-$145.41, and $-$142.01\,MeV\,fm$^3$ for $\Delta
E_{\text{max}}$=10, 20, 30, and 40\,MeV, respectively.
}
\label{FIG29}
\end{figure}

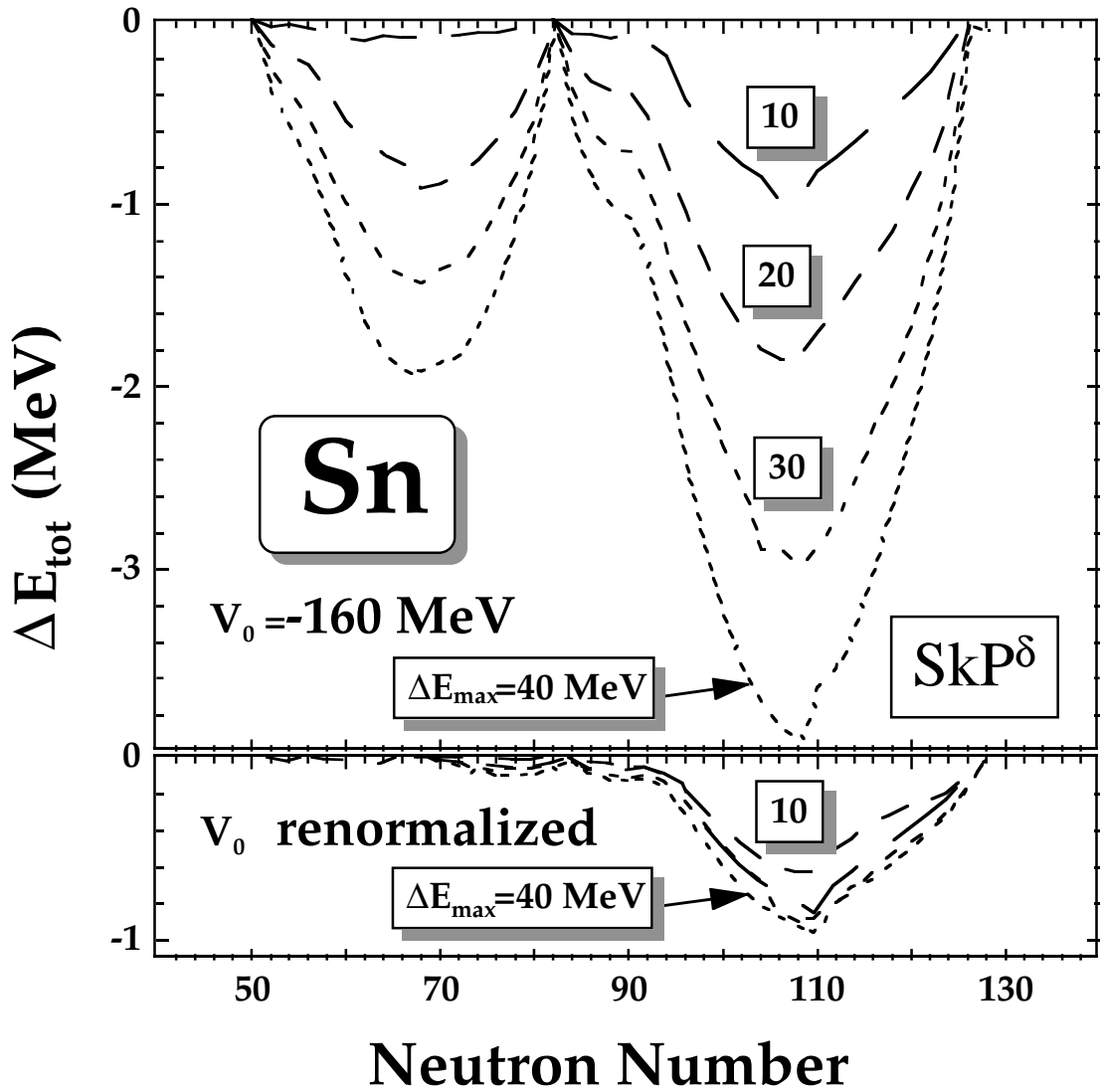
\begin{figure}[ht]
\caption[FF]{
Same as in Fig.{\ }\protect\ref{FIG29}, but for the total energy
relative to that  obtained with $\Delta E_{\text{max}}$=0.
}
\label{FIG30}
\end{figure}

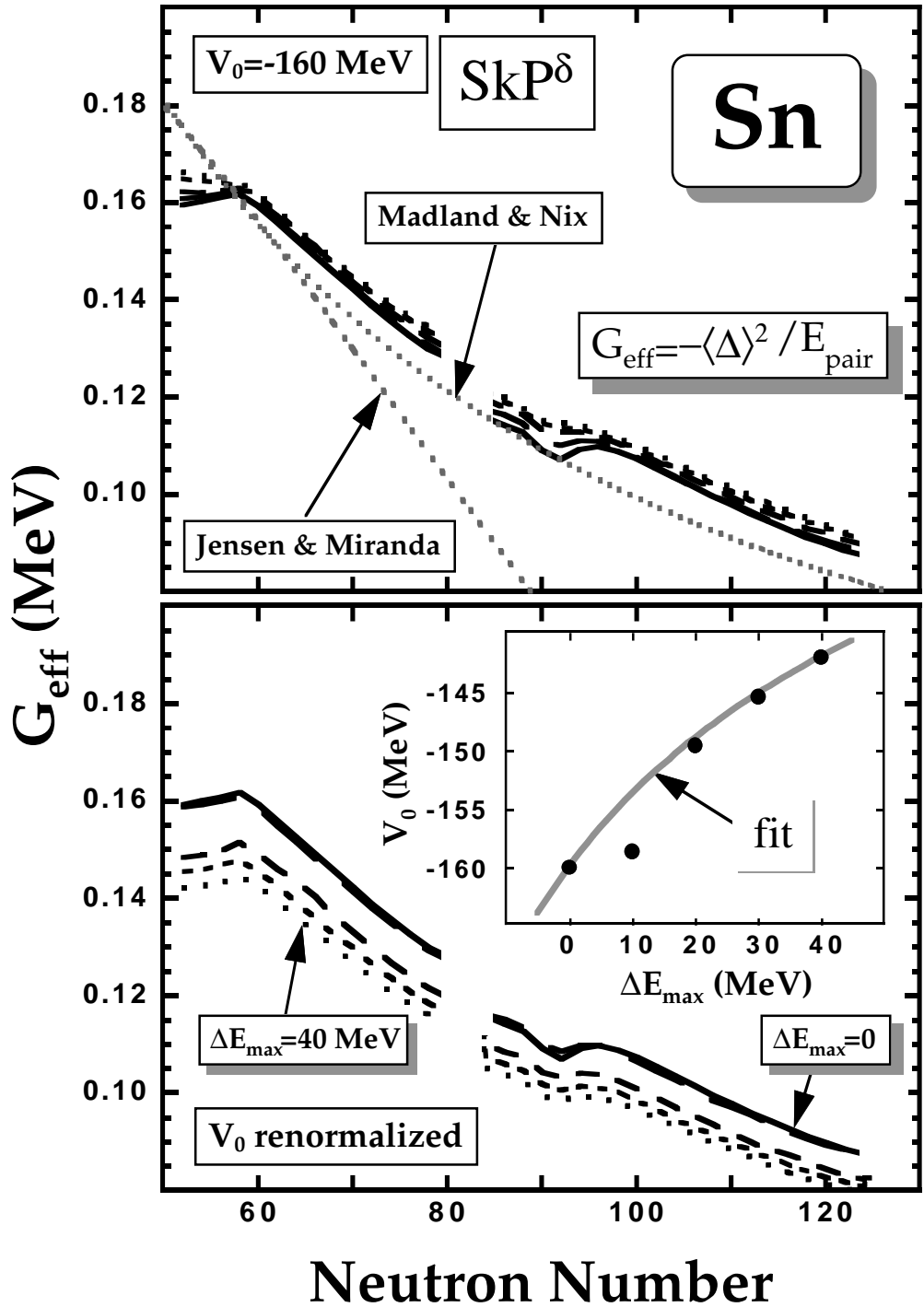
\begin{figure}[ht]
\caption[FF]{
Same as in Fig.{\ }\protect\ref{FIG29}, but for the effective
pairing strength defined as
$G_{\text{eff}}$=$-$$\langle\Delta\rangle^2/E_{\text{pair}}$.
The insert shows the renormalized strength $V_0$ compared with
that given by Eq.{\ }(\protect\ref{V0}).
}
\label{FIG31}
\end{figure}

%
%

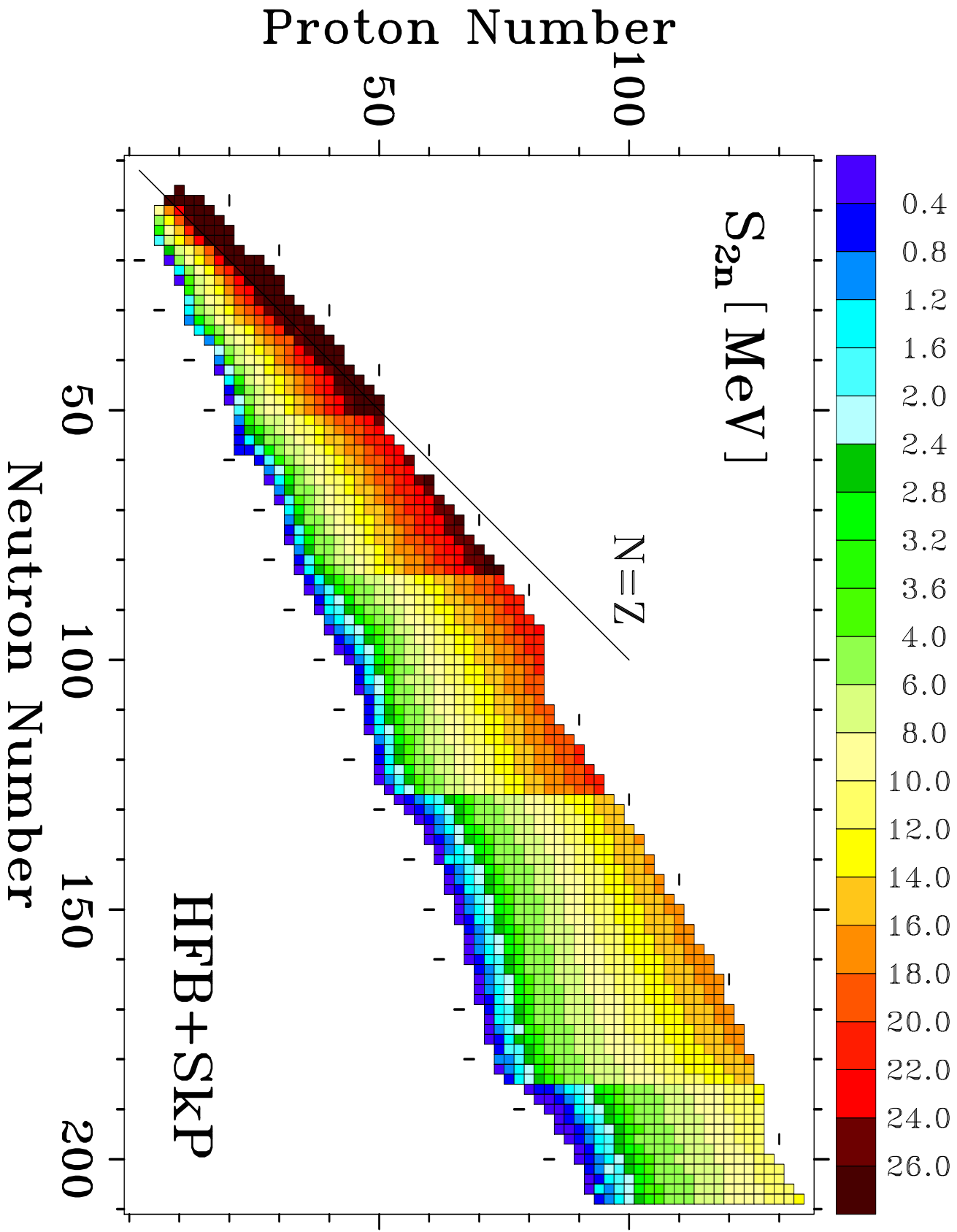
\end{document}